\documentclass[12pt]{iopart}
\usepackage[pdftex]{graphicx,color}
\usepackage{natbib}
\usepackage{indentfirst}
\usepackage{amssymb}
\usepackage{url}
\usepackage{here}
\usepackage{comment}

\eqnobysec

\makeatletter
\def\subsubsection{\@startsection {subsubsection}{3}{\z@}{2.5ex plus -1ex minus-.2ex}
{1ex plus 1.2ex}
{\small\it}
}

\begin{document}

\title[Evolution of the Angular Momentum of Molecular Cloud Cores]{Evolution of the Angular Momentum of Molecular Cloud Cores Formed from Filament Fragmentation}

\author{\small Yoshiaki Misugi$^{1,2}$, Shu-ichiro Inutsuka$^1$, and Doris Arzoumanian$^{3,4}$}
\address{$^1$ Department of Physics, Graduate School of Science, Nagoya University, 464-8602 Nagoya, Japan}
\address{$^2$ Graduate Schools of Science and Engineering, Kagoshima University, 890-8580 Kagoshima, Japan}
\address{$^3$Division of Science, National Astronomical Observatory of Japan, 2-21-1 Osawa, Mitaka, Tokyo 181-8588, Japan}
\address{$^4$ Aix Marseille Univ, CNRS, CNES, LAM, Marseille, France}

\ead{misugi@sci.kagoshima-u.ac.jp}
\vspace{10pt}
\begin{indented}
\item[]Dec 2022
\end{indented}

\begin{abstract}
The angular momentum of molecular cloud cores plays an essential role in the star formation process. However, the time evolution of the angular momentum of molecular cloud cores is still unclear. In this paper, we perform three-dimensional simulations to investigate the time evolution of the angular momentum of molecular cloud cores formed through filament fragmentation. As a result, we find that most of the cores rotate perpendicular to the filament axis. The mean angular momentum of the cores changes by only around 30\% during the initial stage of their formation process and then remains almost constant.  In addition, we analyze the internal angular momentum structure of the cores. Although the cores gain angular momentum with various directions from the initial turbulent velocity fluctuations of their parent filaments, the angular momentum profile in each core converges to the self-similar solution. We also show that the degree of complexity of the angular momentum structure in a core slightly decreases with time. Moreover, we perform synthetic observations and show that the angular momentum profile measured from the synthetic mean velocity map is compatible with the observations when the filament inclination is taken into account. The present study suggests a theory of core formation from filament fragmentation where the angular momentum structures of the cores are determined by the velocity fluctuation along the filaments and both are compatible with the observations.  This theory also provides new insights on the core properties that could be observationally tested.  
\end{abstract}

\section{Introduction}
The angular momentum of molecular cloud cores plays an important role in the star formation process, being linked to the origin of the outflow and the jet, and the protoplanetary disk formation.  The outflow and the jet are launched by the magnetic field lines deformed by the core rotation \citep{Tomisaka2002}, and the angular momentum of the protoplanetary disk is inherited from their parental molecular cloud core. In addition, the angular momentum of molecular cloud cores is one of the important parameters for the multiplicity of a stellar system (single star, binary, multiple stars) and their separation \cite[e.g.,][]{Machida2008}. Moreover, revealing the angular momentum profile in the cores is crucial for the star and planet formation process, because it determines how much of the angular momentum accretes onto the circumstellar disk. Therefore, understanding how molecular cloud cores acquire their angular momentum and how the angular momentum evolves are key questions in the star and planet formation studies. \par
Recent observational results derived from the analysis of Herschel data revealed that most stars are born along molecular filaments \citep{Andre2010,Arzoumanian_2011,Konyves2015,Pineda2022} and the characteristic width of filaments is $\sim 0.1$ pc \citep{Arzoumanian_2011,Arzoumanian2019,Koch2015,Andre2022}. In addition, observations show that prestellar cores and protostars are primarily embedded in thermally critical and supercritical filaments (${\rm M_{\rm line}} \gtrsim {\rm M_{\rm line,crit}}$) \citep{Andre2010,Tafalla_2015}. $ {\rm M_{\rm line,crit}}$ is the critical line mass for isothermal unmagnetized cylindrical filaments \citep{Stodolkiewicz1963,Ostriker1964}. If the line mass is larger than this critical value, the filament cannot be in equilibrium against fragmentation and collapses if there is no support from magnetic field and internal turbulence. These observations indicate that the line mass ${\rm M_{\rm line}}$ of the filaments is an important parameter for the filament fragmentation and hence the star formation process. Theoretically, star forming cores are expected to be formed along a thermally critical/supercritical filament through self-gravitational fragmentation \citep{Inutsuka_Miyama1997}. Therefore, if most of the star forming cores are formed along critical/supercritical filaments, the theory of core formation out of filament fragmentation should explain the origin and time evolution of the angular momentum of the cores. \par
The angular momentum of cores has been derived from molecular line observations using, for example, the ${\rm NH}_3$ transition \citep{Goodman1993} and ${\rm N}_2{\rm H}^+$ line \citep{Caselli2002,Tatematsu2016}. More recently, \cite{Punanova2018} studied a sample of cores in the L1495 filament in the Taurus molecular cloud using ${\rm N}_2{\rm D}^+(2-1), {\rm N}_2{\rm H}^+(1-0),{\rm DCO}^+(2-1)$, and ${\rm H}^{13}{\rm CO}^+(1-0)$. These observational results have mainly shown three properties of the specific angular momentum of cores $j$ (angular momentum per unit of mass); (1) the range of the specific angular momentum of the cores is $j=10^{20-22}{\rm cm^2 s^{-1}}$, (2) it is an increasing function of the core mass $M$, $j \propto M^{0.5-0.9}$, and (3) the core radius $R$, $j \propto R^{1.6-2.4}$. The angular momentum profile in protostellar envelopes has been also derived \citep{Pineda2019,Gaudel2020}.  \cite{Pineda2019} and \cite{Gaudel2020} show that the angular momentum profile in the protostellar envelopes follows the power law $j$-$r$ relation, $j \propto r^{1.80 \pm 0.04}$ and $j \propto r^{1.6 \pm 0.2}$ (for $r$ larger than a few thousands AU), respectively, similar to the relations observed for prestellar cores. While this slope index is often interpreted as being between indexes resulting from turbulence ($=1.5$) and solid-body rotation ($=2.0$), the origin of the $j$-$r$ relation is not well understood.\par
The time evolution and the origin of the core angular momentum have been investigated using numerical simulations which start from molecular cloud scales \citep{Offner2008,Dib2010,Chen2018,Ntormousi2019,Kuznetsova2019,Kuznetsova2020,Arroyo2021}. However, these latter studies do not focus on the filament fragmentation process. They build an initial condition of a molecular cloud with supersonic turbulence and perform three-dimensional simulations. The simulations of \cite{Chen2018} include large scale converging flows. \cite{Dib2010}, \cite{Chen2018}, and \cite{Ntormousi2019} also include the effect of magnetic field. In \cite{Offner2008} and \cite{Dib2010}, their resultant angular momentum of cores is smaller than the observed angular momentum. \cite{Dib2010} claimed that the observations of the angular momentum of cores tend to overestimate the intrinsic core angular momentum by an order of magnitude. However, \cite{Zhang2018} showed that the observations underestimate the core angular momentum within a factor of two or three. On the other hand, \cite{Chen2018} showed the $j$-$R$ relation measured in their simulations, and their results seem to be consistent with observations. \cite{Ntormousi2019} also showed the $j$-$R$ relation, but their resultant angular momentum seems to be larger than the observations. \cite{Kuznetsova2019} claimed that the median of angular momentum is consistent with the observations, but they did not show the $j$-$R$ relation. \cite{Arroyo2021} investigate the time evolution of the angular momentum of the cores using smoothed particle hydrodynamical (SPH) simulations. They showed that the cores evolve to the observed $j$-$R$ relation on the $j$-$R$ diagram as time progresses. They also suggest that the angular momentum is transferred by the Reynolds stress and the pressure gradient terms in the momentum equation. While \cite{Offner2008}, \cite{Dib2010}, \cite{Chen2018}, \cite{Ntormousi2019} consider that the origin of the core angular momentum is the turbulence, \cite{Kuznetsova2019} claimed that the cores acquire their angular momentum through gravitational torques between high density regions.\par
As described above, the time evolution and the origin of the angular momentum of cores are still unclear. Especially, the relation between the core angular momentum and the filamentary structures in the molecular cloud is not well understood because in previous works, the initial condition is a molecular cloud and the focus is not on the core formation through filament fragmentation. In addition, those studies did not investigate the internal structure of angular momentum of cores in detail. Therefore, the evolution of the angular momentum profile in the cores formed  through the filament fragmentation process is still unclear.\par
To investigate the evolution and origin of the core angular momentum, the initial condition of filament fragmentation, for example the statistical properties of the filament density and velocity fluctuations, is important. The observations of molecular clouds reveal that the power spectrum of column density in molecular clouds is consistent with Kolmogorov turbulence \citep{Miville2010,Roy2019}. In addition, The observations also showed that the power spectrum of line mass fluctuations along the filaments also follow Kolmogorov turbulence \citep{Roy2015,Arzoumanian2021,Arzoumanian2022}. Moreover, \cite{Inutsuka2001} showed that the observed core mass function can be explained analytically using Press-Schechter formalism if the power spectrum of the line mass fluctuations in the filament follows a Kolmogorov-like slope. These results indicate that the Kolmogorov turbulence is appropriate for the initial condition of filamentary molecular clouds.\par
\cite{Misugi2019} focused on the filament fragmentation process and investigated the origin of the core angular momentum resulting from that fragmentation. As a result, they showed that if the initial turbulent velocity field along the filament follows the Kolmogorov turbulence, the resultant core angular momentum is compatible with the observations. They also suggested that the initial sub (tran) sonic velocity dispersion for a critical filament ($M_{\rm line} \sim {\rm M}_{\rm line, crit}$) can explain the observed core angular momentum. This velocity dispersion is consistent with observations of filaments with line masses close to the critical value \citep{Hacar2011,Arzoumanian2013,Hacar2016}. However, since \cite{Misugi2019} assumed the conservation of angular momentum during the core formation phase to calculate the core angular momentum analytically, the time evolution of the core angular momentum during the filament fragmentation was not studied. Therefore, in this paper, we investigate the various properties and the evolution of the angular momentum of cores formed through filament fragmentation using SPH simulations.\par
The structure of the paper is as follows; the method of our calculation is given in Section 2. In Section 3 we show our results. Section 4 presents a discussion of our results. We summarize this paper in Section 5.

\section{Numerical Setup}
We use the Godunov-type smoothed particle hydrodynamical (SPH) method \citep{Inutsuka2002} to solve the following equation of motion:
\begin{eqnarray}
\frac{d {\bi v}}{d t}=-\frac{1}{\rho} \nabla P+\nabla \int d x^{\prime 3} \frac{G \rho\left(x^{\prime}\right)}{\left|x-x^{\prime}\right|},
\label{eq:eom}
\end{eqnarray}
where $\nabla $ is the derivative with respect to $x$. We have used the Framework for Developing Particle Simulator \cite[FDPS: ][]{Iwasawa2016} in our SPH code to accelerate the calculation. We have applied a periodic boundary condition in the $z$-direction which is parallel to the filament axis. We adopt an isothermal equation of state. We use the Barnes-Hut tree algorithm \citep{Barnes1986} with opening angle of 0.4 to solve the gravity. We adopt the hydrostatic equilibrium density profile of a filament as initial condition:
\begin{eqnarray}
\rho(r)=\rho_{\rm c0}\left[1+\left(\frac{r}{H_{0}}\right)^{2}\right]^{-2},
\label{eq:ostini}
\end{eqnarray}
where $\rho_{\rm c0}$ is the density along the axis, $r$ is the radius in the cylindrical coordinate system, and $H_{0}$ is the scale height defined by
\begin{eqnarray}
H_{0} \equiv \sqrt{\frac{2 c_{\rm s}^{2}}{\pi G \rho_{\rm c0}}},
\label{eq:scaleheight}
\end{eqnarray}
where $c_{\rm s}=0.2\  {\rm km \ s^{-1}}$ is the sound speed for $T=10 \  {\rm K}$. We consider an unmagnetized isothermal model filament with a width of 0.1 pc \citep{Arzoumanian_2011,Arzoumanian2019} and a line mass equal to the critical line mass, ${\rm M}_{\rm line,crit}=18 \ {\rm M_{\odot} pc^{-1}}$ for $T=10$ K. In this paper we choose $H_{0} = 0.05\ {\rm pc}$. 
To mimic the periodic boundary condition in the $z$-direction, we put $N$ copies of the filament on both sides. In this paper, we use $N=4$. 

First we numerically generate the velocity field $\bi{v}$ in the filament following the method described in, e.g., \citet{Dubinski1995}
\begin{eqnarray}
\bi{v}(\bi{x})=\sum_{\bi{k}} \bi{V}(\bi{k}) \exp(\rmi \bi{k}\cdot \bi{x}),
\label{eq:velofieldfourier}
\end{eqnarray}
where $\bi{V}(\bi{k})$ is the Fourier transform, $\bi{k}$ is the wave vector. The detailed description is shown in \cite{Misugi2019}. We use $\sigma=c_{\rm s}$ for our fiducial model, where $\sigma$ is the three-dimensional velocity dispersion. Note that this transonic velocity turbulence is consistent with the recent measurement of fluctuations of line of sight velocity along critical filaments in nearby molecular clouds \citep{Hacar2011,Arzoumanian2013,Hacar2016}. Misugi et al (2019) concluded that, if the filament is embedded in a molecular cloud with 3D Kolmogorov turbulence, the resultant core angular momentum is consistent with the observations, and the velocity power spectrum along the filament axis follows the 1D Kolmogorov power spectrum. Hence, we adopt the Kolmogorov turbulence as the initial velocity field in this paper. Although Misugi et al. (2019) found that a Kolmogorov-like, anisotropic power spectrum model was also compatible with the observed angular momentum of the cores, in this paper we focus on the isotropic Kolmogorov turbulence, namely we assume that $P(k)dk= \left<|V(k)|^2\right> dk \propto k^{-5/3} dk$ in the filament.

\section{Results}

\begin{figure}[t]
\centering
\includegraphics[width=18cm]{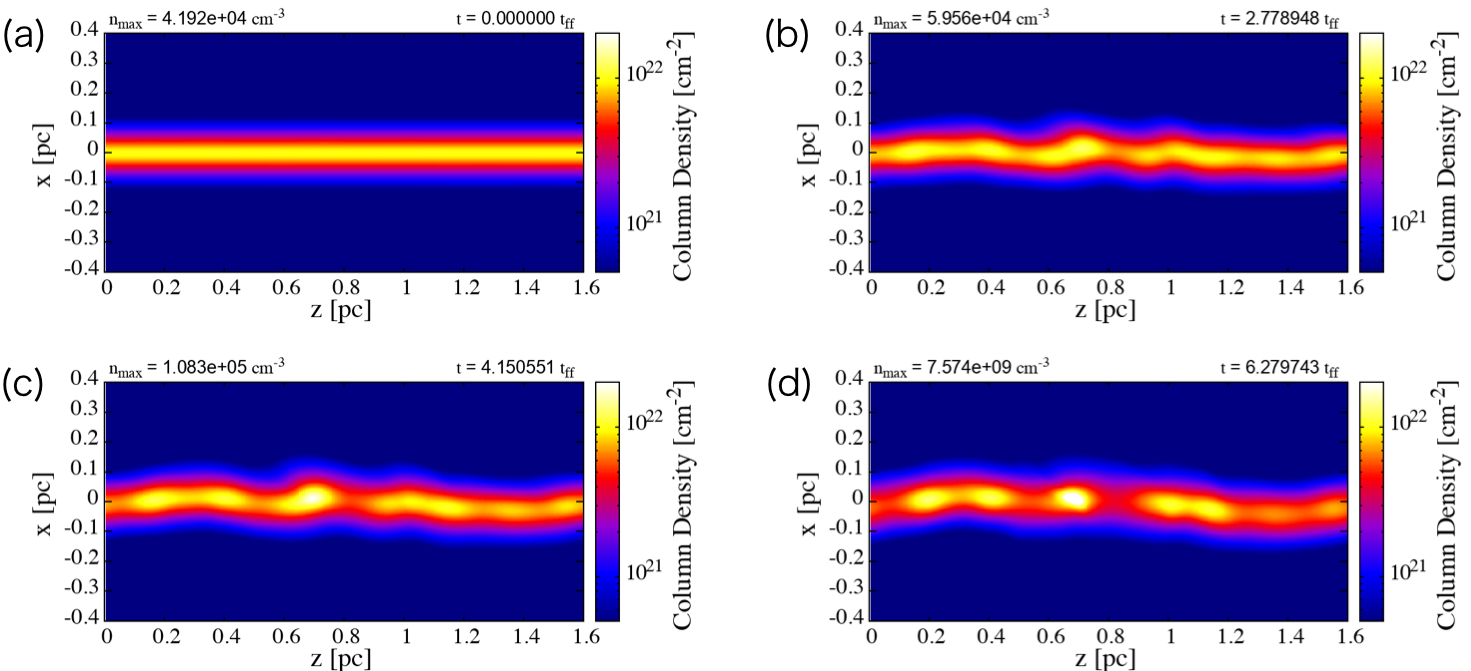}
\caption{Time sequence of the fragmentation of a filament (from panel (a) to (d)). The color scale is the column density derived by integrating the density along the $y$-axis. The elapsed time and maximum density are denoted in each panel, where $t_{\rm ff}$ is the free fall time defined by $t_{\rm ff}=(4\pi G \rho_{\rm c0})^{-1/2}$.}
\label{filtimeevo}
\end{figure}

\subsection{Overview}
We run a filament fragmentation simulation until the maximum density along the filament reaches $\rho_{\rm crit}=2.8\times 10^{-14} \ {\rm g \ cm^{-3}}$ ($n_{\rm crit}=7.3\times 10^{9} \ {\rm cm^{-3}}$), which corresponds to $\rho_{\rm crit}=2\times 10^5\rho_{\rm c0}$, where $\rho_{\rm c0}=1.4\times 10^{-19} \ {\rm g \ cm^{-3}}$ ($n_{\rm c0}=3.4\times 10^{4} \ {\rm cm^{-3}}$) is the initial peak density of the filament. Therefore, our simulation is stopped before the first core, i.e., a protostar, forms (for $n \simeq 5 \times 10^{10} \ {\rm cm^{-3}}$) \citep{Masunaga2000}. First, we describe the filament fragmentation due to the growth of initial perturbations. Figure \ref{filtimeevo} displays the time sequence of the fragmentation of a filament. The $z$-axis is parallel to the filament axis, and the $x$ and $y$-axis are perpendicular to the filament axis.  The color scale is the column density derived by integrating the density along the $y$-axis. The elapsed time and maximum density are denoted in each panel, where $t_{\rm ff}$ is the free fall time defined by $t_{\rm ff}=(4\pi G \rho_{\rm c0})^{-1/2}$.

\subsection{Core Angular Momentum} 
\begin{figure}[t]
\centering
\includegraphics[width=18cm]{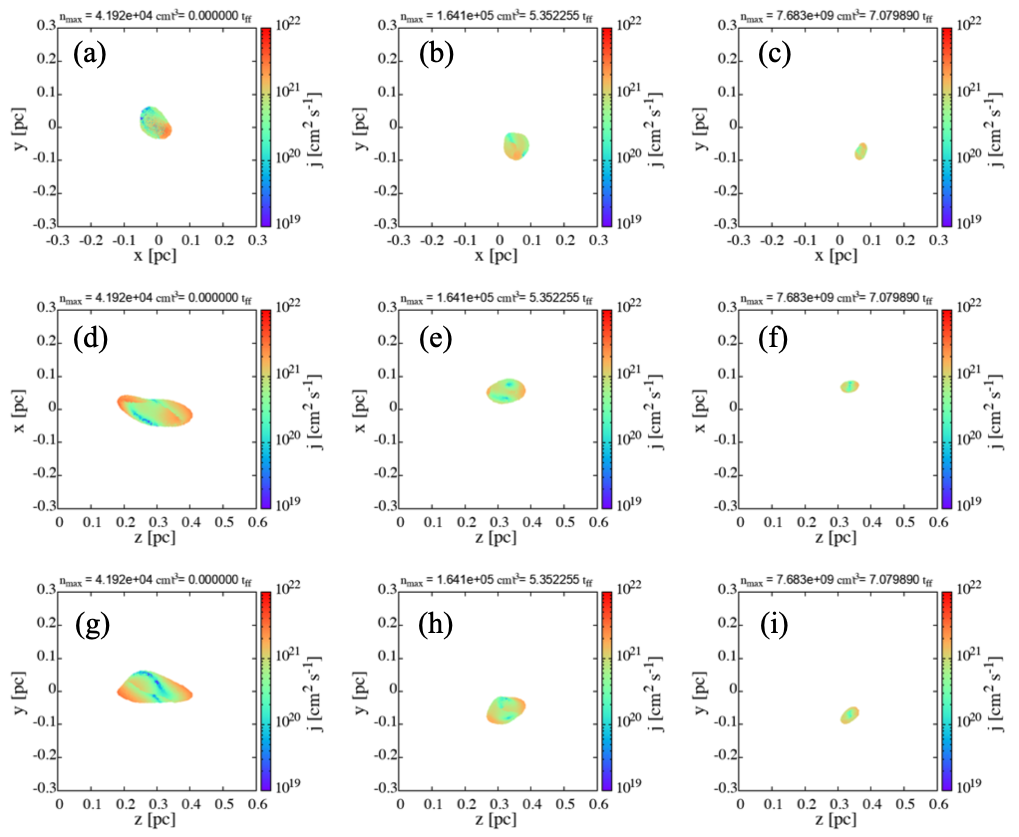}
\caption{Time sequence of the morphology and the specific angular momentum of a core from left to right. The first, second, and third rows are on the $x$-$y$,  $x$-$z$ plane, and $y$-$z$ plane, respectively. The color scale is the specific angular momentum of the SPH particles. This core is defined by the density contour enclosing a mass of $1 \ {\rm M}_{\odot}$ at the final time step. The panels from left to right are snapshots at stages of $n=4.2 \times 10^4\ {\rm cm^{-3}}$, $n=1.6 \times 10^5 \ {\rm cm^{-3}}$, and $n=7.7 \times 10^9\ {\rm cm^{-3}}$, respectively. The elapsed time and maximum density are denoted in each panel.}
\label{corelag}
\end{figure}

\begin{figure}[t]
\begin{tabular}{cc}
\begin{minipage}[t]{.35\textwidth}
\centering
\includegraphics[width=8cm]{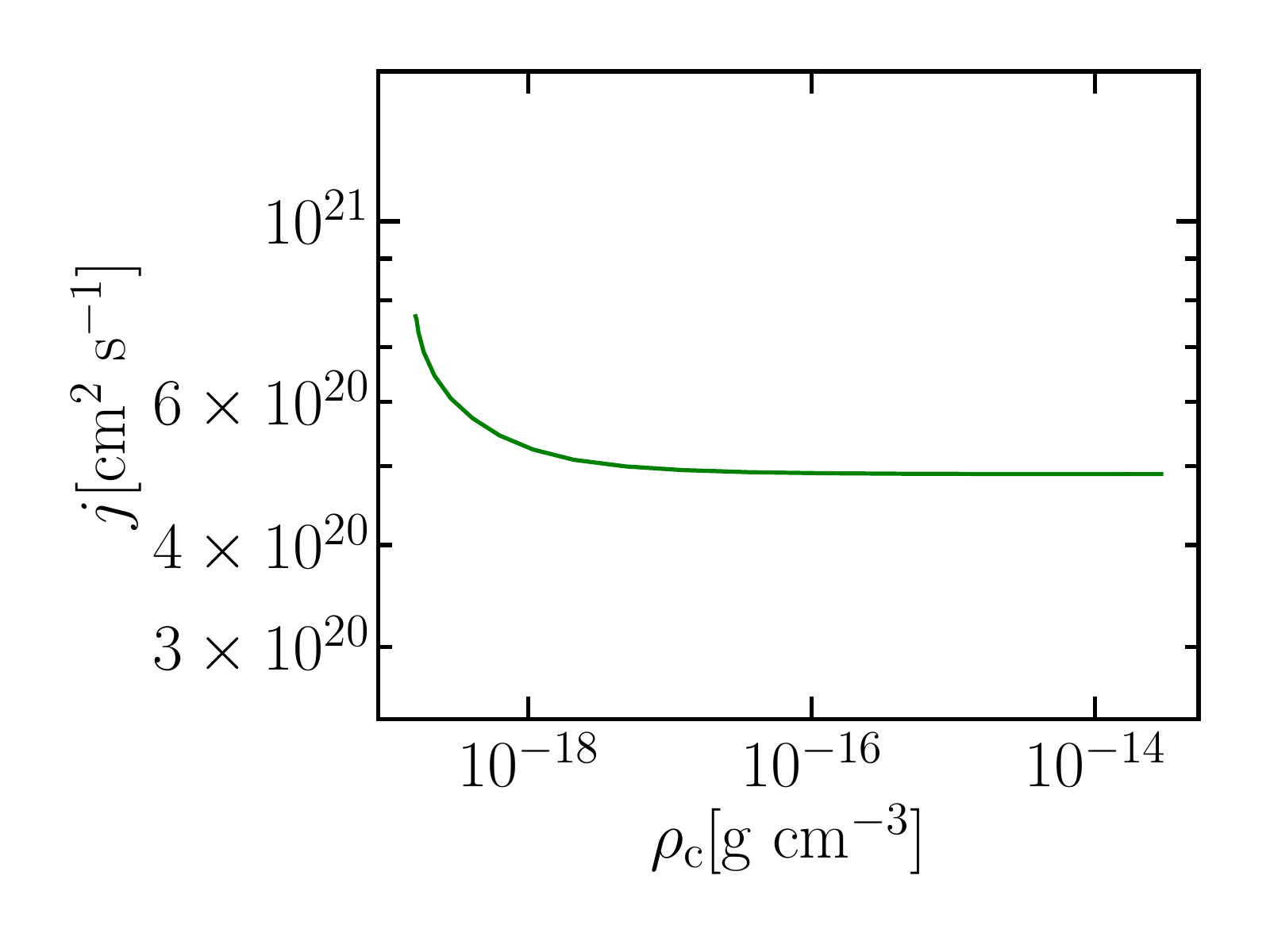}
\end{minipage}
\begin{minipage}{.15\textwidth}
\hspace{10mm}
\end{minipage}

\begin{minipage}[t]{.35\textwidth}
\centering
\includegraphics[width=8cm]{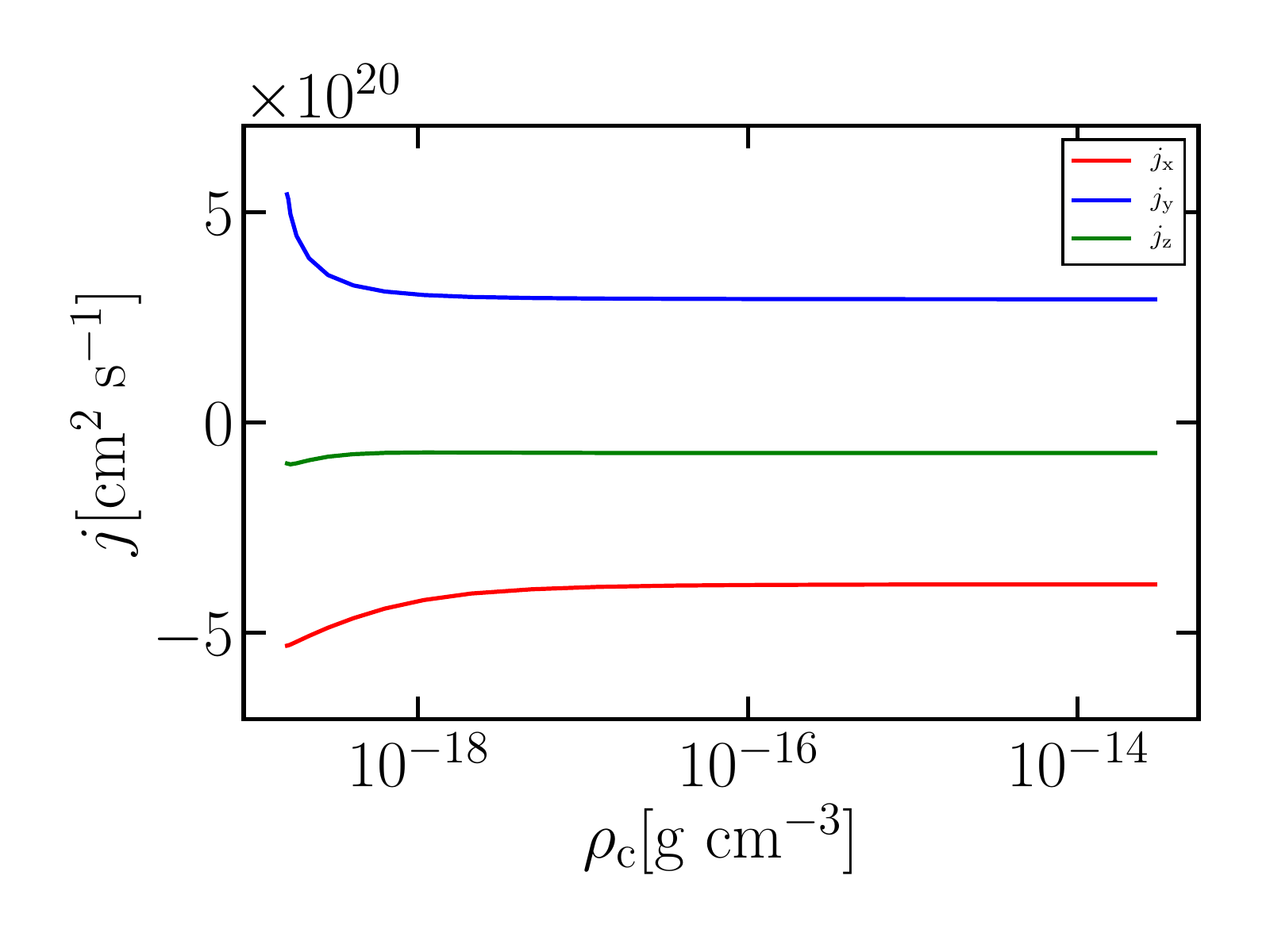}
\end{minipage}
\end{tabular}
\caption{ (left) Time evolution of the absolute value of the specific angular momentum of the core shown in Figure \ref{corelag}. The horizontal axis is the maximum density of the core. The density evolution traces the time evolution as the core collapses. (right) Time evolution of the component of the specific angular momentum of the core shown in Figure \ref{corelag}. The red, blue, and green solid lines are the $x$, $y$, and $z$ components of the specific angular momentum. The vertical axis is logarithmic in the left panel but linear in the right panel.} 
\label{fig:amtimeevo}
\end{figure}

In this subsection, we measure the core angular momentum and investigate its time evolution. As we mentioned in Section 3.1, we stop the simulations before the first core forms from the fragmentation of the filament  where the density reaches $n \simeq 5 \times 10^{10}\  {\rm cm^{-3}}$. Then, we identify and analyze the core around that high density peak. That core corresponds to the fastest collapsing core along the filament. We define the core using a density contour threshold and we measure the enclosed mass in the contour. We repeat this procedure changing the density threshold until the enclosed mass reaches $1\ {\rm M}_{\odot}$. Thus, in this analysis, all cores have the same mass, $1\ {\rm M}_{\odot}$. The core finding algorithm takes into account the periodic boundary conditions. Then, we trace the trajectories of the SPH particles from the final state to the initial state of the simulation. Figure \ref{corelag} shows the time sequence of the morphology and the specific angular momentum of a core from our simulations. The color scale is the specific angular momentum of the SPH particles around the barycenter of the core. The panels from left to right are snapshots at the stages of $n=4.2 \times 10^4 \ {\rm cm^{-3}}$, $n=1.6 \times 10^5 \ {\rm cm^{-3}}$, and $n=7.7 \times 10^9 \ {\rm cm^{-3}}$, respectively. The elapsed time and maximum density are denoted in each panel. Next, we derive the specific angular momentum of the core with respect to the center of mass using the definition of the core explained above:
\begin{eqnarray}
\bi{J}=\sum_{i} m_i (\bi{x}_i-\bi{x}_{\rm c}) \times (\bi{v}_i-\bi{v}_{\rm c}),
\label{eq:angcal}
\end{eqnarray}
where $\bi{x}_i$ and $\bi{v}_i$ are the position and velocity vectors of the SPH particles, respectively, $\bi{x}_{\rm c}$ and $\bi{v}_{\rm c}$ are the position and velocity of the center of mass of the core, respectively. The specific angular momentum of a core is defined as follows:
\begin{eqnarray}
\bi{j}=\frac{|\bi{J}|}{M},
\label{eq:samdef}
\end{eqnarray}
where $M$ is the total mass of the core (here $M=1\ {\rm M}_{\odot}$). Figure \ref{fig:amtimeevo} displays the time evolution of the specific angular momentum of the core shown in Figure \ref{corelag}. The horizontal axis is the elapsed time. The left panel is for the absolute value of the specific angular momentum, and the right panel shows each component of the specific angular momentum. The red, blue, and green solid lines in the right panel are the $x$, $y$, and $z$ components of the specific angular momentum. Note that the angular momentum should be conserved in the absence of interaction with surrounding matter. As one can see from Figure \ref{fig:amtimeevo}, the core angular momentum changes by around 30\%. In the right panel of Figure \ref{fig:amtimeevo}, $|j_z|$ is smaller than $|j_x|$ and  $|j_y|$. Statistically, for a large sample of filaments simulated with random turbulent seeds, $j_x$ and $j_y$ follow the same distribution because the velocity field is isotropic and the filament is axisymmetric, thus the ensemble average of $j_x$ and $j_y$ are the same. The reason why $|j_z|$ is smaller than $|j_x|$ and $|j_y|$ will be explained in Section 3.3. 

To analyze this evolution in more detail, we calculate the torque exerted on the core. The gravitational torque ($\bi{T}_{\rm g}$) and the pressure torque  ($\bi{T}_{\rm f}$) are derived separately as follows:

\begin{eqnarray}
\bi{T}_{\rm g}=\sum_{i} (\bi{x}_i-\bi{x}_{\rm c}) \times \bi{F}_{{\rm g},i} ,
\label{eq:tgcal}
\end{eqnarray}

\begin{eqnarray}
\bi{T}_{\rm f}=\sum_{i} (\bi{x}_i-\bi{x}_{\rm c}) \times \bi{F}_{{\rm f},i} ,
\label{eq:tfcal}
\end{eqnarray}

\begin{figure}[t]
\begin{tabular}{cc}
\begin{minipage}[t]{.35\textwidth}
\centering
\includegraphics[width=8cm]{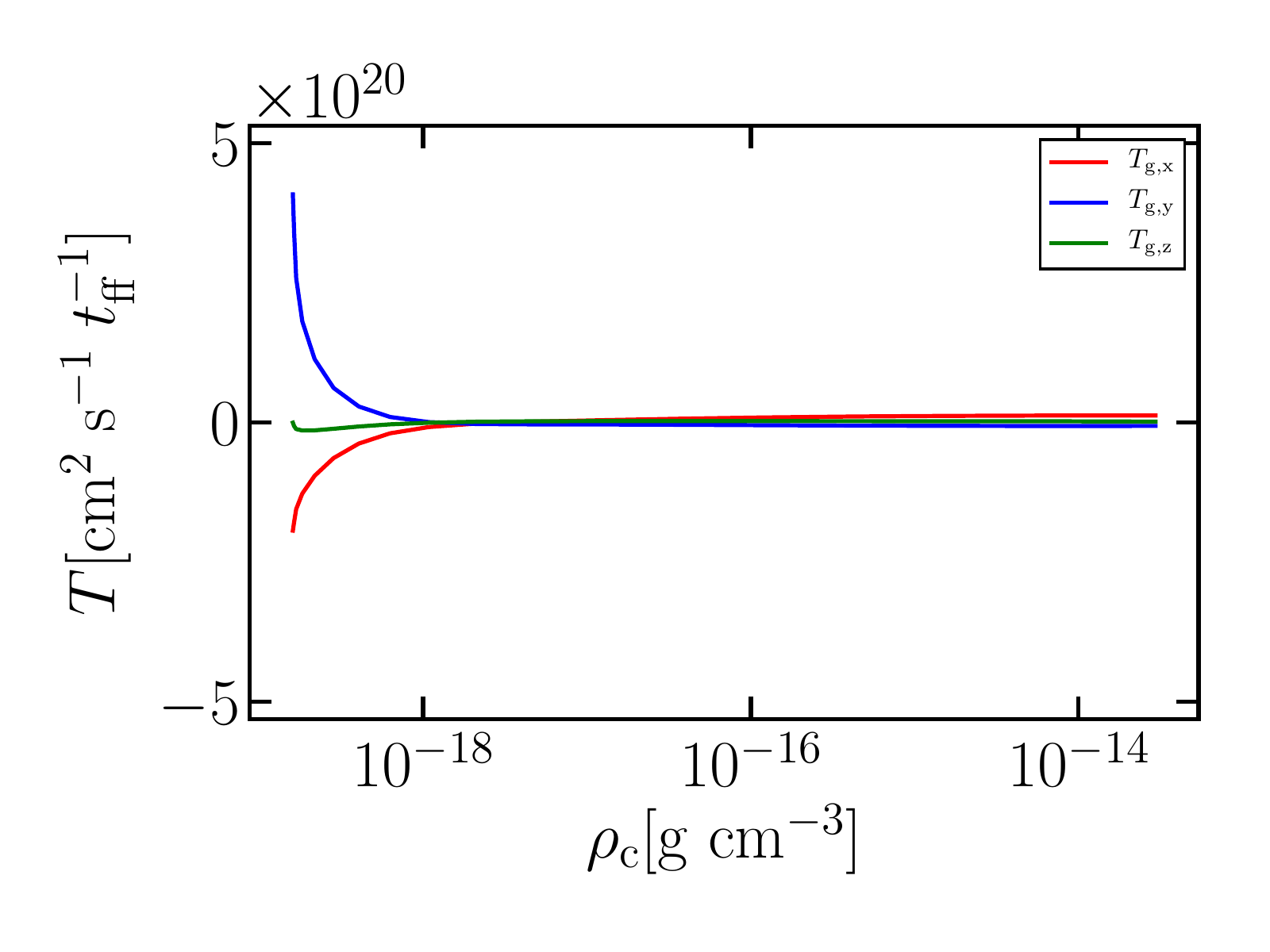}
\end{minipage}
\begin{minipage}{.15\textwidth}
\hspace{10mm}
\end{minipage}

\begin{minipage}[t]{.35\textwidth}
\centering
\includegraphics[width=8cm]{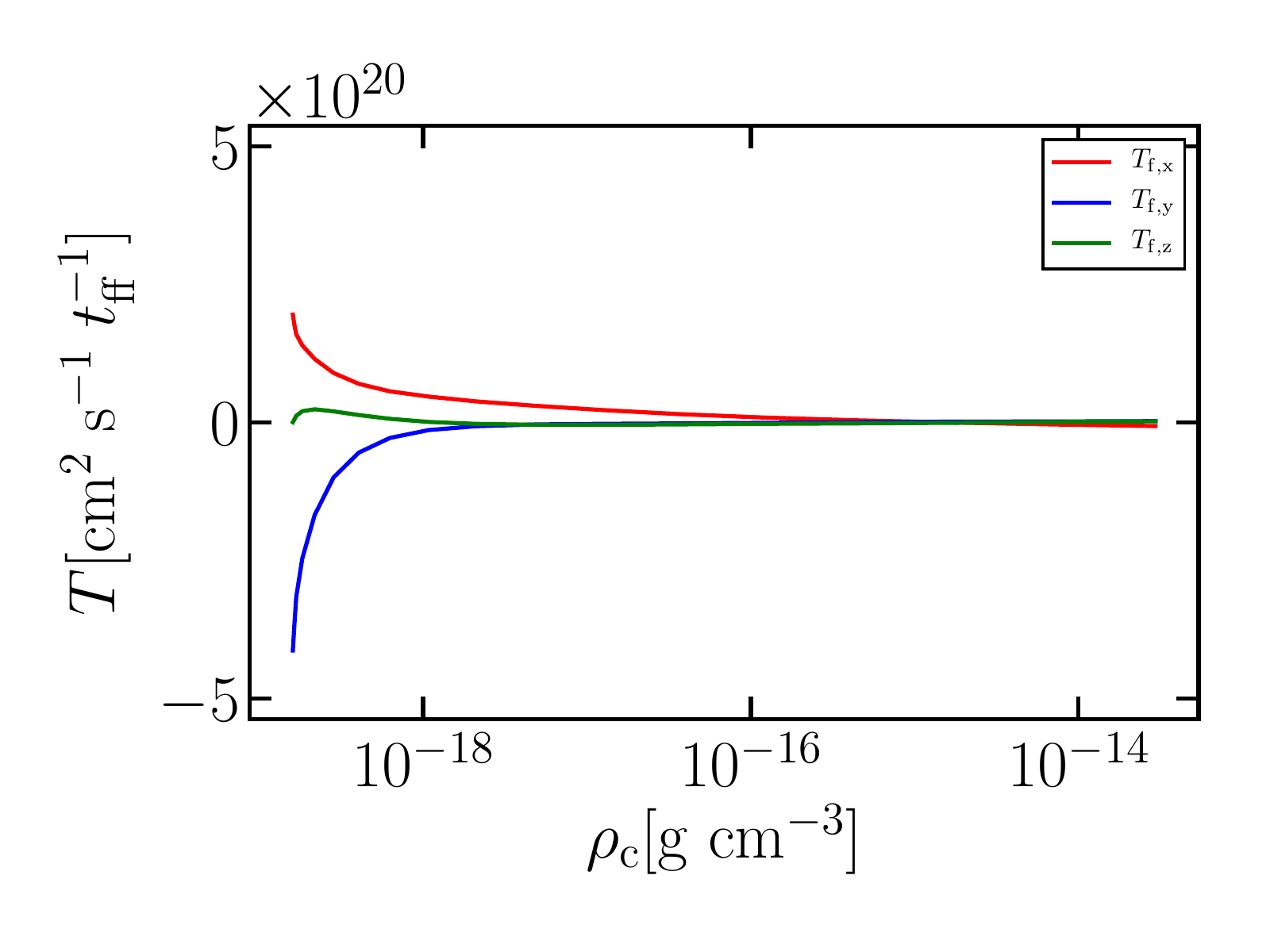}
\end{minipage}

\end{tabular}
\caption{Gravitational torque ($T_{\rm g}$, left) and pressure torque ($T_{\rm f}$, right) exerted on the core. The horizontal axis is the maximum density of the core and the vertical axis is the torque. The red, blue, and green curves are the $x$, $y$, and $z$ components of the torque.}
\label{fig:torcomp}
\end{figure}

\begin{figure}[t]
\centering
\includegraphics[width=8cm]{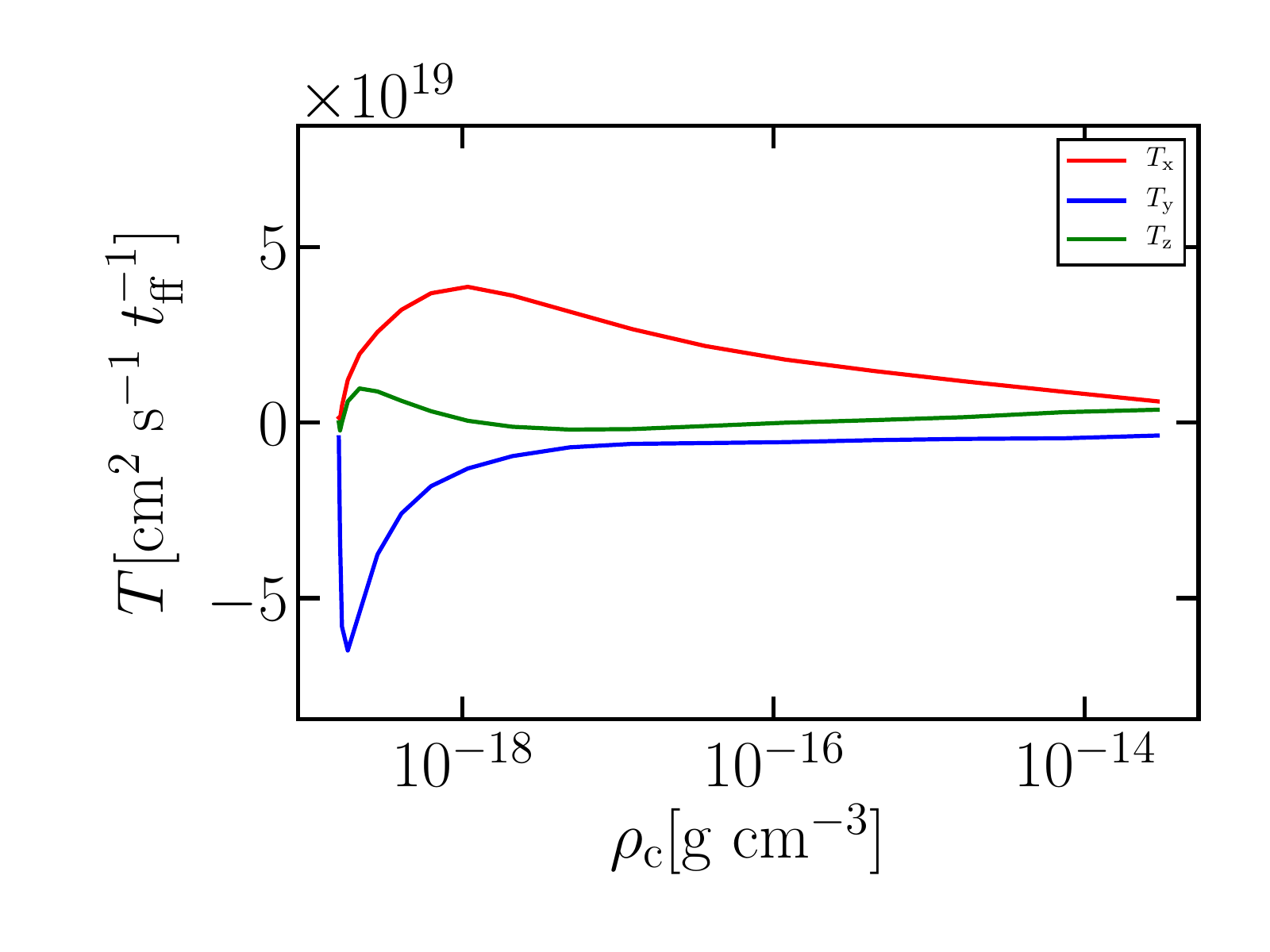}
\caption{Total torque exerted on the core. The horizontal axis is the maximum density of the core and the vertical axis is the total torque $ \bi{T}= \bi{T}_{\rm f} + \bi{T}_{\rm g}$. The red, blue, and green curves are the $x$, $y$, and $z$ components of the total torque.}
\label{fig:tor}
\end{figure}

where $\bi{F}_{{\rm f},i}$ and $\bi{F}_{{\rm g},i}$ are the pressure force (the first term of the right hand side of Equation \ref{eq:eom}) and the gravity (the second term of the right hand side of Equation \ref{eq:eom}) of $i$th particle, respectively. The gravitational torque and the pressure torque are shown in Figure \ref{fig:torcomp}. The total torque $ \bi{T}= \bi{T}_{\rm f} + \bi{T}_{\rm g}$ is plotted in Figure \ref{fig:tor}. The horizontal axis is the maximum density of the core, that also corresponds to the time evolution. Since our initial filament is in  hydrostatic equilibrium, the total torque exerted on the core is zero at $t=0$. As time progresses, both the gravitational and the pressure torques decrease in magnitude as shown in Figure \ref{fig:torcomp}. This is because the core evolves to the sphere-like shape as shown in Figure \ref{corelag}. The specific angular momentum of the core changes by  30\% due to the fluctuations of gravity and pressure generated by the initial Kolmogorov velocity fluctuations. Note that the angular momentum decreases by around 30\% at the early stage and is almost constant at the later stage. The reason why the angular momentum of the cores does not change at the later stage will be explained in the next subsection (Section 3.3 e.g., see Equation \ref {eq:djest}). \par

\subsection{Statistical Properties of the Total  Angular Momentum of the Core}

\begin{figure}[t]
\centering
\includegraphics[width=12cm]{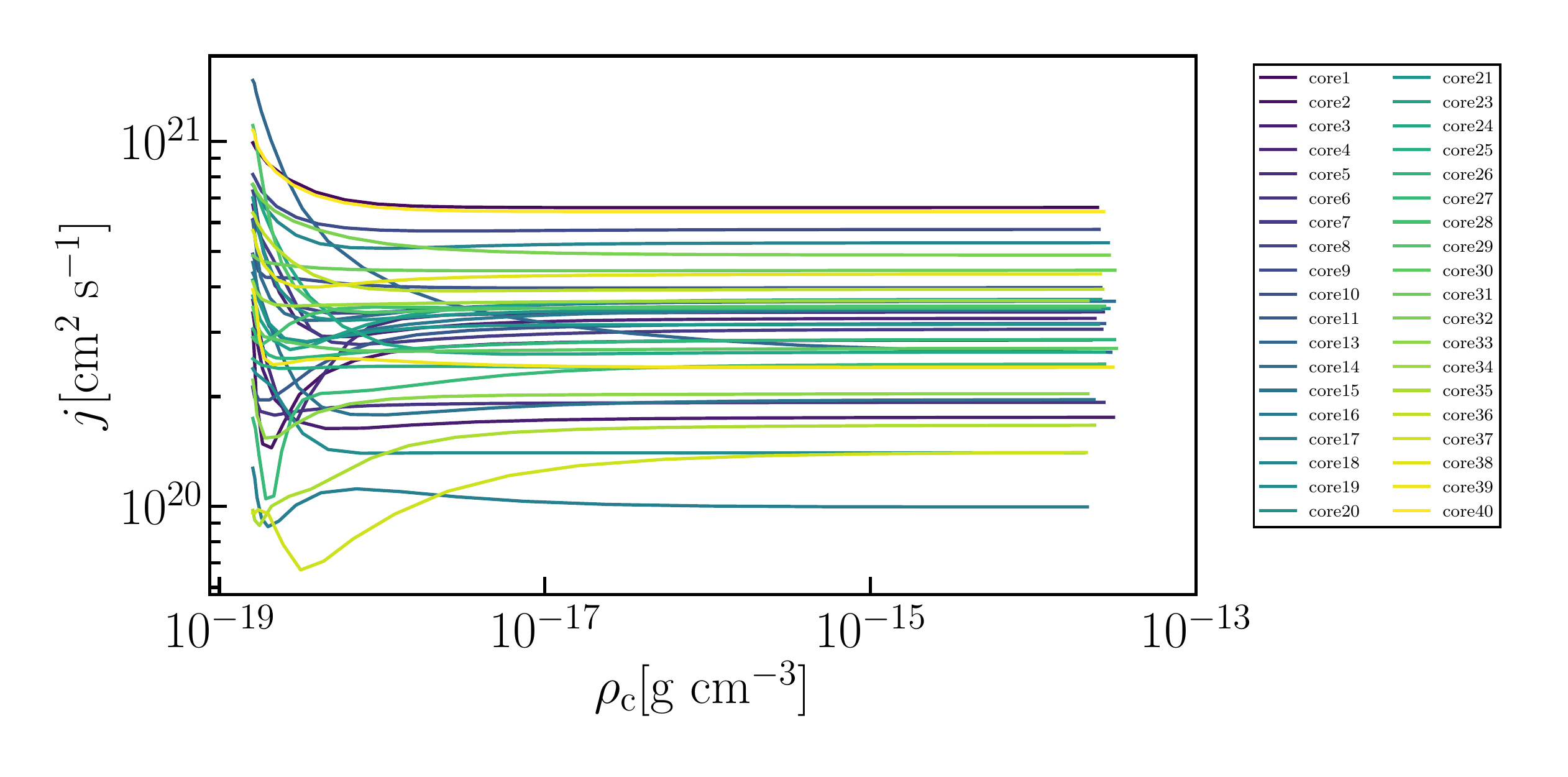}
\caption{Time evolutions of the specific angular momentum of the 38 cores. The horizontal axis is the maximum density of the cores. The different colors of solid lines correspond to the different cores derived from the 38 runs. }
\label{fig:allcoreang}
\end{figure}

\begin{figure}[t]
\centering
\includegraphics[width=12cm]{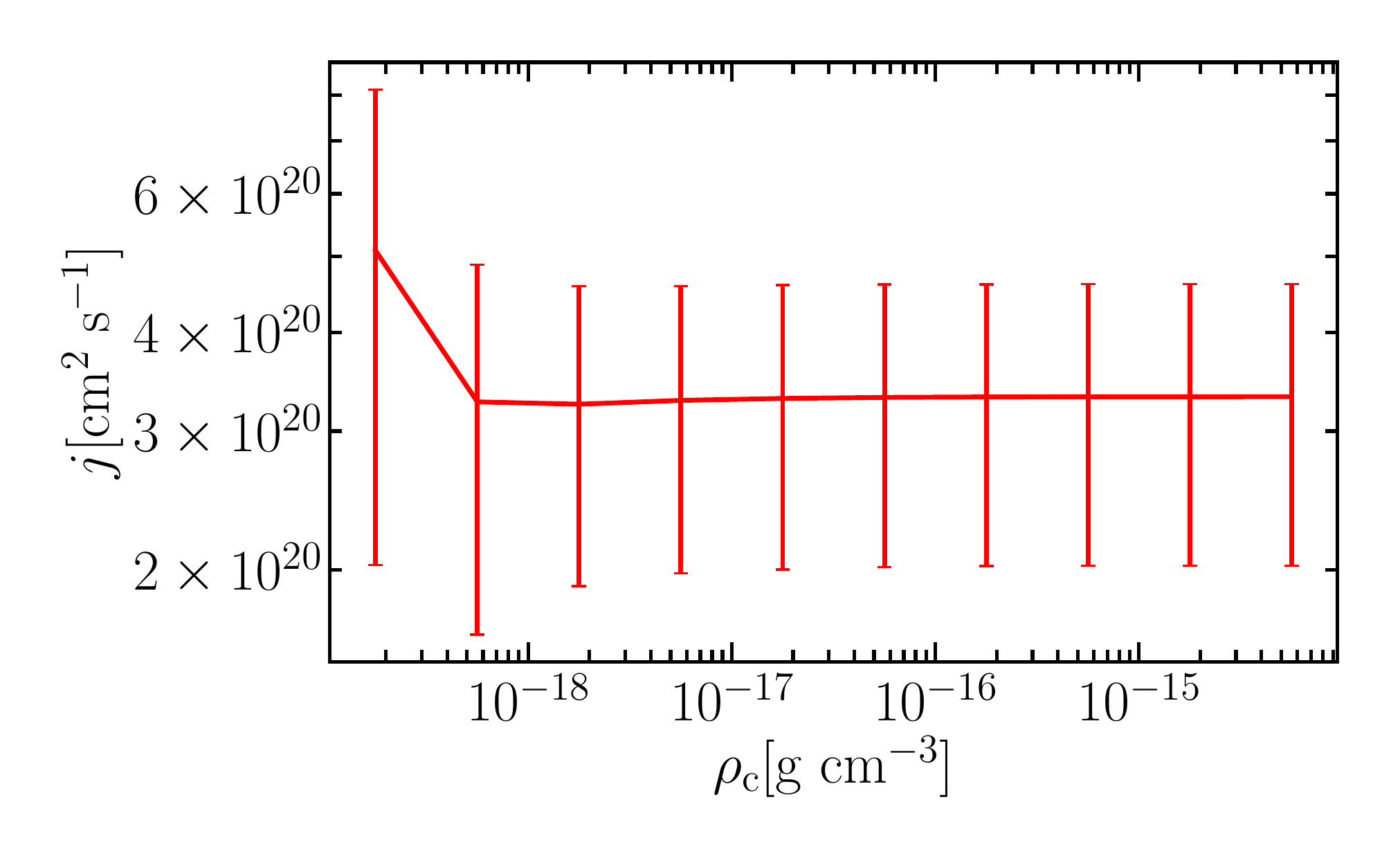}
\caption{Averaged time evolution of the specific angular momentum of the cores. The red solid line is the averaged specific angular momentum of the 38 cores shown in Figure \ref{fig:allcoreang}. The vertical bars show the dispersion. }
\label{fig:jave}
\end{figure}

\begin{figure}[t]
\centering
\includegraphics[width=12cm]{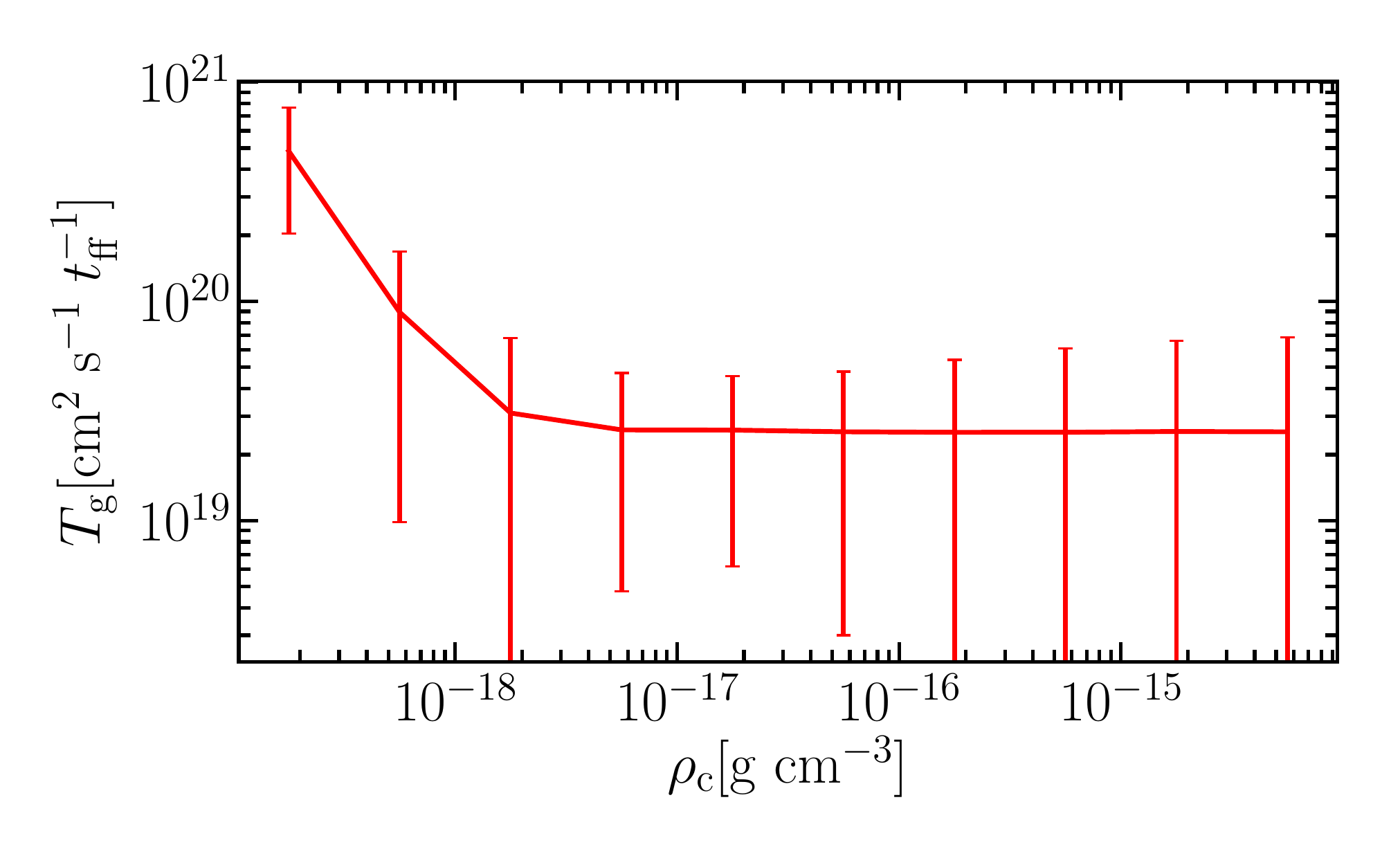}
\caption{Averaged time evolutions of the gravitational torque exerted on the cores. The vertical axis is the gravitational torque. The horizontal axis is the maximum density of the core. The vertical bars show the dispersion. }
\label{fig:tgave}
\end{figure}

We run 40 sets of simulations using different turbulence seeds and statistically analyze the properties of the core angular momentum. The analysis described in the previous subsections (Section 3.2) is applied to all the resultant filaments in these simulations. We build a sample of 40 cores selecting the fastest growing core of each simulation. The time evolution of 38 cores are shown in Figure \ref{fig:allcoreang}. Because two cores have two peaks within their $1 \ {\rm M}_{\odot}$ contour, we exclude these two cores from the analysis in this section. The horizontal axis of Figure \ref{fig:allcoreang} is the maximum density of the cores. The different colors of the solid lines correspond to different cores. Figure \ref{fig:jave} is the averaged time evolution of the specific angular momentum of all 38 cores shown in Figure \ref{fig:allcoreang}. As we can see from Figures \ref{fig:allcoreang} and \ref{fig:jave}, the core angular momentum is almost constant during the runaway collapse phase. This result can be understood as follows. First, the change of the angular momentum can be estimated by integrating the gravitational torque with respect to time. Figure \ref{fig:tgave} displays the averaged time evolution of the gravitational torque exerted on the cores. Figure \ref{fig:tgave} shows that the averaged gravitational torque during the runaway collapse phase is $T_{\rm g}\sim 2.5\times 10^{19}\ {\rm cm^2 \ s^{-1} \ t_{\rm ff}^{-1}}$. Using this value, the change of the angular momentum during the runaway collapse phase is estimated as follows: 
\begin{eqnarray}
\Delta j &= \int T_{\rm g} dt \nonumber \\
& = T_{\rm g} [ t_{\rm ff} (\rho_{\rm r}) - t_{\rm ff}(\infty) ]  \nonumber \\
& \sim 2.5 \times 10^{19}\sqrt{ \frac{\rho_{\rm c0}}{\rho_{\rm r}} } \ {\rm cm^2 \ s^{-1}},
\label{eq:djest}
\end{eqnarray}
where $\rho_{\rm c0}$ is the initial peak density of the filament, and $\rho_{\rm r}$ is the lower limit of the integration. In this analysis, we use $\rho_{\rm r}=10^{-17} \ {\rm g \ cm^{-3}}$ because we confirm that the collapse follows the runaway collapse when $\rho_{\rm r} \gtrsim 10^{-17} \  {\rm g \ cm^{-3}}$. If we substitute $\rho_{\rm r}=10^{-17} \ {\rm g \ cm^{-3}}$ into Equation \ref{eq:djest}, the resultant change of the angular momentum is $\Delta j \simeq 2.5 \times 10^{18} \ {\rm cm^2 \ s^{-1}}$. Since this is smaller by two orders of magnitude compared to the averaged specific angular momentum at the later stages ($ j \simeq 3 \times 10^{20} \ {\rm cm^2 \ s^{-1}}$) as shown in Figure \ref{fig:allcoreang}, we can conclude that the total angular momentum of the cores is almost constant during the runaway collapse phase. \par

\begin{figure}[t]
\begin{tabular}{cc}
\begin{minipage}[t]{.35\textwidth}
\centering
\includegraphics[width=7cm]{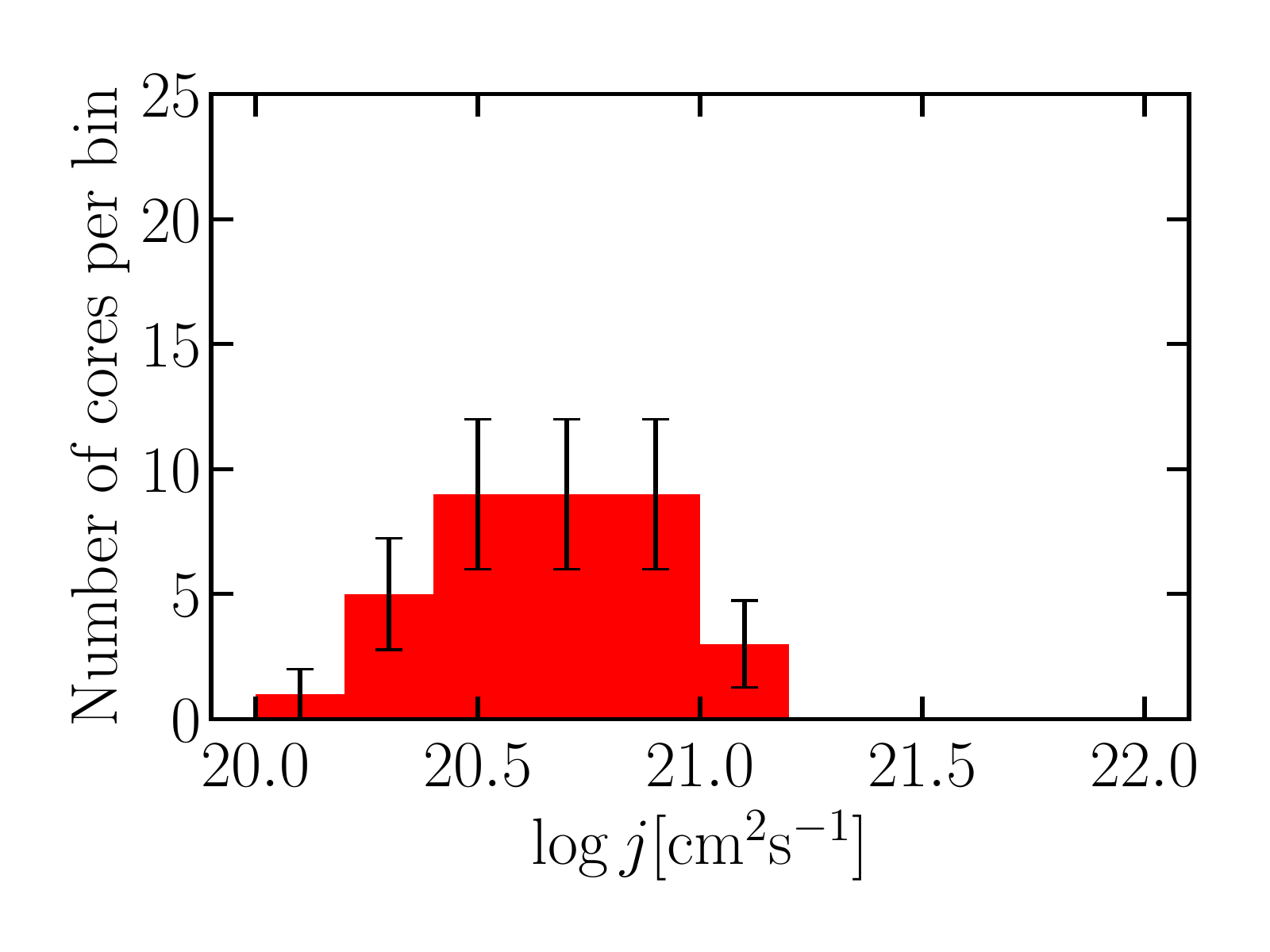}
\end{minipage}
\begin{minipage}{.10\textwidth}
\hspace{10mm}
\end{minipage}

\begin{minipage}[t]{.35\textwidth}
\centering
\includegraphics[width=7cm]{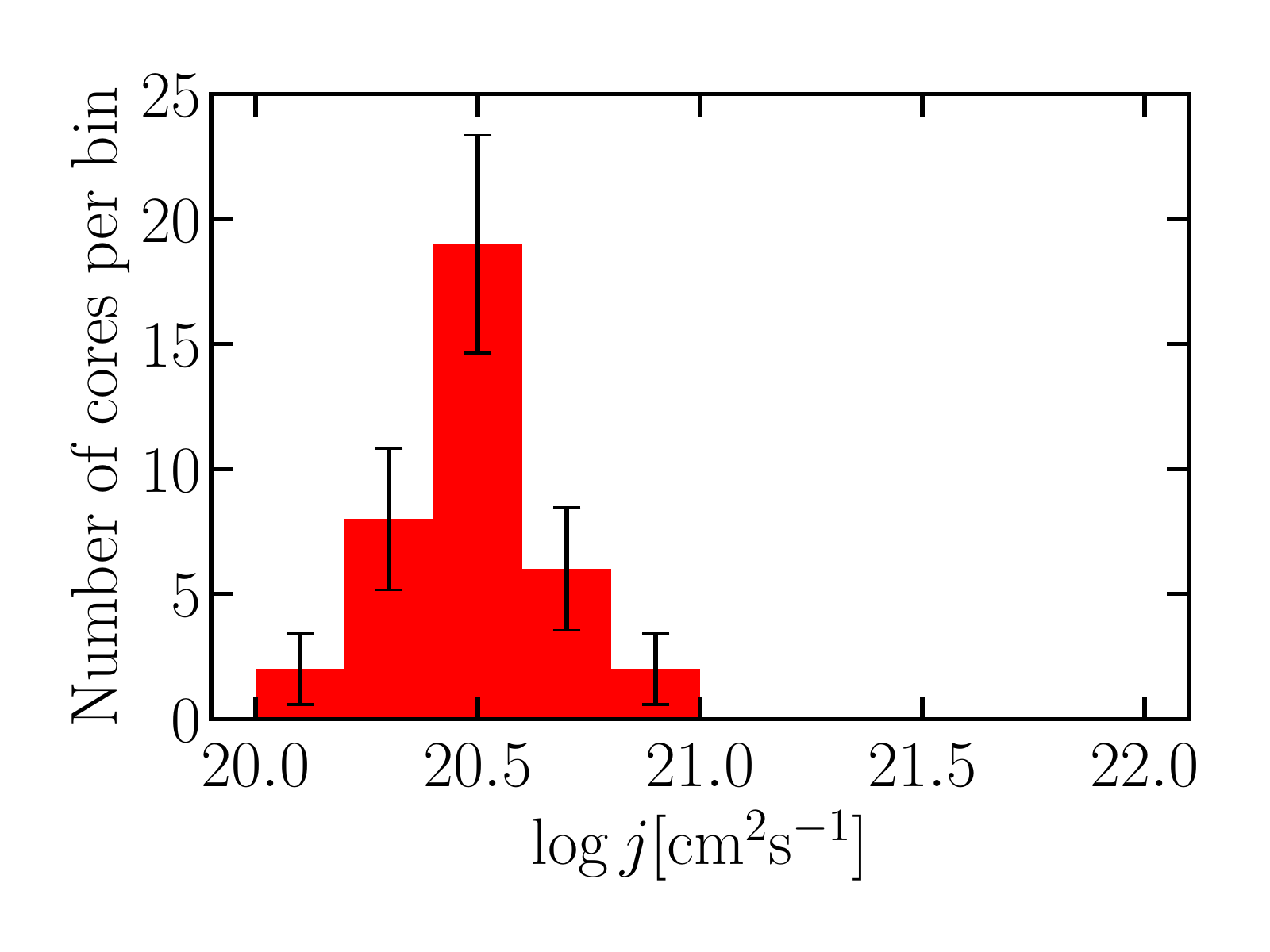}
\end{minipage}

\end{tabular}
\caption{Histograms of the specific angular momentum of the cores at the initial state (left) and at the final state (right) with statistical error bars. The vertical axis is the number of cores per $\log j$ bin. The average of the specific angular momentum at the initial state and the final state are $5.0 \times 10^{20}\ {\rm cm^2 s^{-1}}$ and $3.3 \times 10^{20}\ {\rm cm^2 s^{-1}}$, respectively.  }
\label{fig:amhist}
\end{figure}

Figure \ref{fig:allcoreang} also shows that the specific angular momentum of the cores decreases initially by around 30\% as shown in Section 3.2. To investigate the time evolution of the specific angular momentum of the cores statistically, we show the histograms for the specific angular momentum of the cores at both the initial stage and at the final stage (Figure \ref{fig:amhist}). The average of the specific angular momentum at the initial state and the final state are $5.0 \times 10^{20}\ {\rm cm^2 s^{-1}}$ and $3.3 \times 10^{20}\ {\rm cm^2 s^{-1}}$, respectively. Figure \ref{fig:amhist} shows that the initial specific angular momentum decreases by 30\% as shown in Figure \ref{fig:amtimeevo} and Figure \ref{fig:allcoreang}.\par
The typical value of the core angular momentum can be estimated by using the method shown in \cite{Peebles1969}. At the initial stage, the shape of the cores tend to be ellipsoidal rather than spherical (Figure \ref{corelag}). The angular momentum of the initial ellipsoidal core can be written formally as follows: 
\begin{eqnarray}
\bi{J} =\int  \rho  \bi{r} \times \bi{v} d^3x.
\label{eq:Jana}
\end{eqnarray}
This integration is calculated as follows (See Appendix A for more details):
\begin{eqnarray}
\bi{J} = -\frac{M}{5}\sum_{\bi{k}}\bi{V}(\bi{k}) \times\bi{k}''f(y) \cos \phi_{\bi{k}}
\label{eq:Janafinal}
\end{eqnarray}
where $\bi{k}''=(k_xa_1^2,k_ya_2^2,k_za_3^2)$, $y=|\bi{k}'|$, $\bi{k}'=(k_xa_1,k_ya_2,k_za_3)$. $M=4\pi \rho a_1 a_2 a_3 /3$ and $a_i$ $(i=1,2,3)$ are the mass and the principal axes of the ellipsoid, respectively. $\phi_{\bi{k}}$ is the phase of the initial velocity field. For simplicity, we assume the density is constant. $f(y)$ is defined as follows:
\begin{eqnarray}
f(y)=45\left(\frac{\sin y}{y^{5}}-\frac{\cos y}{y^{4}}-\frac{\sin y}{3 y^{3}}\right).
\label{eq:fdef}
\end{eqnarray}
Then we can derive the specific angular momentum
\begin{eqnarray}
j &\equiv \frac{ \sqrt{\left<  \bi{J}^2 \right>} }{M} \\
& = \sqrt{\frac{1}{75} \sum_{\bi{k}}P(k)\bi{k}''^2f(y)^2},
\label{eq:spejana}
\end{eqnarray}
where $\left< \right>$ represents the ensemble average. $P(k)$ is the power spectrum of the velocity field. If we substitute the typical values found in our simulation $a_3=0.085$ pc, $a_1=a_2=a_3/2=0.0425$ pc into Equation \ref{eq:spejana}, the specific angular momentum of the core becomes
\begin{eqnarray}
j  = 5.6 \times 10^{20}{\rm cm^2 s^{-1}}.
\label{eq:spejtypvalue}
\end{eqnarray}
This is consistent with the peak position of Figure \ref{fig:amhist} (left).\par

\begin{figure}[t]
\begin{tabular}{cc}
\begin{minipage}[t]{.35\textwidth}
\centering
\includegraphics[width=7cm]{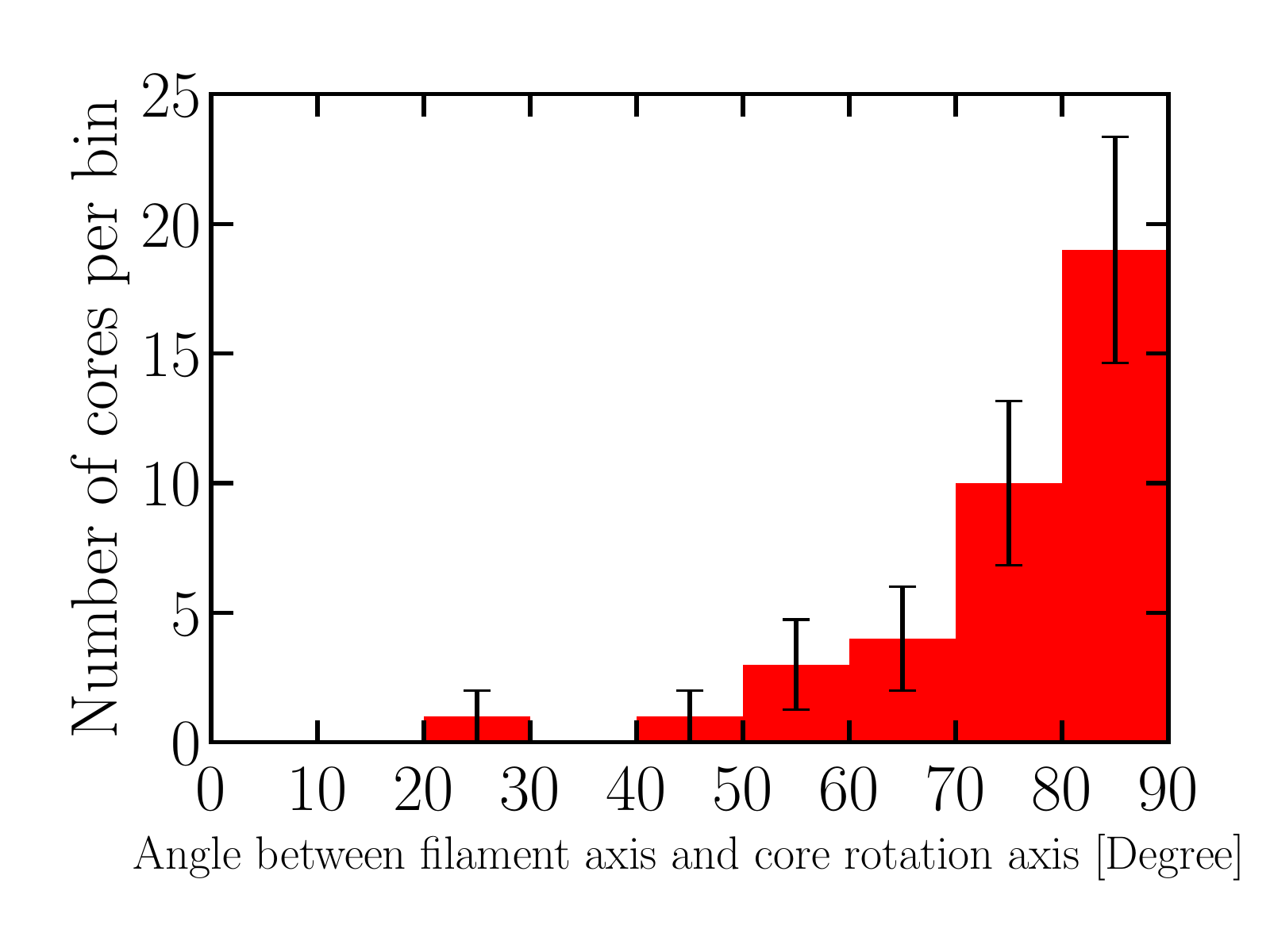}
\end{minipage}
\begin{minipage}{.10\textwidth}
\hspace{10mm}
\end{minipage}

\begin{minipage}[t]{.35\textwidth}
\centering
\includegraphics[width=7cm]{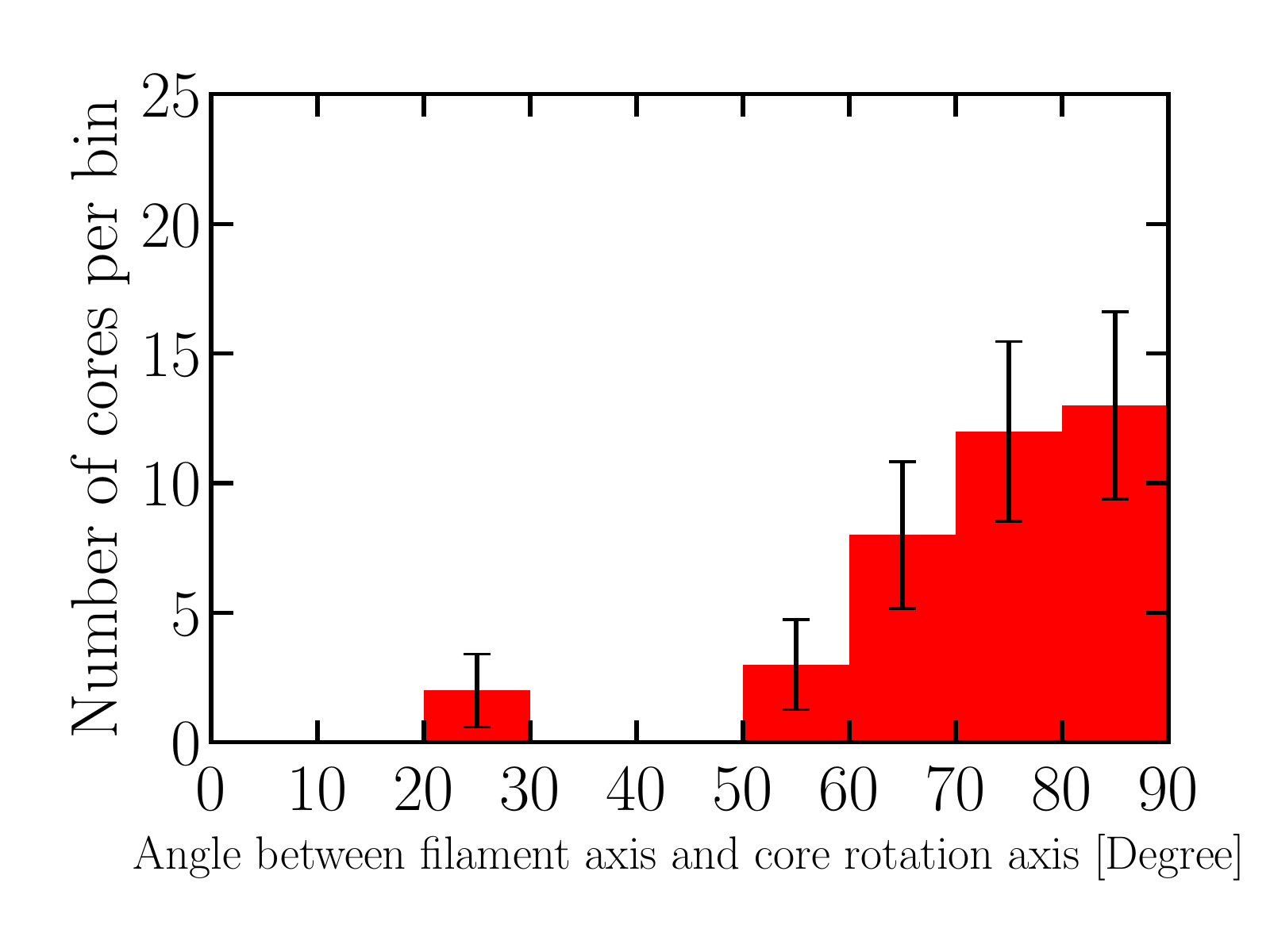}
\end{minipage}

\end{tabular}
\caption{Histogram of the angle $\theta$ between the filament axis and core angular momentum axis at the initial state (left) and at the final state (right) with statistical error bars. The vertical axis is the number of cores per $\theta$ bin.}
\label{fig:thetahist}
\end{figure}

Additionally, We also derived the time evolution of the angle between the filament axis and the core angular momentum axis. The angle between the filament axis and the core angular momentum axis is defined as $\cos \theta=J_z/J$. The time evolution of $\theta$ is plotted in Figure \ref{fig:thetahist}. The vertical axis is the number of cores per $\theta$ bin. Figure \ref{fig:thetahist} describes that most of the cores rotate nearly perpendicular to the filament axis. Since the core is formed through filament fragmentation, the core is elongated along the $z$-axis (filament longitudinal direction) at the initial state as shown in Figure \ref{corelag}. Equation \ref{eq:angcal} shows that the angular momentum of the core depends on the position vector from the barycenter of the core. Only the $x$ and $y$ components of the angular momentum are affected by the $z$ component of the position vector from the barycenter of the core. Therefore, most cores rotate nearly perpendicular to the filament axis (along the $z$-axis) even though the initial velocity turbulence field is isotropic. We stress that this trend can be observed even at the final state of the simulation (right panel of Figure \ref{fig:amtimeevo} and right panel of Figure \ref{fig:thetahist}).\par

\begin{figure}[t]
\centering
\includegraphics[width=9cm]{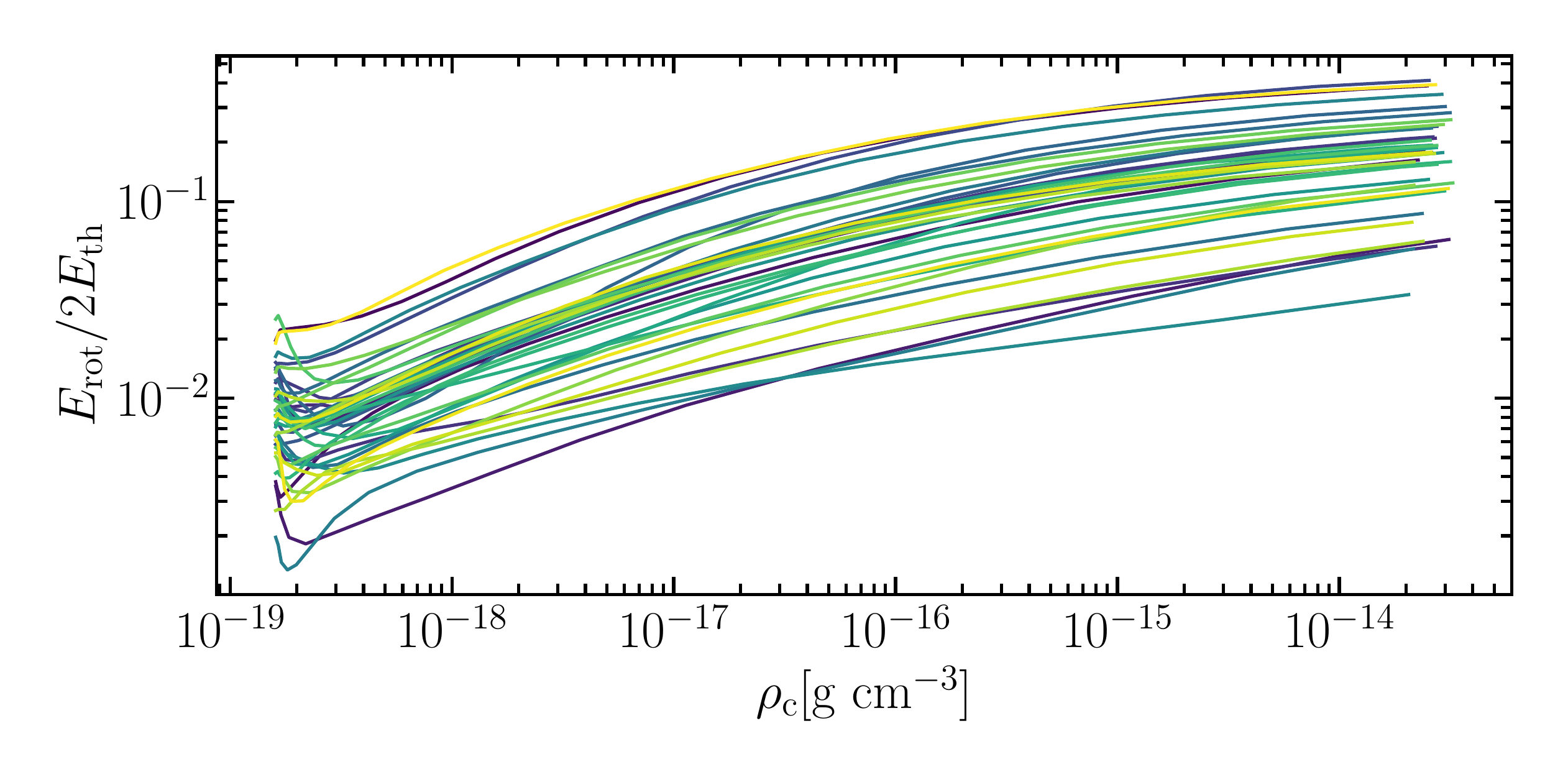}
\caption{Time evolution of the rotational energy normalized by the thermal energy.  The horizontal axis is the maximum density of the core. Each color corresponds to a given core.}
\label{fig:erot_eth}
\end{figure}

Figure \ref{fig:erot_eth} shows the time evolution of the rotational energy normalized by the thermal energy, $E_{\rm rot}/2E_{\rm th} $. The rotational energy $E_{\rm rot}$ is calculated as follows:
\begin{eqnarray}
E_{\rm rot}  = \sum_{i} \frac{m_i}{2}  \left[ \frac{ (\bi{x}_i - \bi{x}_{\rm c} ) \times ( \bi{v}_i - \bi{v}_{\rm c} )}{ | \bi{x}_i - \bi{x}_{\rm c} | } \right]^2,
\label{eq:erot}
\end{eqnarray}
where $i$ is a subscript of an SPH particle. Figure \ref{fig:erot_eth} shows that the rotational energy decreases during the core formation stage and increases at the later stage. This evolution can be understood as follows. Since the angular momentum of a core decreases initially during the core formation stage, the rotational energy also decreases when the central density $10^{-19} \ {\rm g \ cm^{-3}} \lesssim \rho_{\rm c} \lesssim 10^{-18} \ {\rm g \ cm^{-3}}$. 
Since we trace the trajectories of particles in this analysis, the mass $M$ defined in Equation \ref{eq:erot} is constant, the rotational energy $E_{\rm rot}  \sim I\omega^2 \sim J^2/MR^2 \propto R^{-2}$. Therefore, the rotation energy increases during the runaway collapse phase. As we can see from Figure \ref{fig:erot_eth},  $E_{\rm rot}/2E_{\rm th} \sim 0.01$ at the initial state, then  $E_{\rm rot}/2E_{\rm th}$ increases up to  $E_{\rm rot}/2E_{\rm th} \sim 0.1$ at the final state, just before the first core formation.\par

\begin{figure}[t]
\centering
\includegraphics[width=12cm]{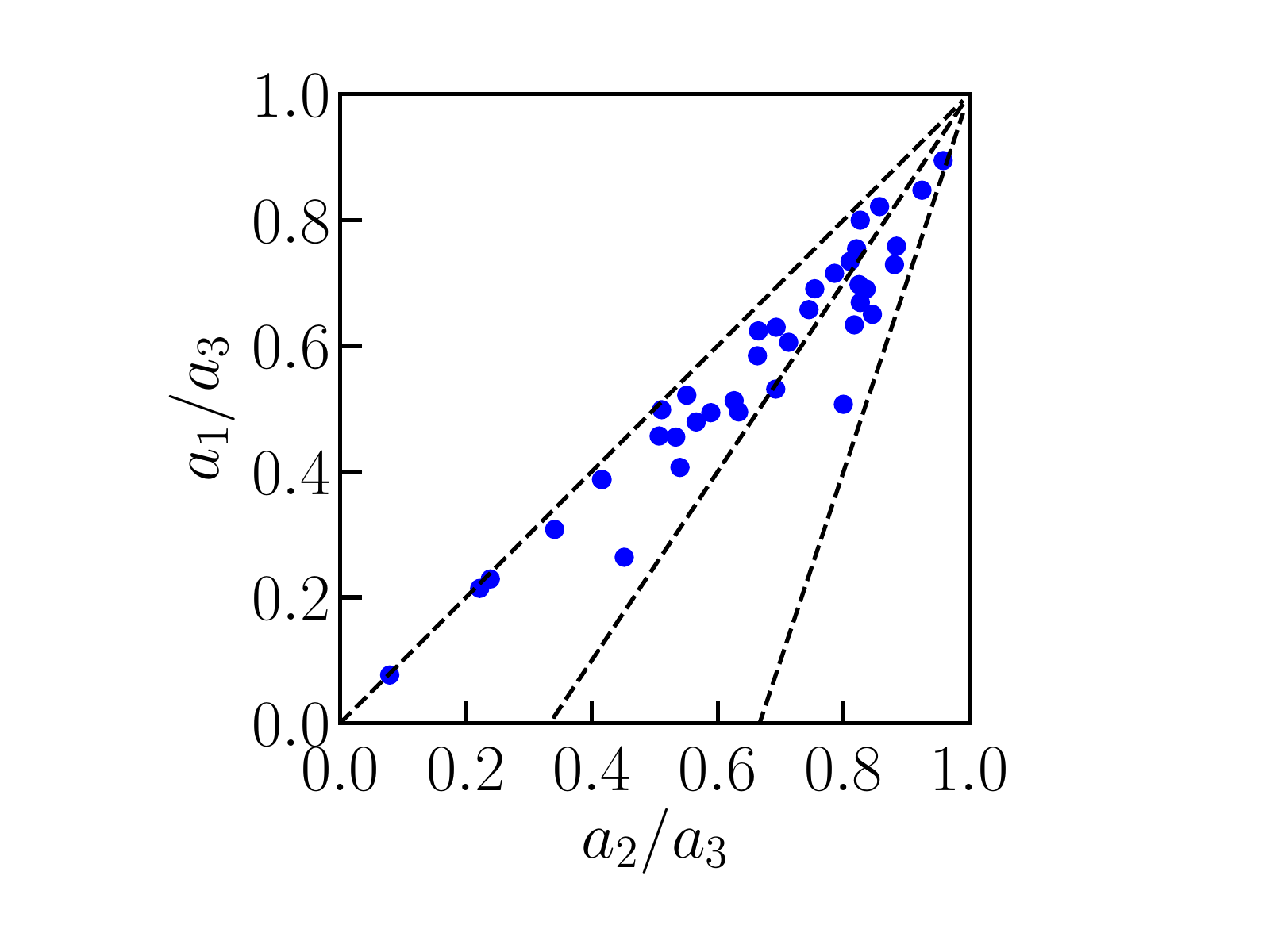}
\caption{Distribution of axis ratios of all cores at the final state. Each blue circle corresponds to a given core. The axis of a core are the eigenvalue of inertia moment, and $a_1<a_2<a_3$. The three dashed lines are the boundaries between prolate, triaxial, and oblate shapes from left to right. We define the prolate, triaxial, and oblate regions as $4a_2/3a_3-1/3 < a_1/a_3 < a_2/a_3$, $ 5a_2/3a_3-2/3< a_1/a_3 < 4a_2/3a_3-1/3 $, and $0 < a_1/a_3 < 5a_2/3a_3-2/3$, respectively.}
\label{fig:coreshape}
\end{figure}

Figure \ref{fig:coreshape} displays the distribution of the axis ratios of all cores at the final state. To estimate the shape of a core, we calculate the moment of coordinates,
\begin{eqnarray}
K_{ln}=\sum_{i} m_i (x_{l,i}-x_{l, {\rm c}}) (x_{n,i}-x_{n, {\rm c}}).
\label{eq:inertiak}
\end{eqnarray}
The shape of the core is estimated from the three principal axes ($a_1 < a_2 < a_3$) defined by the square roots of three eigenvalues of $K_{ln}$. We define the prolate, triaxial, and oblate regions as $4a_2/3a_3-1/3 < a_1/a_3 < a_2/a_3$, $ 5a_2/3a_3-2/3< a_1/a_3 < 4a_2/3a_3-1/3 $, and $0 < a_1/a_3 < 5a_2/3a_3-2/3$, respectively. This analysis is essentially the same as that used in \cite{Matsumoto2011}.  Figure \ref{fig:coreshape} shows that most cores formed through filament fragmentation tend to have prolate shapes.

\subsection{Internal Structure of Core Rotation}
Observations show that the internal motion of protostellar cores is not always coherent \cite[e.g.,][]{Gaudel2020}. Moreover, warped disks are discovered in young stellar objects \cite[e.g.,][]{Sakai2019}. Some observations even reported that the outflows/jets were driven in multiple directions, which was interpreted as the incoherent angular momentum axis distribution inside a parental dense core \cite[e.g.,][]{Okoda2021}. To elucidate the formation mechanism of observed complex structures, studying the internal structure of the cores at the core formation stage from filament fragmentation is very important. In Section 3.3, we provide the results for the time evolution of the total angular momentum of the cores. In this subsection, we focus on the internal angular momentum structure of a core.

\subsubsection{Angular Momentum Profile inside a Core}

\begin{figure}[t]
\centering
\includegraphics[width=12cm]{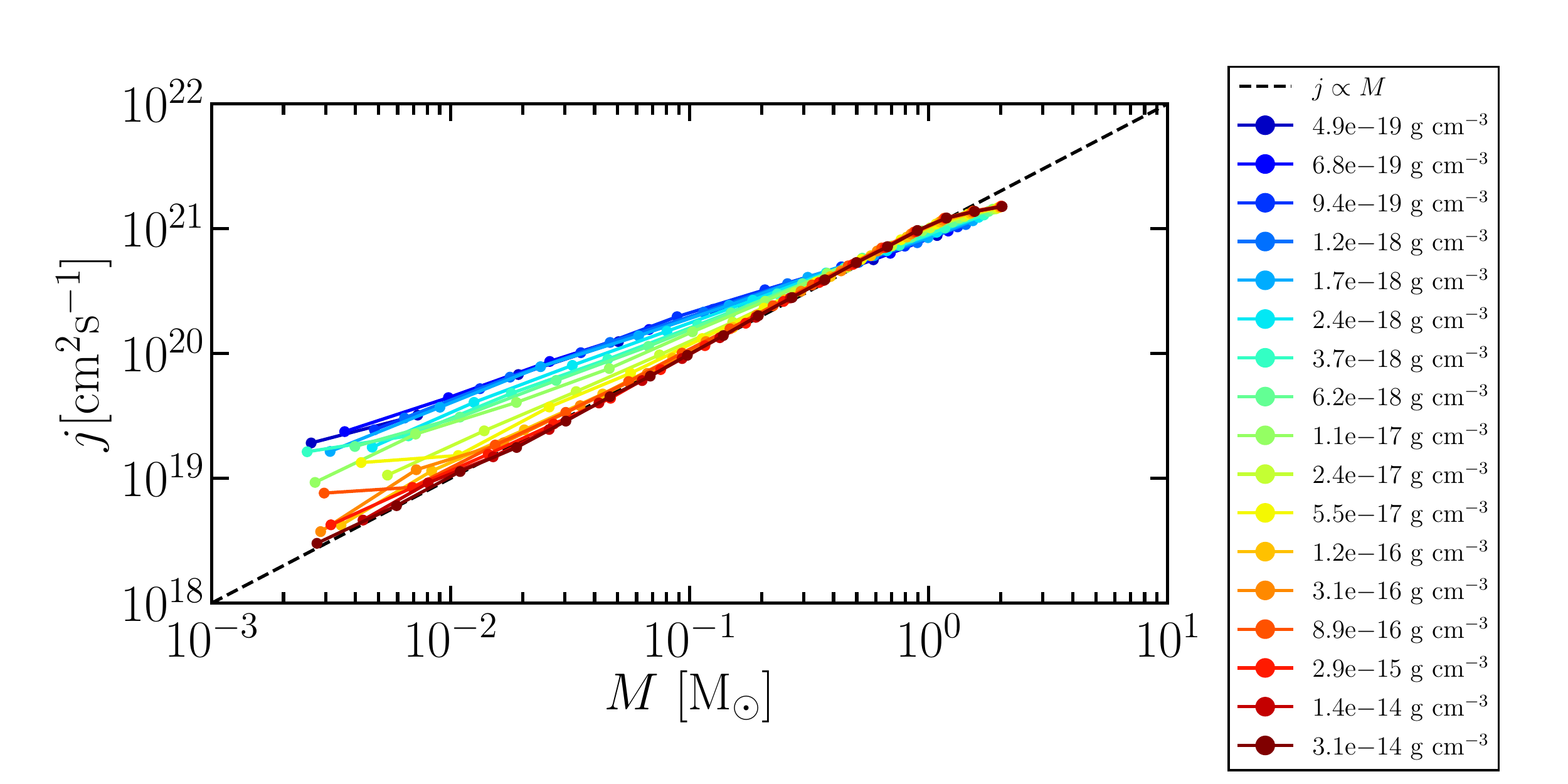}
\caption{Time evolution of $j$-$M$ profile in a single core (core 18). The horizontal axis is the enclosed mass, which corresponds to the distance from the density peak. The black dashed line is $j \propto M$. The $j$-$M$ profiles of the core evolve from the blue line to the red line. The density shown in the legend corresponds to the maximum density of the cores at each time step. }
\label{ssomegaevo}
\end{figure}

\begin{figure}[t]
\centering
\includegraphics[width=10cm]{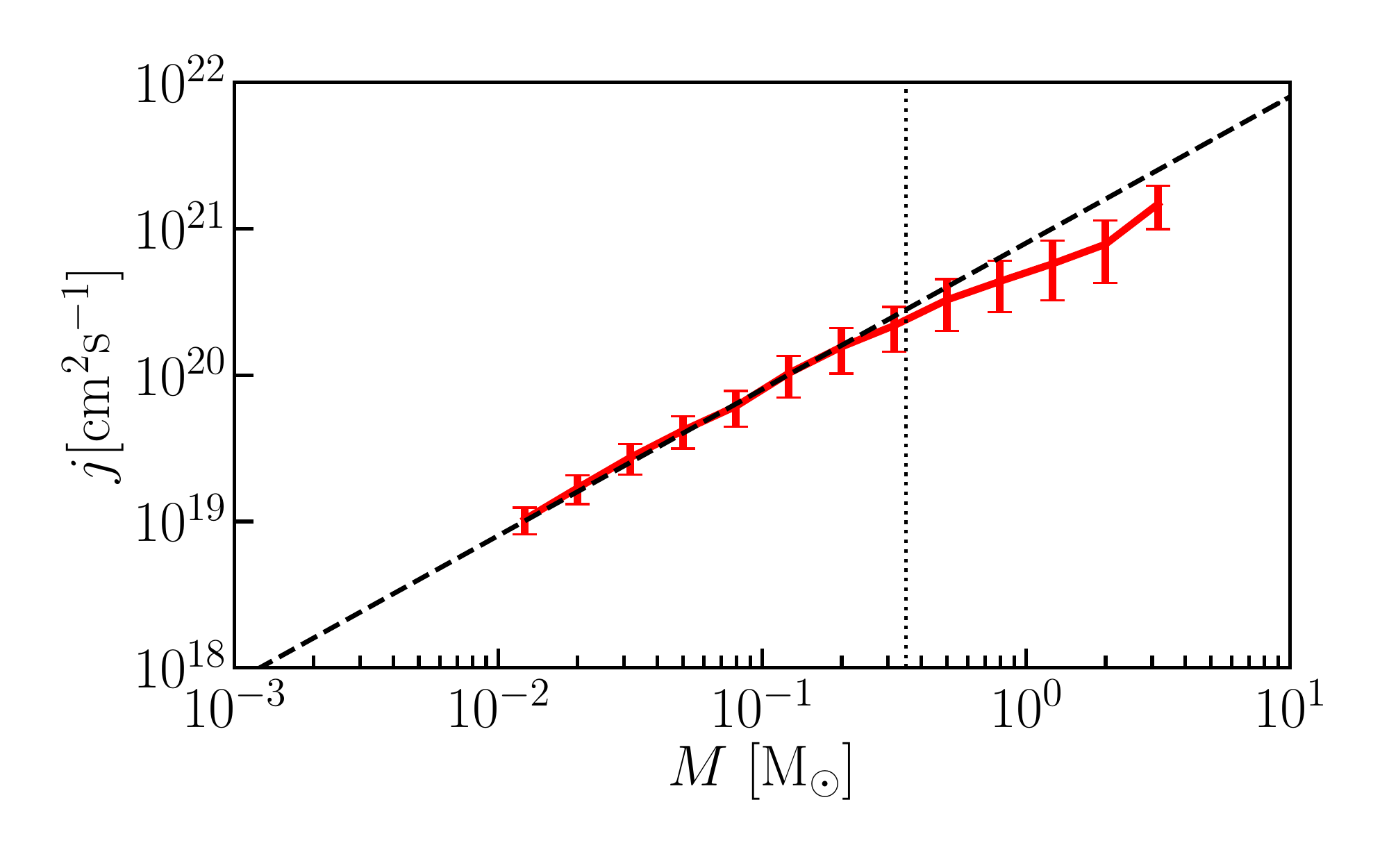}
\caption{Averaged $j$-$M$ profile for the 38 cores. The red solid line is the averaged $j$-$M$ profile using all $j$-$M$ profiles of the cores at the final stage as shown in Figure \ref{fig:interAMprof}. The black dashed line is $j \propto M$. The vertical dotted line is the analytical estimate of the boundary of the self-similar solution derived from Equation \ref{eq:selfsimcross} (see more details in Sect. 4.1). }
\label{fig:jMpro}
\end{figure}

First, we divide the core in concentric shells at each time step of the simulations. We measure the angular momenta of the shells around the density maxima as follows: 
\begin{eqnarray}
\bi{J}_{\rm shell}=\sum_{i \in {\rm shell}, \  j } m_i (\bi{x}_i-\bi{x}_{\rm \rho max}) \times (\bi{v}_i-\bi{v}_{\rm \rho max}),
\label{eq:angcalpeak}
\end{eqnarray}
where $\bi{x}_{\rm \rho max}$ and $\bi{v}_{\rm \rho max}$ are the position and velocity of the density maxima, respectively. The summation is calculated for all SPH particles contained in each shell $j$. Figure \ref{ssomegaevo} displays the time evolution of $j$-$M$ profile in a core. The horizontal axis is the enclosed mass, which corresponds to the distance from the density peak. The black dashed line represents $j \propto M$. The $j$-$M$ profiles evolve from the blue line to the red line, i.e., from a shallower slope to a relation close to $j \propto M$. In addition, Figure \ref{fig:jMpro} displays the averaged $j$-$M$ profile at the final stage for the 38 cores of our study. The red solid line is the averaged $j$-$M$ profile using all $j$-$M$ profiles of the cores shown in Figure \ref{fig:interAMprof}. Figures \ref{ssomegaevo} and \ref{fig:jMpro} show that the $j$-$M$ profile evolves to $j\propto M$ with time during the runaway collapse phase. $j\propto M$ is the self-similar solution discussed in \cite{Saigo1998} and \cite{Basu1997}, and we will explain this later. This convergence to $j \propto M$ was reported in \cite{Tomisaka2000} where they investigated the evolution of the angular momentum profile in collapsing magnetized molecular cloud cores without turbulence. \par

\begin{figure}[t]
\begin{tabular}{c}
\begin{minipage}[t]{.35\textwidth}
\centering
\includegraphics[width=9cm]{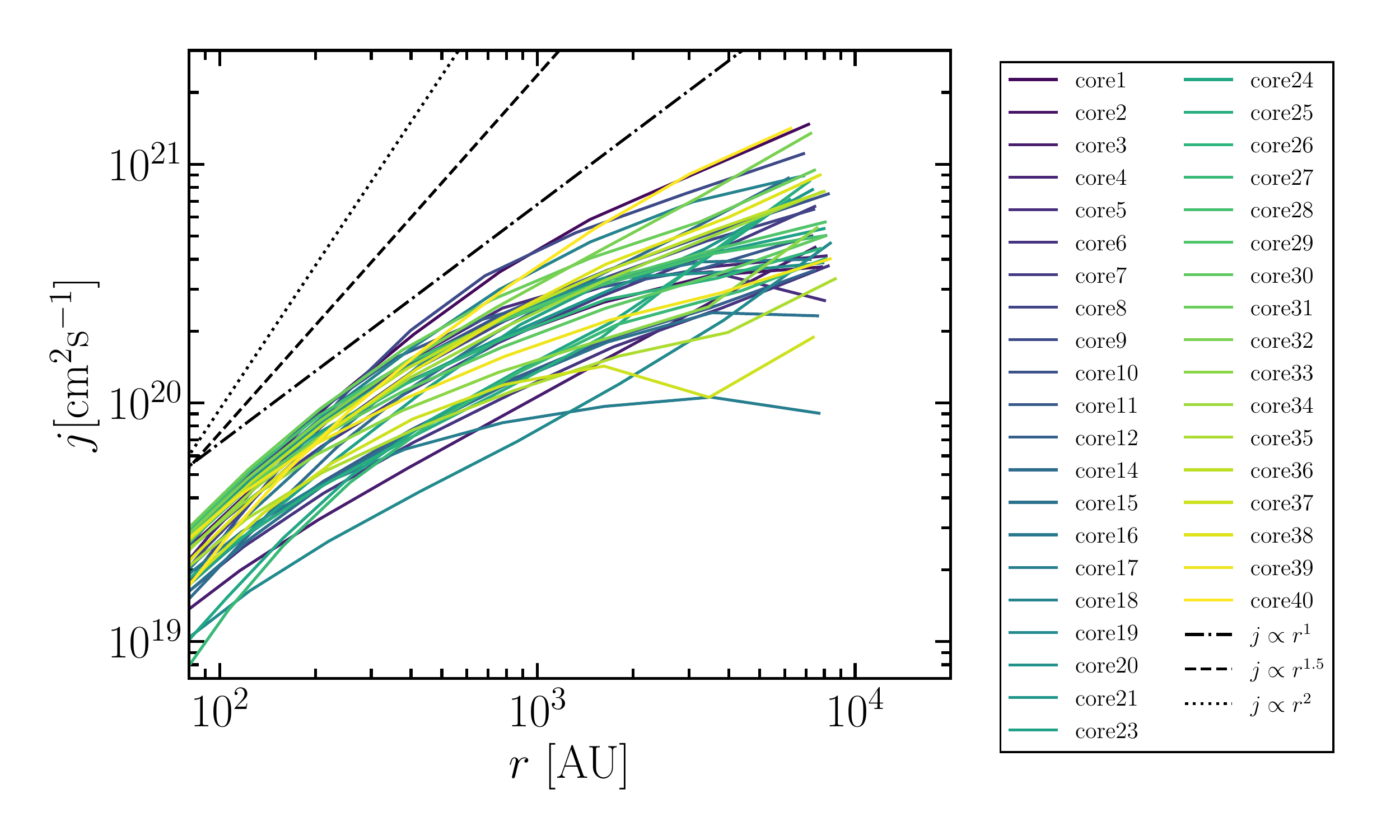}
\end{minipage}
\begin{minipage}{.2\textwidth}
\hspace{10mm}
\end{minipage}

\begin{minipage}[t]{.35\textwidth}
\centering
\includegraphics[width=9cm]{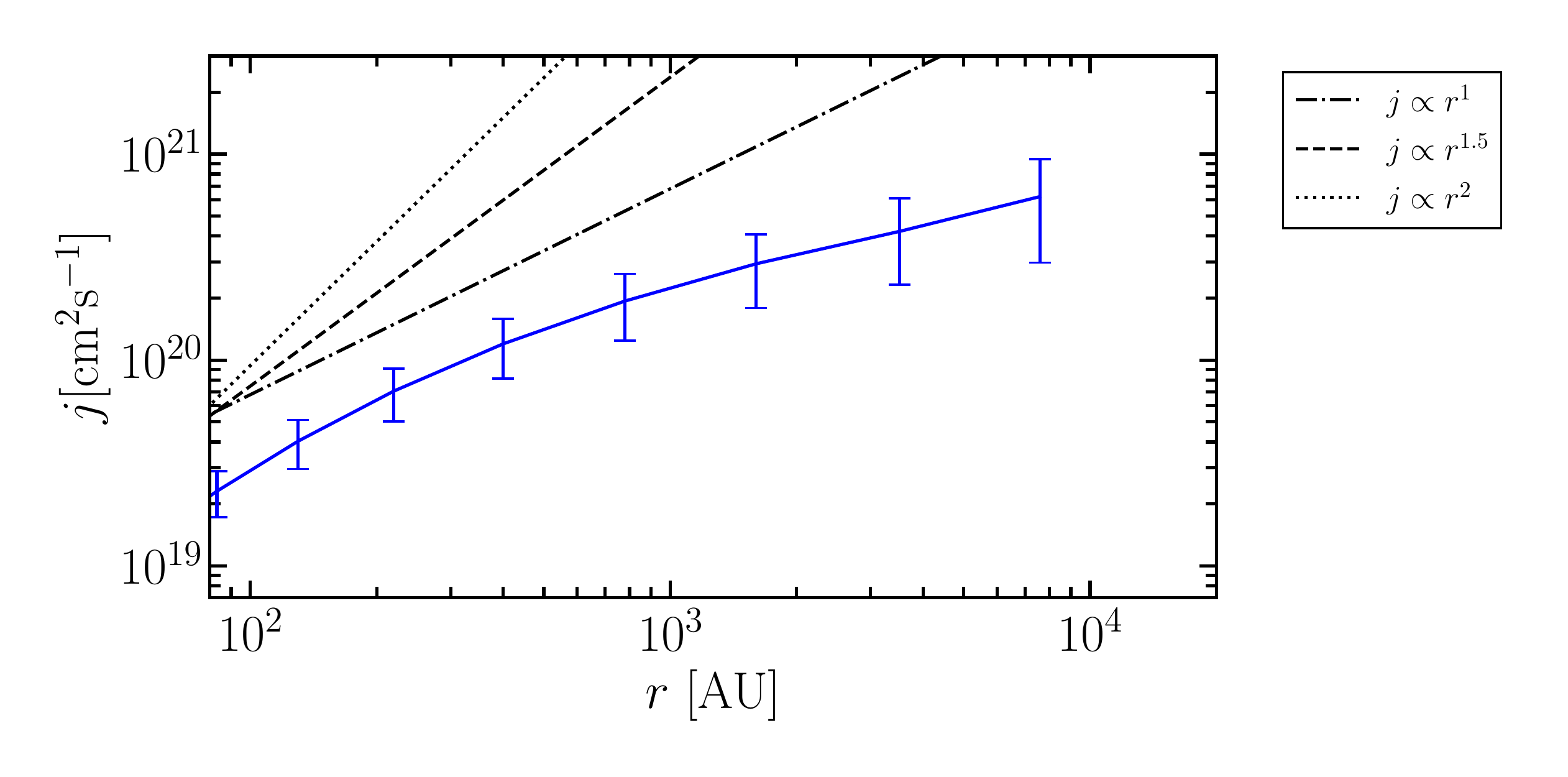}
\end{minipage}

\end{tabular}
\caption{Circularly averaged angular momentum as a function of core radius.  The vertical axis is the specific angular momentum of a core, and the horizontal axis is the distance from the density peak of the core. In the left panel, each color corresponds to a given core. The black dash-dotted line, dashed line, and dotted line are $j \propto r^{1}$, $j \propto r^{1.5}$, and $j \propto r^{2}$, respectively. The blue line in the right panel is the averaged $j$-$r$ profile of the 38 cores shown in the left panel.}
\label{fig:interAMradiusprof}
\end{figure}

Figure \ref{fig:interAMradiusprof} shows the $j$-$r$ profile in the cores, where $r$ is the distance from the density peak. As a reference, $j \propto r$, $j \propto r^{1.5}$, and $j \propto r^{2}$ lines are also shown. The right panel gives the averaged $j$-$r$ profile of the cores shown in the left panel.  The mean $j$-$r$ relation in the inner region of the cores is compatible with $j \propto r$. This is because Larson-Penston solution \citep{Larson1969,Penston1969} has a density profile of $\rho \propto r^{-2}$, which means $M\propto r$. Using this relation and $j \propto M$ (Figure \ref{ssomegaevo}), we can derive the $j \propto r$ relation. We note that the slope index of the $j$-$r$ relation in the inner region of the core expected from the self-similar solution ($j \propto r$) is different from the slope index expected from Larson's law ($j \propto r^{1.5}$).\par

\begin{figure}[t]
\centering
\includegraphics[width=9cm]{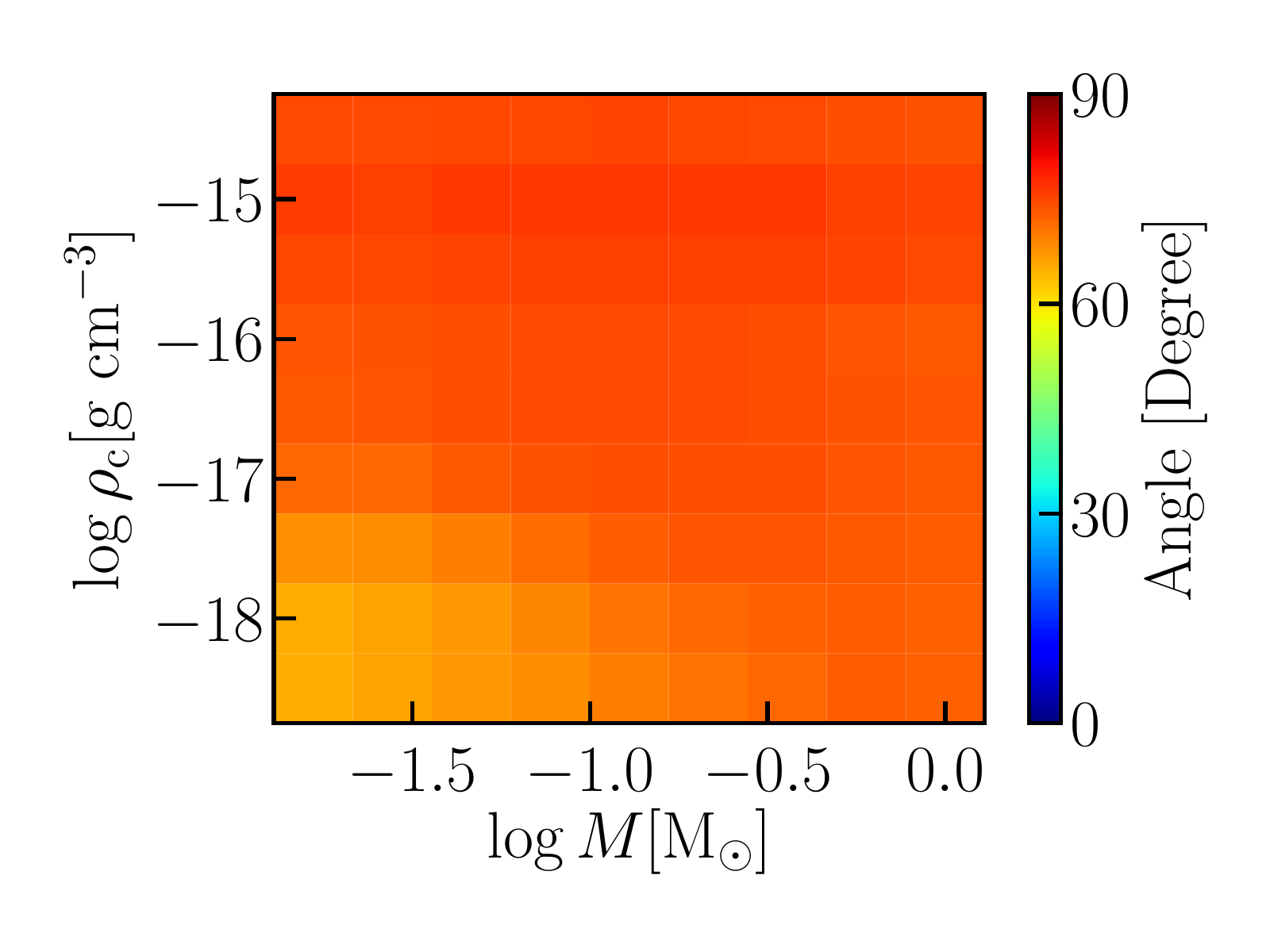}
\caption{Time evolution of the angle between the rotation axis of the inner region of the core and the filament axis averaged for all the cores. The horizontal axis is  the mass of the inner region. The inner region is defined as the spherical region enclosed by any given spherical shell. The vertical axis is the maximum density of a core. }
\label{fig:jfil_angle_colormap}
\end{figure}

Figure \ref{fig:jfil_angle_colormap} displays the time evolution of the angle between the rotation axis of the inner region of the core and the filament axis. Here, the inner region is defined as the spherical region enclosed by any given spherical shell. The mass of the inner region corresponds to the horizontal axis of Figure \ref{fig:jfil_angle_colormap}. The color scale represents the angle between the rotation axis of the inner region of the core and the filament axis averaged over all the cores. The vertical axis is the maximum density. The horizontal axis is the enclosed mass within a shell and corresponds to the distance from the center of the core. Figure \ref{fig:jfil_angle_colormap} indicates that the angle between the rotation axis and the filament axis is almost constant over time and nearly perpendicular to the filament axis over the whole mass range, even in the inner region of the core. Note that $\theta$ of the inner region of the core slightly changes with time, while the outer region is almost constant, as shown in Equation \ref{eq:djest}.  The reason why the angular momentum of the inner region of the core continues to evolve in the early stage of the runaway collapse is discussed in the next paragraph. At the initial stage, the rotation axis of the inner region of the core tends to be less perpendicular to the filament axis compared to the outer region. This is because the inner region of the core is less affected by the filament geometry compared to the outer region. \par

\begin{figure}[t]
\centering
\includegraphics[width=9cm]{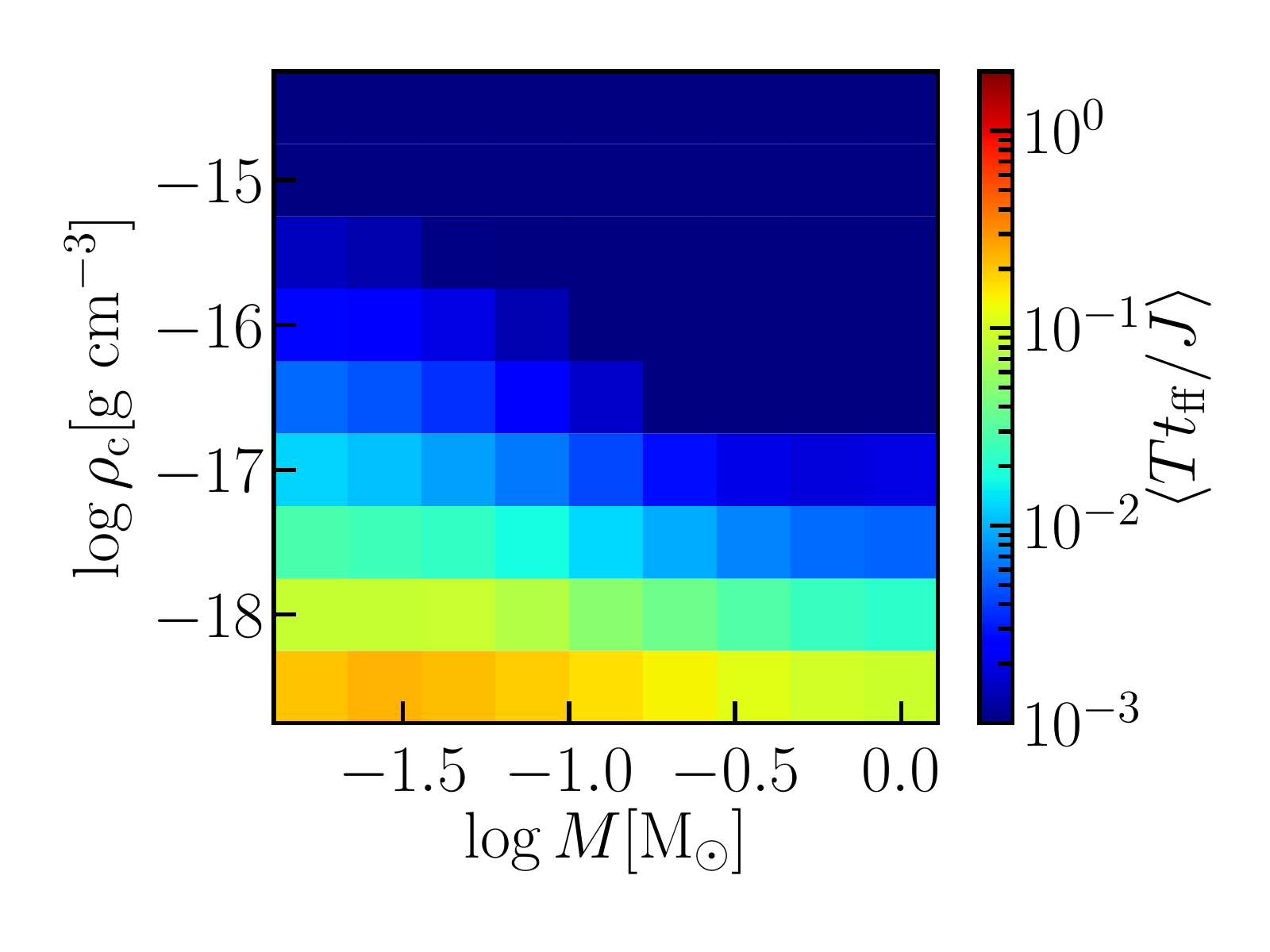}
\caption{Efficiency of the angular momentum transfer due to the total torque during the runaway collapse phase averaged for all the cores. The vertical axis is the maximum density of the core. The horizontal axis is the enclosed mass within the shell. The color shows the efficiency of the angular momentum transfer which is defined by the ratio of the torque acting on the shell during the collapse $T t_{\rm ff}$ and the angular momentum of the shell $J$, where $t_{\rm ff}$ is the free fall time at each time step, and  $\left< \ \right>$ represents the ensemble average for the 38 cores.}
\label{fig:torquetimescale_fg}
\end{figure}

\begin{figure}[t]
\begin{tabular}{c}

\begin{minipage}[t]{.35\textwidth}
\centering
\includegraphics[width=9cm]{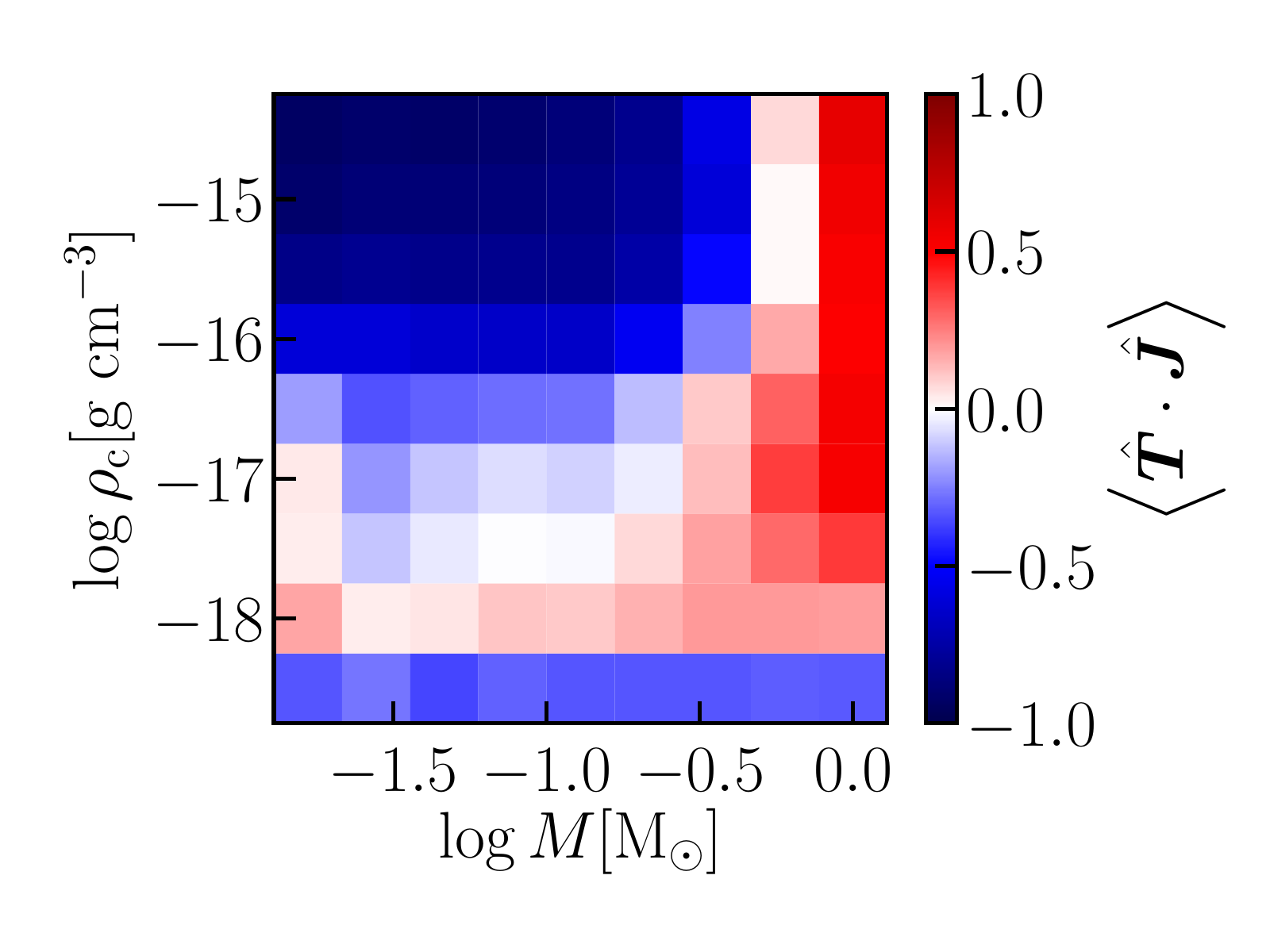}
\end{minipage}\\

\begin{minipage}[t]{.35\textwidth}
\centering
\includegraphics[width=9cm]{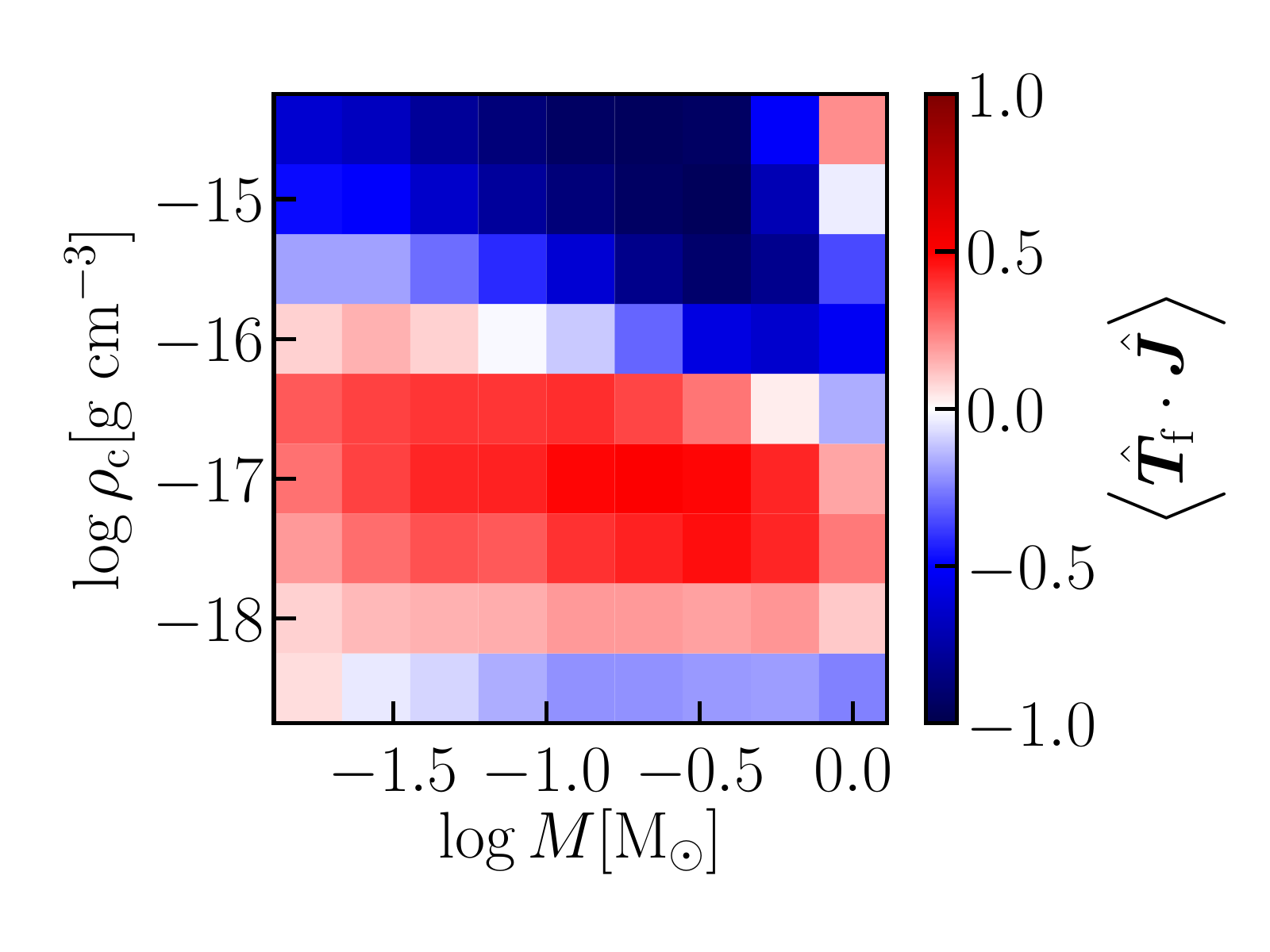}
\end{minipage}
\begin{minipage}{.2\textwidth}
\hspace{10mm}
\end{minipage}

\begin{minipage}[t]{.35\textwidth}
\centering
\includegraphics[width=9cm]{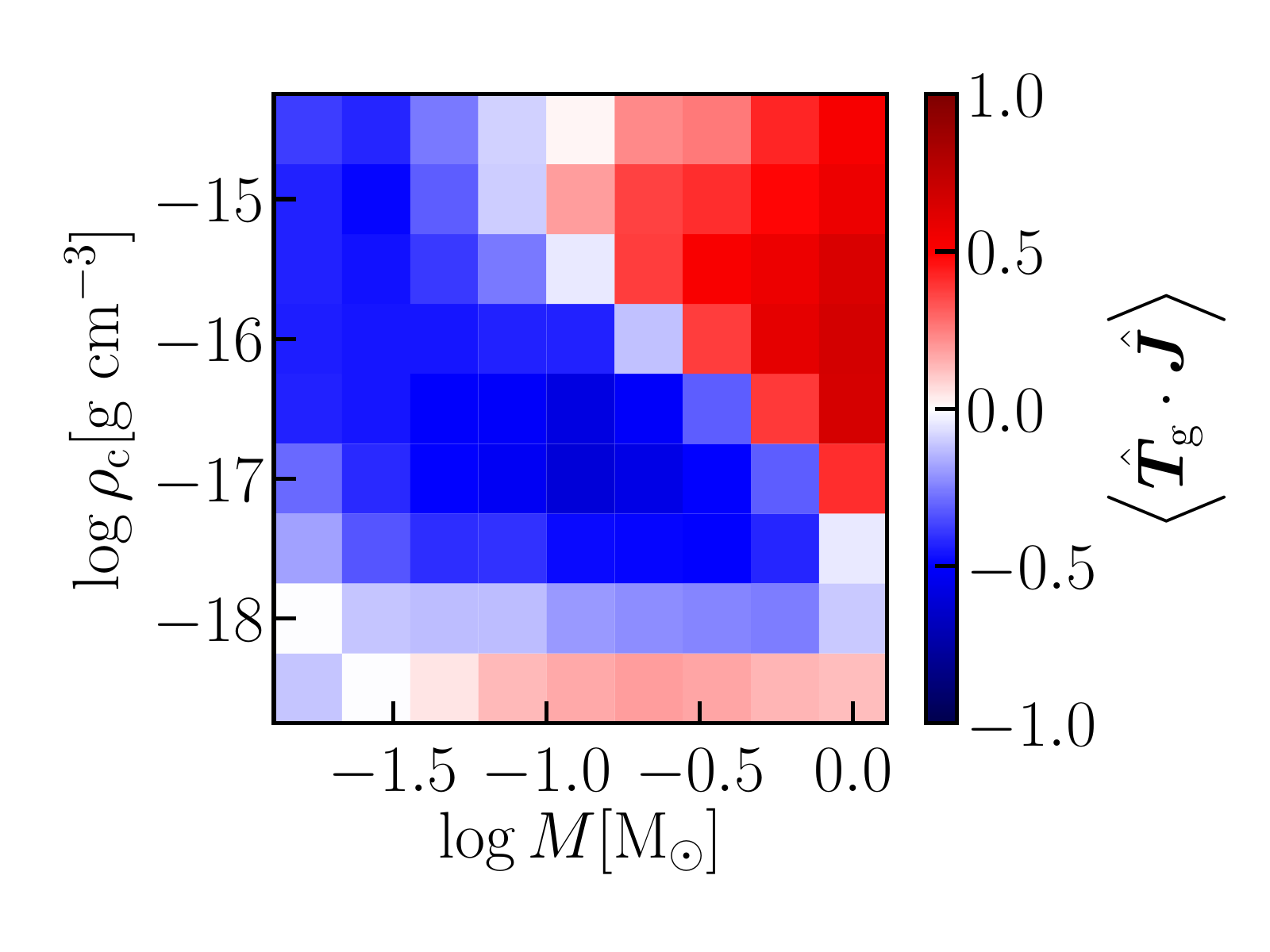}
\end{minipage}

\end{tabular}
\caption{Sign of the total (upper panel), pressure (lower left panel), and gravitational (lower right panel) torques with respect to the angular momentum during the runaway collapse phase averaged for all the cores. The vertical axis is the maximum density of a core. The horizontal axis is the total mass within the shell. The hat symbol means the unit vector. The color shows the inner product between the unit vector of angular momentum and the unit vector of the total (upper panel), pressure (lower left panel), and gravitational (lower right panel) torques.}
\label{fig:torquerate_fg}
\end{figure}

In the following, we study the mechanism of angular momentum transfer in the cores. First, we evaluate the pressure and  gravity torques as follows:
\begin{eqnarray}
\bi{T}_{\rm f,shell}=\sum_{i \in {\rm shell},\ j} (\bi{x}_i-\bi{x}_{\rm \rho max}) \times \bi{F}_{{\rm f},i} ,
\label{eq:tfcalshell}
\end{eqnarray}
\begin{eqnarray}
\bi{T}_{\rm g,shell}=\sum_{i \in {\rm shell},\ j} (\bi{x}_i-\bi{x}_{\rm \rho max}) \times \bi{F}_{{\rm g},i}.
\label{eq:tgcalshell}
\end{eqnarray}
In Figure \ref{fig:torquetimescale_fg}, we show the efficiency of the angular momentum transfer due to the total torque during the runaway collapse phase, averaged for the 38 cores. The color shows the efficiency of the angular momentum transfer which is defined by the ratio of $T t_{\rm ff}$ and the angular momentum of the shell $J$, where $t_{\rm ff}$ is the free fall time at each time step, $T$ is the total torque exerted on the shell.  Figure \ref{fig:torquetimescale_fg} shows that the transfer of the angular momentum due to the total torque is efficient to change the angular momentum only when the central density $\lesssim 10^{-17} \ {\rm g \ cm^{-3}}$. Since most observed starless cores have densities lower than $10^{-17} \ {\rm g \ cm^{-3}}$, they are in the regime where the transfer of angular momentum is effective. In our simulations, once the central density reaches $\rho_{\rm c} \sim 10^{-17} \ {\rm g \ cm^{-3}}$, the evolution of the central density is well described by the self similar solution. Therefore, the angular momentum transfer occurs almost entirely before the collapse converges to the self similar solution.  The angular momentum is an increasing function of the enclosed mass as shown in Figure \ref{ssomegaevo}. This is the reason why the total torque acting on a core is more efficient in the inner region than in the outer region until the start of the runaway collapse phase.
Figure \ref{fig:torquerate_fg} displays the sign of the total, pressure, and gravitational torques with respect to the angular momentum during the runaway collapse phase. The vertical and the horizontal axes are the same as in Figure \ref{fig:torquetimescale_fg}. The hat symbol denotes the unit vector. The color shows the inner product between the unit vector of the angular momentum and the unit vector of the total, pressure, and gravitational torques. The upper panel of Figure \ref{fig:torquerate_fg} indicates that the total torque exerted on the cores reduces the angular momentum of the cores. This is because the initial shape of the core has a complex structure as shown in Figure \ref{corelag}. The reason why the negative pressure torque acts on the core in the initial evolutionary stage is as follows. At the initial state, the major axis of the core is not aligned with the filament axis as shown in Figure \ref{corelag} (d). However, as time progresses, the core major axis tends to line up with the filament axis (Figure \ref{corelag} (e)). In Figure \ref{corelag}(d)-(f), the position angle of the core major axis changes in the counterclockwise direction, then the major axis of the core passes through the $z$-axis. This is the typical evolution observed in our simulations. This rotational direction is preferred to gather the mass of a core. Since the initial filament is in hydrostatic equilibrium, the direction of the pressure torque and the gravitational torque are opposite. However, in the runaway collapse phase, the cores forget the initial filament geometry, and both torques become negative with respect to the rotation direction as shown in Figure \ref{fig:torquerate_fg}.

\subsubsection{Complexity of the Internal Angular Momentum Structure}

\begin{figure}[t]
\begin{tabular}{c}
\begin{minipage}[t]{.95\textwidth}
\centering
\includegraphics[width=13cm]{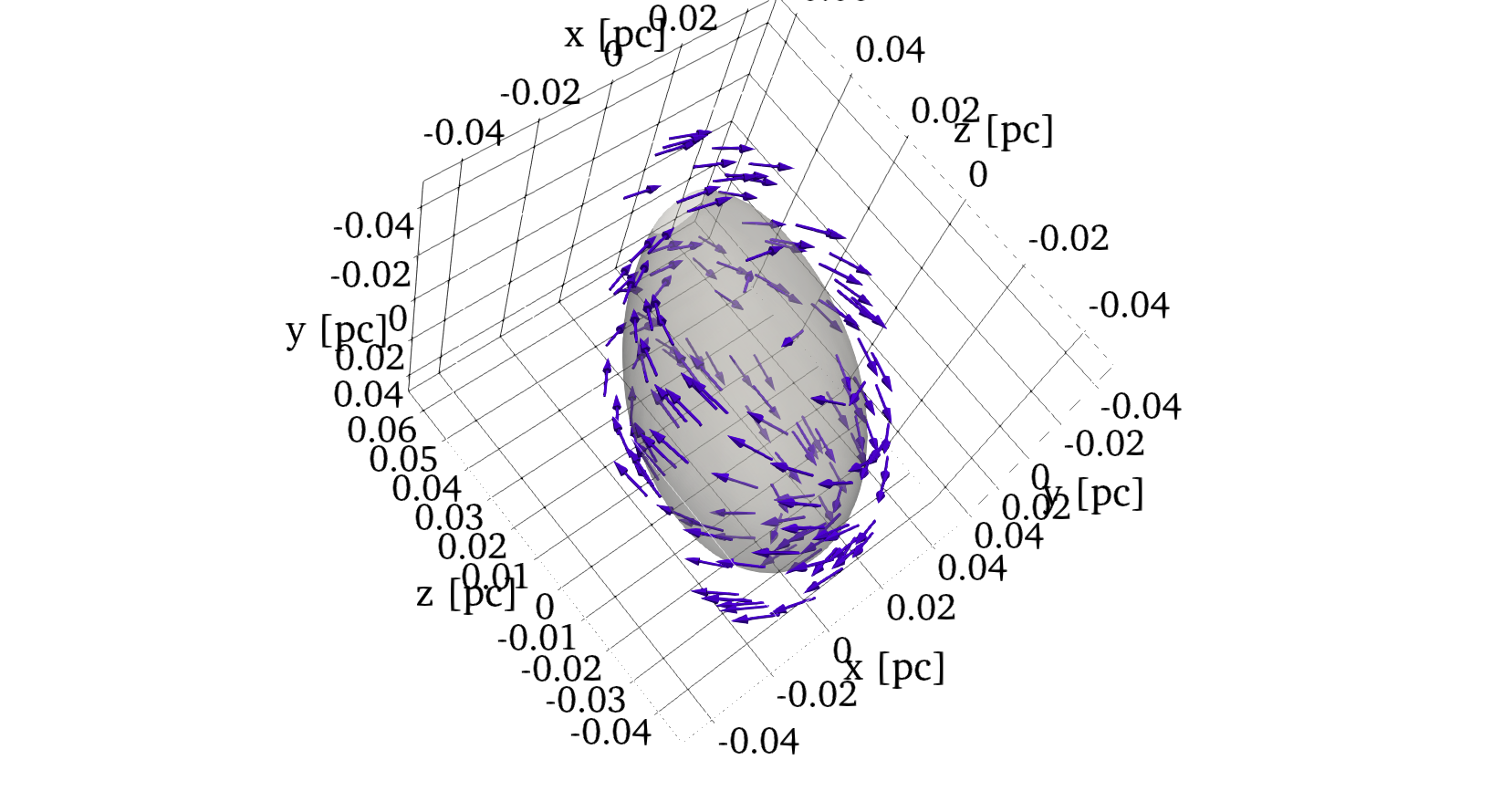}
\end{minipage}\\

\begin{minipage}[t]{.95\textwidth}
\centering
\includegraphics[width=13cm]{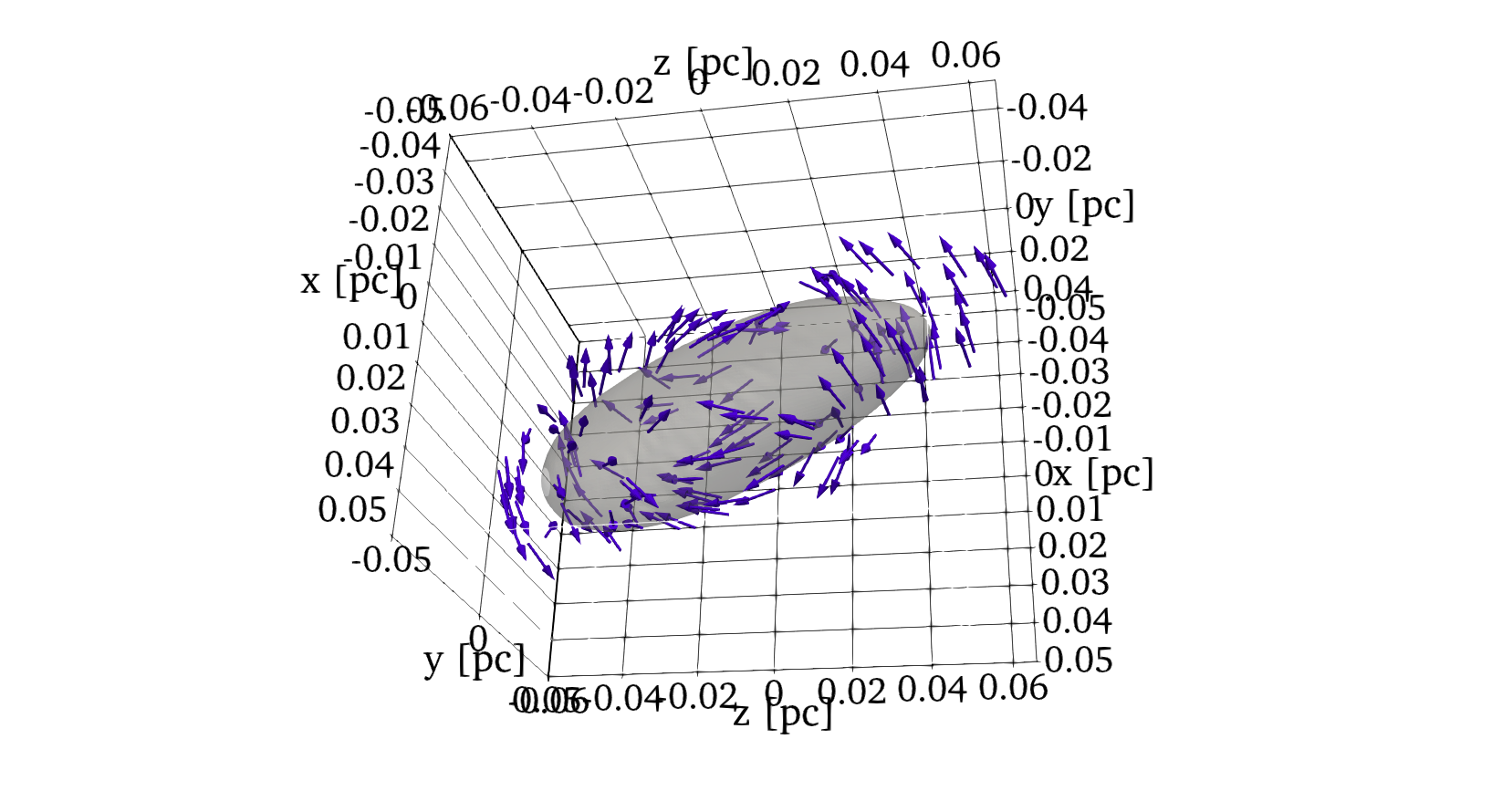}
\end{minipage}

\end{tabular}
\caption{3D plots of the velocity structure in two cores formed in our simulations. The upper and lower panels are smooth and complex velocity structure cases, respectively. The isosurfaces represent the isodensity surfaces, $\rho=3.0\times 10^{-19} \ {\rm g \ cm^{-3}}$ ($n=7.8\times 10^{4} \ {\rm cm^{-3}}$) (upper panel) and $\rho=3.5 \times 10^{-19} \ {\rm g \ cm^{-3}}$ ($n=9.1\times 10^{4} \ {\rm cm^{-3}}$) (lower panel). The blue arrows are the direction of the rotation velocity. The central densities of the cores are $\rho=8.0\times 10^{-19} \ {\rm g \ cm^{-3}}$ ($n=2.0\times 10^{5}\ {\rm cm^{-3}}$) (upper panel) and $\rho=1.2 \times 10^{-18} \ {\rm g \ cm^{-3}}$ ($n=3.1\times 10^{5}\ {\rm cm^{-3}}$) (lower panel)}
\label{fig:vel_paraview}
\end{figure}

Figure \ref{fig:vel_paraview} is a comparison between a clean rotation pattern (nearly uniform angular momentum distribution in the core) and a complex rotation pattern, both found in our simulations. The upper and lower panels show smooth and complex velocity structure examples, respectively. The isosurfaces represent the isodensity surfaces, $\rho=3.0\times 10^{-19} \ {\rm g \ cm^{-3}}$ (upper panel) and $\rho=3.5 \times 10^{-19}\  {\rm g \ cm^{-3}}$ (lower panel). The blue arrows are the direction of the rotation velocity. The rotation velocity $v_{\rm rot}$ is defined as follows:
\begin{eqnarray}
{\bi v}_{\rm rot} = {\bi v} - {\bi v}_{\rho {\rm max} }- [({\bi v} - {\bi v}_{\rho {\rm max}}) \cdot {\bi r}]{\bi r}/r^2,
\label{eq:vrot}
\end{eqnarray}
where $ {\bi v}_{\rho {\rm max}}$ is the velocity of the density peak, and ${\bi r}$ is the distance from the density peak. The central densities of the cores are $\rho=8.0\times 10^{-19}\ {\rm g \ cm^{-3}}$ (upper panel) and $\rho=1.2 \times 10^{-18} \ {\rm g \ cm^{-3}}$ (lower panel). Although a smooth rotation pattern can be seen in the upper panel of Figure \ref{fig:vel_paraview}, a complex velocity structure can be observed in the lower panel. In the lower panel, the rotation velocity pattern around the central region is different from the rotation velocity pattern of the outer shell of the core. This kind of complex rotation structure appears due to the combination of different phases of the initial turbulent velocity field. 

\clearpage

\begin{figure}[t]
\centering
\includegraphics[width=10cm]{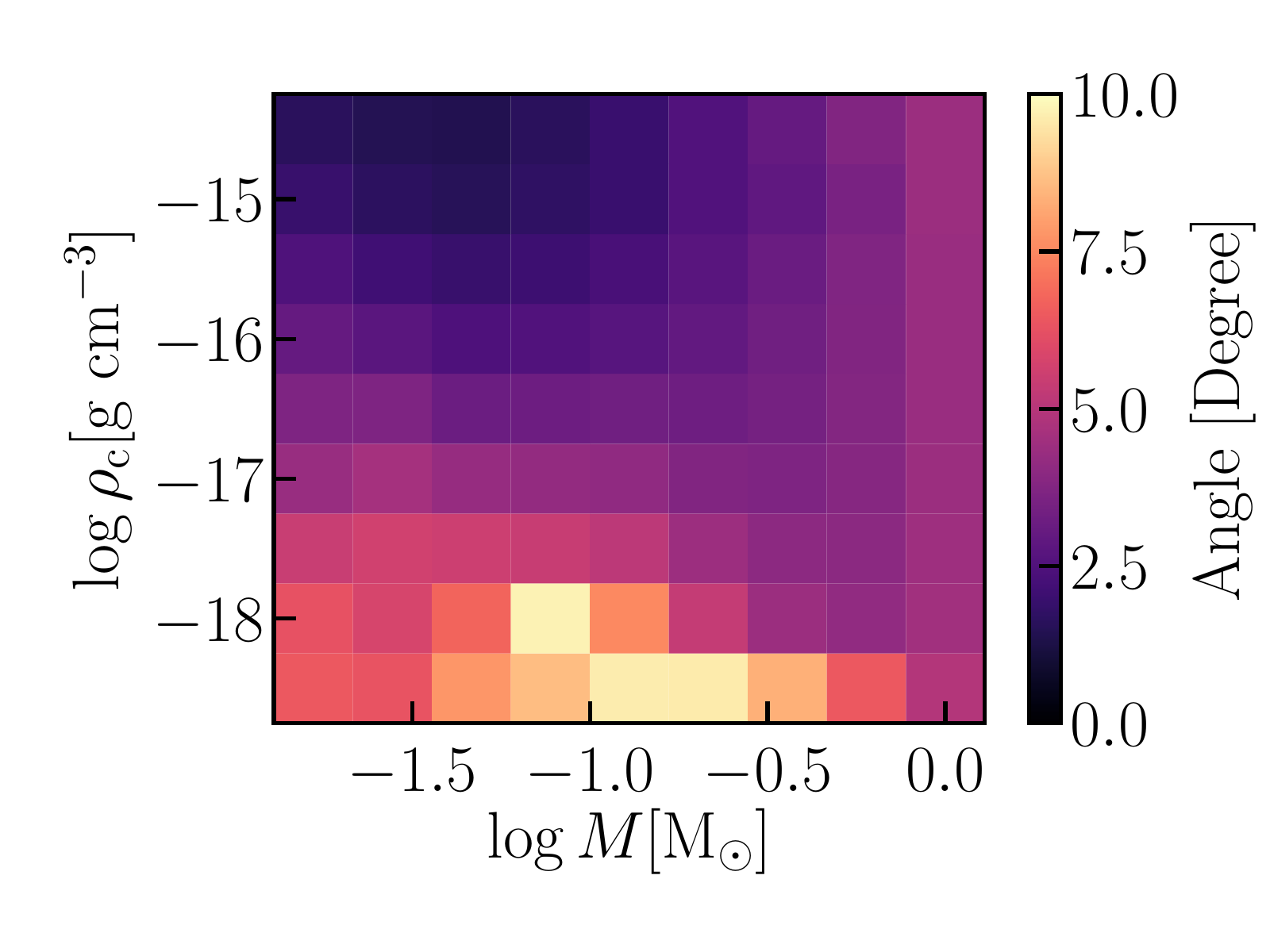}
\caption{Angle between the angular momentum vector of the inner region and the outer shell. Here, we define the inner region as the region enclosed by the outer shell where the inner region and the outer shell have both the same mass. The color scale represents the angle between the rotation axis of the inner region and the outer shell averaged over all the cores. The horizontal axis is the total mass within the shell. The vertical axis is the maximum core density.}
\label{shellinc}
\end{figure}

\begin{figure}[t]
\vspace{20mm}
\begin{tabular}{cc}
\begin{minipage}[t]{.35\textwidth}
\centering
\includegraphics[width=8cm]{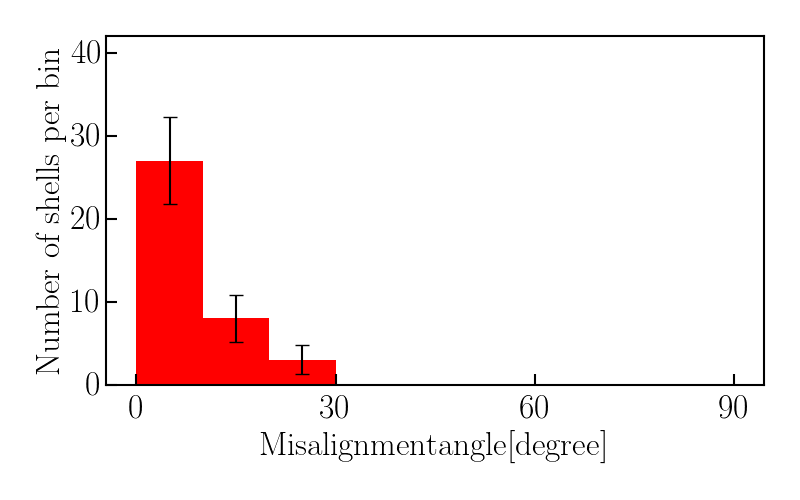}
\end{minipage}
\begin{minipage}{.15\textwidth}
\hspace{10mm}
\end{minipage}

\begin{minipage}[t]{.35\textwidth}
\centering
\includegraphics[width=8cm]{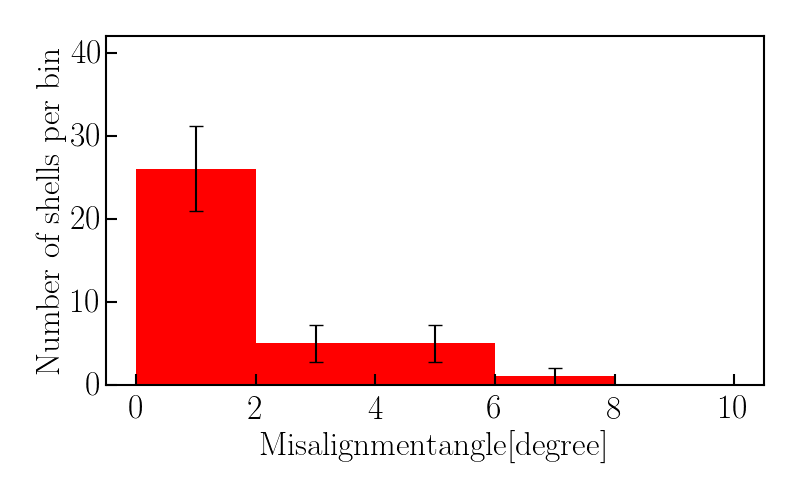}
\end{minipage}

\end{tabular}
\caption{Histograms of the angles between the angular momentum of the inner region with a mass of $0.02 \ {\rm M}_{\odot}$ and the outer shell with the same mass for the 38 cores at the initial state (left) and at the final state (right) with statistical error bars. Note that the scales of the $x$-axis are not the same on both panels.}
\label{fig:shell_ang_hist}
\end{figure}

Figure \ref{shellinc} displays the angle between the rotation axis of the inner region and the outer shell both with the same mass. Here, we define the inner region as the region enclosed by the outer shell. The color scale represents the angle between the rotation axis of the inner region and the outer shell averaged over all the cores. Figure \ref{shellinc} indicates that the complexity of the angular momentum structure of the cores slightly decreases with time. This trend also can be slightly seen in Figure \ref{fig:jfil_angle_colormap}. However, the angular momentum of the outer shell of the core ($\sim 0.5 \ {\rm M}_{\odot}$) has a relatively large inclination with respect to the angular momentum of the inner region even at the final stage of our simulation as shown in Figure \ref{shellinc}. This is because the outer shell of the core does not have enough time to transfer its large angular momentum during the runaway collapse phase. Figure \ref{fig:shell_ang_hist} shows histograms of the angles between the angular momentum of the inner region with a mass of $0.02 \ {\rm M}_{\odot}$ and that of the outer shell with the same mass of $0.02 \ {\rm M}_{\odot}$ for the 38 cores at the initial state (left) and at the final state (right). The averaged inclination angle of inner region of $0.02 \ {\rm M}_{\odot}$ evolves from $6^{\circ}$ at $\rho_c = 10^{-18} \ {\rm g \ cm^{-3}}$ to $2^{\circ}$ at $\rho_c = 10^{-14} \ {\rm g \ cm^{-3}}$. The dispersion of the histogram evolves from $9^{\circ}$ at $\rho_c = 10^{-18} \ {\rm g \ cm^{-3}}$ to $2^{\circ}$ at $\rho_c = 10^{-14} \ {\rm g \ cm^{-3}}$. These histograms also show that the complexity of the core angular momentum structure slightly decreases as time progresses. However, it still remains at the final state of our simulation especially in the outer region of the cores, just before the first core formation. Therefore, this kind of complex angular momentum structure might be related to the warped (or misaligned) disk around the protostar  observed by e.g., \cite{Sakai2019}. In the future, we will perform long term and high resolution simulations to investigate the formation of such misaligned disks.

\subsection{Angular Momentum in the Central Region of the Cores}
\begin{figure}[t]
\centering
\includegraphics[width=12cm]{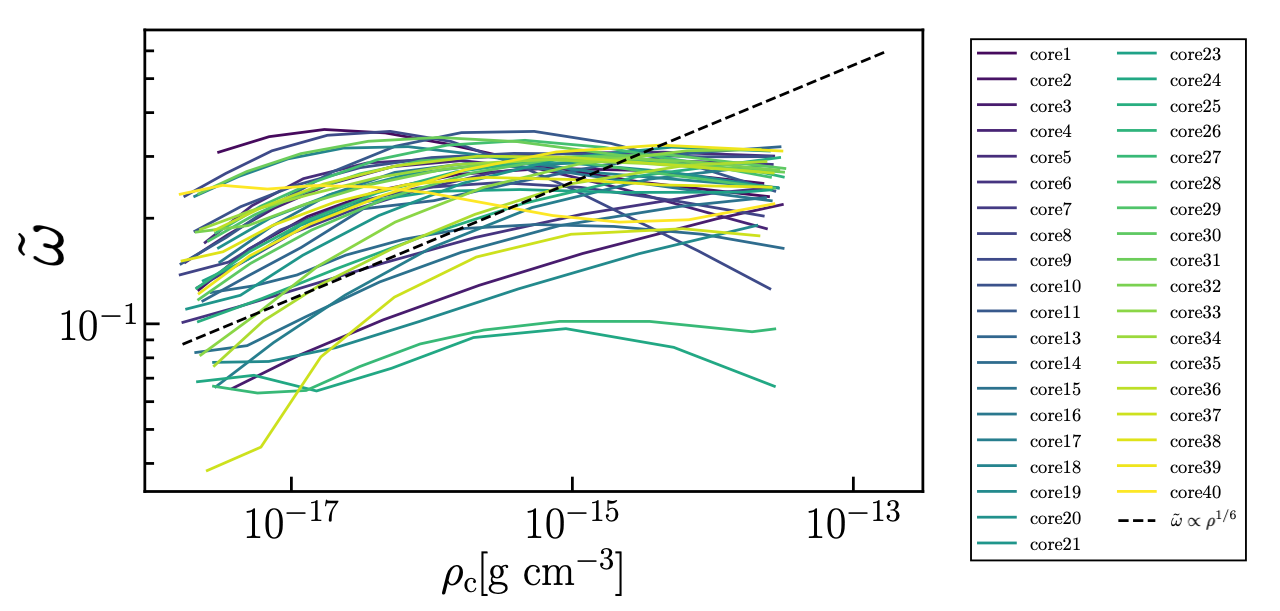}
\caption{Time evolution of the angular velocity derived from Equation \ref{eq:omega01}. The vertical axis is the normalized angular velocity of the central region of the cores. The horizontal axis is the maximum density of the cores. The different colors of the solid lines correspond to the different cores. The black dashed line is $\tilde{\omega} \propto \rho_{\rm c}^{1/6}$ discussed in \cite{Hanawa1997}.}
\label{centomega_evo}
\end{figure}

\begin{figure}[t]
\centering
\includegraphics[width=10cm]{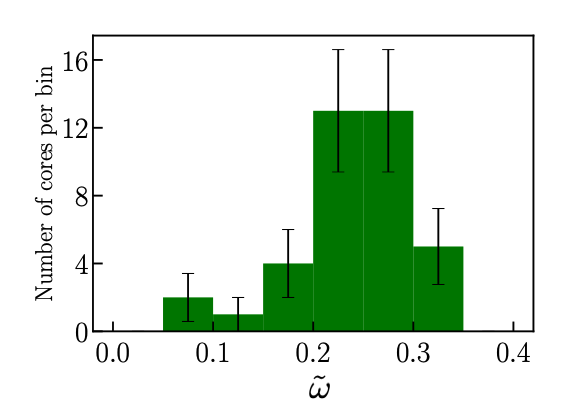}
\caption{Histogram of the angular velocity derived from Equation \ref{eq:omega01} at the final stage. The horizontal axis is the normalized angular velocity of the central region of the cores. The error bars refer to the standard deviation. }
\label{centomega}
\end{figure}

The rotation and shape of the central high density region of a core are important for the formation of multiple systems \citep{Matsumoto2003,Machida2005} and the subsequent formation of protostar-disk systems. In this subsection, we analyze the central high density region of the cores. In this analysis, the central high density region is defined as the gas with $\rho>0.1\rho_{\rm max}$. We calculate the angular momentum and the moment of inertia of the central region of a core as follows:
\begin{eqnarray}
\bi{J}_{01}=\sum_{i, \rho_i>0.1\rho_{\rm max}} m_i (\bi{x}_i-\bi{x}_{\rm 01, c}) \times (\bi{v}_i-\bi{v}_{\rm 01, c}),
\label{eq:angcalcen}
\end{eqnarray}
and 
\begin{eqnarray}
I_{ln,01}=\sum_{i, \rho_i>0.1\rho_{\rm max}} m_i [(\bi{x}_{i}-\bi{x}_{{\rm 01, c}})^2 \delta_{ln}- (x_{l,i}-x_{l, {\rm 01, c}}) (x_{n,i}-x_{n, {\rm 01, c}})],
\label{eq:inertia01}
\end{eqnarray}
$\bi{x}_{\rm 01, c}$ and $\bi{v}_{\rm 01, c}$ are the position and velocity of the center of mass of the core central region. We also estimate the angular speed of the core central region as follows:
\begin{eqnarray}
\omega_{01,l}=\sum_{n}I^{-1}_{01,ln} J_{01,n},
\label{eq:omega01}
\end{eqnarray}
where $I^{-1}_{01,ln}$ is the inverse matrix of $I_{01,ln}$. Figure \ref{centomega_evo} displays the time evolution of the normalized angular speed of the central region, $\tilde{\omega}_{01}={\omega}_{01}t_{\rm ff}$, where  $t_{\rm ff}=1/\sqrt{4\pi G \rho_{\rm c}}$ is the free fall time of the central region at each time step. $\omega_{01}$ is the magnitude of the angular velocity vector. The black dashed line is $\tilde{\omega} \propto \rho_{\rm c}^{1/6}$, which corresponds to the growth rate of $\tilde{\omega}_{01}$ discussed in \cite{Hanawa1997}. In \cite{Hanawa1997}, they did the linear analysis for Larson-Penston solution and found that there is a spin-up mode which can grow slowly during the self-similar collapse. The spin-up suggested by  \cite{Hanawa1997} is not clearly observed in our simulations since the angular velocity is large even at the early evolutionary stage. Figure \ref{centomega} is the histogram of $\tilde{\omega}_{01}$ at the final state. Figure \ref{centomega} shows that the peak position of the histogram is $\tilde{\omega}_{01}\sim 0.2$. This result indicates that the central region of the cores converges to the self-similar solution characterized by a rotation consistent with the values discussed in \cite{Matsumoto1997} and \cite{Matsumoto2003}.

\section{Discussion}

\subsection{An Origin for the Specific Angular Momentum Profile in a Core}
As presented in Section 3.4 and 3.5, the rotation profile in the cores converges to the self-similar solution, $j \propto M$.  As discussed in \cite{Saigo1998}, the self-similar solution with rotation can be derived analytically by assuming a geometrically thin disk and symmetry around the rotation axis (see also \cite{Basu1997}).
The origin of the $j \propto M$ relation can be understood as follows. The mass of the central high density region is given by  
\begin{eqnarray}
M \sim \rho_{\rm c}\lambda^3 \propto \rho_{\rm c}^{-1/2},
\label{eq:Mrhoc}
\end{eqnarray}
where $\lambda$ is the Jeans length of the central region. Here, the central high density region corresponds to the flat inner part in the density profile of a collapsing core before protostar formation. The specific angular momentum of the central region can be estimated as follows:
\begin{eqnarray}
j \sim \lambda^2 \omega \propto \rho_{\rm c}^{-1/2},
\label{eq:jrhoc}
\end{eqnarray}
where we used $\omega \propto \rho_{\rm c}^{1/2}$. This is because the time scale of the system is only the free-fall time.
Using Equation \ref{eq:Mrhoc} and \ref{eq:jrhoc}, we can derive
\begin{eqnarray}
j \propto M.
\label{eq:jM}
\end{eqnarray}

The radial extent of the region in which the $j \propto M$ scaling is expected is determined by whether $\tilde{\omega}$ reaches 0.2 or not. \cite{Matsumoto1997} shows that  $\tilde{\omega}$ increases due to the spin up during the collapse and eventually it reaches $\tilde{\omega}=0.2$. Once  it reaches $\tilde{\omega}=0.2$,  $\tilde{\omega}$ remains constant although some oscillations can be observed in Figure \ref{centomega_evo} and in \cite{Matsumoto1997}. To estimate the radial extent of $j \propto M$ scaling, it is important to know when $\tilde{\omega}$ is saturated since we used $\omega \propto \rho_{\rm c}^{1/2}$ to derive Equation \ref{eq:jM}. In Figure \ref{centomega_evo}, most cores reaches $\tilde{\omega} \sim 0.1$ even at the early stage of the collapse ($\rho_{\rm c} \sim 10^{-18} \ {\rm g \ cm^{-3}}$).  Hence, for simplicity, the radial extent of $j \propto M$ scaling is estimated from the intersection of the self-similar profile ($j \propto M$) and $j$-$M$ profile at the initial state. At the initial state, we can estimate the specific angular momentum as follows. The specific angular momentum of the core gained from the initial Kolmogorov turbulent velocity field at the initial state is written as

\begin{eqnarray}
j = \frac{2}{5}\sigma(r) r,
\label{eq:inijsimple}
\end{eqnarray}
where $\sigma(r)$ is the velocity dispersion and the factor comes from the inertia moment. In Equation \ref{eq:inijsimple}, we assume a uniform density, $\rho_{c0}$ for the core. This assumption is justified since the cores form close to the crest of a filament that has a plummer density profile with flat central part (see Section 2 and Equation (2.2)). The core thus forms from a uniform density medium and has a uniform $\rho_{\rm c0}$. After some calculations, we can derive the specific angular momentum profile at the initial state as follows:

\begin{eqnarray}
j = 3.5 \times 10^{20} {\rm cm^2 \ s^{-1} }  \left(\frac{\sigma_0}{0.2 \ {\rm km \ s^{-1}}}  \right)  \left(\frac{M}{ {\rm M}_{\odot}}   \right)^{4/9} ,
\label{eq:jinisimle2}
\end{eqnarray}
where we used $\sigma(r) = \sigma_0 (r/1.6 \ {\rm pc})^{1/3}$ and $M=4 \pi \rho_{c0} r^3 /3$. We adopt $\sigma_0 = 0.2 \ {\rm km \ s^{-1}}$ equivalent to the sound speed for $T=10 \ {\rm K}$.  The self-similar profile can be  roughly estimated as follows: 

\begin{eqnarray}
j = \frac{2}{5}r^2 \omega.
\label{eq:selfsimamp}
\end{eqnarray}

Using $\omega= 0.2 \sqrt{4\pi {\rm G} \rho_{\rm c}}$ and the surface density $\Sigma = \sqrt{2c_{\rm s}^2 \rho_c/\pi G}$, we can derive the self-similar profile,

\begin{eqnarray}
j = 6.3 \times 10^{20}  {\rm cm^2 \ s^{-1} }  \left(\frac{M}{ {\rm M}_{\odot}}   \right) .
\label{eq:selfsimamp2}
\end{eqnarray}

The intersection of Equation \ref{eq:jinisimle2} and Equation \ref{eq:selfsimamp2} defines the boundary of the self-similar solution (radial extent), 

\begin{eqnarray}
M_{\rm ss} = 0.35 \  {\rm M}_{\odot}   \left(\frac{\sigma_0}{0.2 \ {\rm km \ s^{-1}}}  \right)^{9/5} .
\label{eq:selfsimcross}
\end{eqnarray}

This is compatible with the result shown in Figure \ref{fig:jMpro}. Using Larson-Penston solution $M = 8.86 c_{\rm s}^2 r/{\rm G}$, we can derive the radius as follows:

\begin{eqnarray}
r_{\rm ss} = 870 \ {\rm AU}  \left(\frac{\sigma_0}{0.2 \ {\rm km \ s^{-1}}}  \right)^{9/5}   .
\label{eq:selfsimradius}
\end{eqnarray}

\subsection{Impact of the Filamentary Structure on the Angular Momentum of Cores}
As discussed in Section 1, observations show that the filaments are ubiquitous in molecular clouds, and that molecular cloud cores lie along these filaments. This indicates that the cores (and stars) are formed in filamentary structures. This scenario is often referred to as the ``filament paradigm'' \cite[e.g.,][]{Andre2014, Pineda2022}. In this subsection, we summarize the role of the filament paradigm in the evolution of the angular momentum of cores.\par
In Section 3.4.1 and 4.1, we discussed the convergence of the internal angular momentum profile to the self-similar solution. This tendency has been already reported in, for example, \cite{Matsumoto2003} in which they adopted a spherical core as the initial condition. Therefore, our results about the internal angular momentum profile agree with the previous works although we adopt the filamentary molecular cloud with subsonic (transonic) turbulent velocity field as the initial condition.\par
On the other hand, the orientation of the angular momentum of cores is affected by the filament geometry. As shown in Section 3.3, the rotation direction of the cores tend to be perpendicular to the longitudinal axis of the filament since the initial shape of the cores is elongated along the longitudinal axis of the filament. In observations, the relation between the core rotation axis and the filament axis has been studied using the angle between the outflow and the filament axis. For example, \cite{Kong2019} reported that the outflow orientation tends to be perpendicular to the filament axis using Atacama Large Millimeter/submillimeter Array (ALMA) CO(2-1) observations in G28.37+0.07 . More recently, \cite{Xu2022} showed a preference for alignment of observed outflow axes with magnetic field directions measured by Planck observations of dust polarization in the nearby molecular clouds Ophiuchus, Taurus, Perseus, and Orion. Since the magnetic fields are perpendicular to the filamentary structures, they conclude that the outflows tend to be perpendicular to the filament axes.  However, \cite{Stephens2017} and \cite{Baug2020} claimed that the outflow is randomly oriented with respect to the filament axes using Submillimeter Array (SMA) and ALMA, respectively. \cite{Stephens2017}, \cite{Kong2019}, and \cite{Baug2020} focused on relatively massive star forming regions, especially the samples analyzed in \cite{Baug2020} are associated with HII regions. \cite{Baug2020} suggested that the outflow directions might depend on the evolutionary stage of a star forming cluster. In \cite{Feddersen2020}, they showed that the distribution of the angle between the outflow and the filament axis is a random distribution in the full sample of outflows using CARMA-NRO Orion survey data. However, they also showed that the outflow direction is moderately perpendicular to the filament axis in the most reliable subsample. Hence, the observations of distribution of the angle between the outflow and the filament axis is still under debate. In addition, \cite{Machida2020} investigated the effect of misalignment of the initial core rotation axis with respect to the initial magnetic field orientation and showed that the outflow direction changes with time. This theoretical study indicates that we cannot simply define the angle between the outflow direction and the filament axis as the angle between the rotational axis of the core and the filament axis. Note that the anisotropy of turbulent velocity field in a filament might affect the rotation direction of a core with respect to the filament axis. As discussed in \cite{Misugi2019}, the anisotropy of the turbulent velocity field in filaments might be created at the time of filament formation from the accumulation of matter along magnetic field lines perpendicular to the filament axis within a flattened layer formed by a large scale compression \cite[e.g.,][]{Arzoumanian2018,Arzoumanian2022,Inoue2018,Abe2021}. Although the anisotropy of velocity field may affect the resultant core rotation direction. We plan to address this effect in future studies.\par

\subsection{Synthetic Maps and Comparison with Observations}

In the previous section, we derived various properties of the three-dimensional angular momentum of the cores. However, in observations, the specific angular momentum is measured using the line of sight velocity in two-dimensional plane \citep{Goodman1993, Caselli2002, Tatematsu2016, Punanova2018}. To mimic a line of sight velocity map, we integrate the velocity along the line of sight direction as follows:
\begin{eqnarray}
v_{\rm los}(x,z) =  \frac{1}{\Sigma(x,z)}\int \rho v_{\rm y} dy,
\label{eq:vlos}
\end{eqnarray}
where $\Sigma(x,z)$ is the column density defined as
\begin{eqnarray}
\Sigma(x,z) = \int \rho dy.
\label{eq:sigma}
\end{eqnarray}

\begin{figure}[t]
\begin{tabular}{cc}
\begin{minipage}[t]{.35\textwidth}
\centering
\includegraphics[width=9cm]{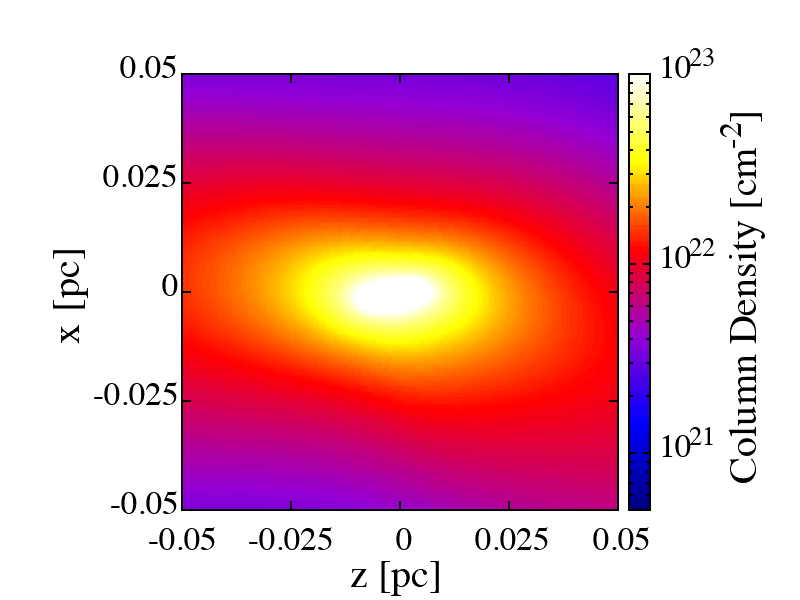}
\end{minipage}
\begin{minipage}{.20\textwidth}
\hspace{10mm}
\end{minipage}

\begin{minipage}[t]{.35\textwidth}
\centering
\includegraphics[width=9cm]{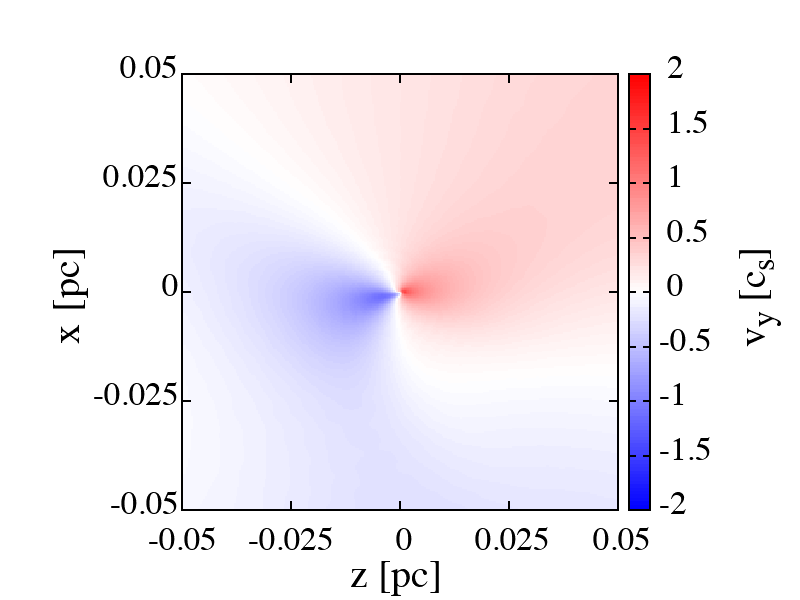}
\end{minipage}

\end{tabular}
\caption{Column density map (left) and line of sight velocity map (right) around the core at the final state of the simulation. In these plots, the longitudinal axis of the filament is parallel to the plane of the sky.}
\label{fig:obsmap}
\end{figure}

\begin{figure}[t]
\begin{tabular}{cc}
\begin{minipage}[t]{.35\textwidth}
\centering
\includegraphics[width=9cm]{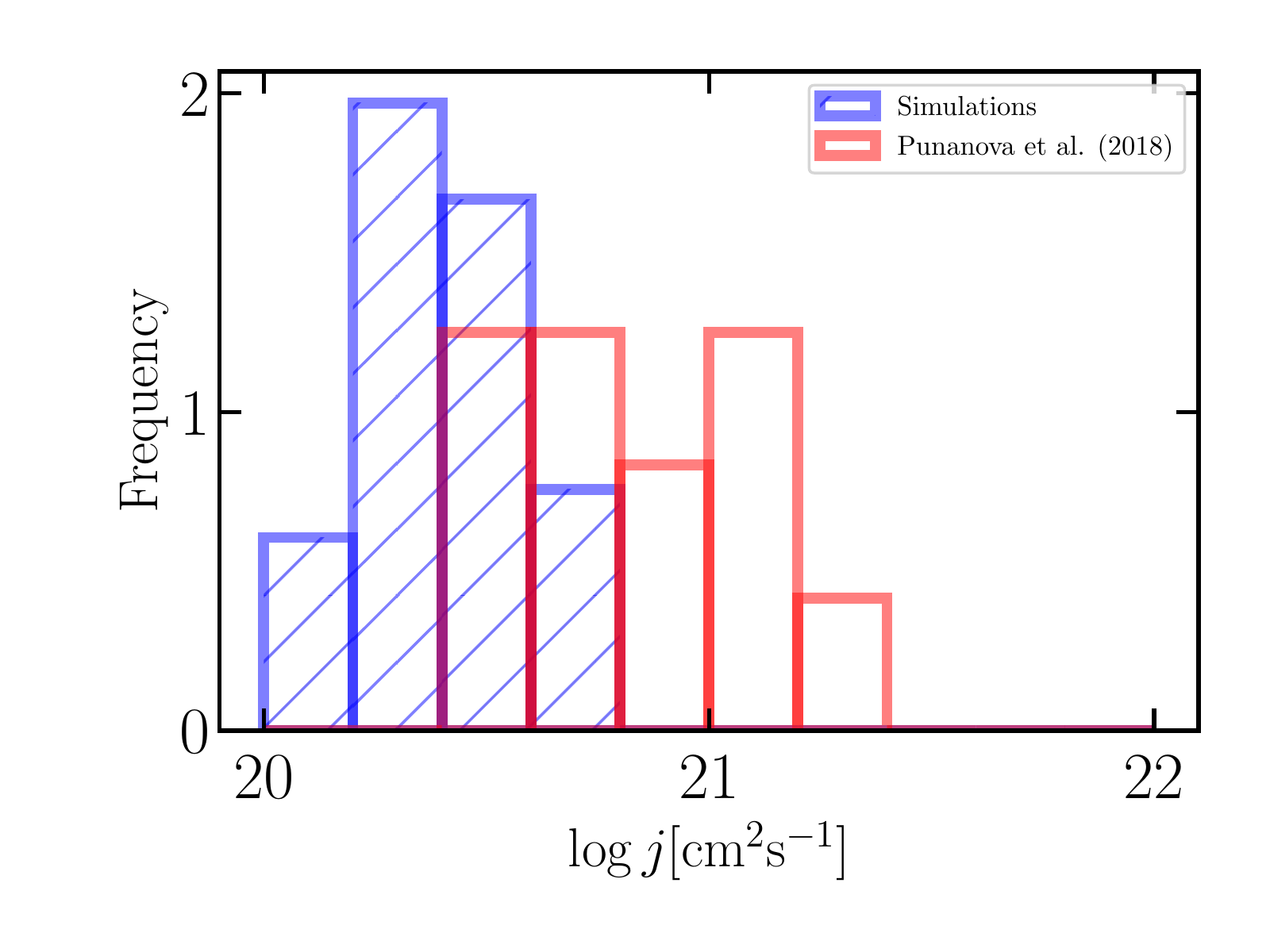}
\end{minipage}
\begin{minipage}{.20\textwidth}
\hspace{10mm}
\end{minipage}

\begin{minipage}[t]{.35\textwidth}
\centering
\includegraphics[width=9cm]{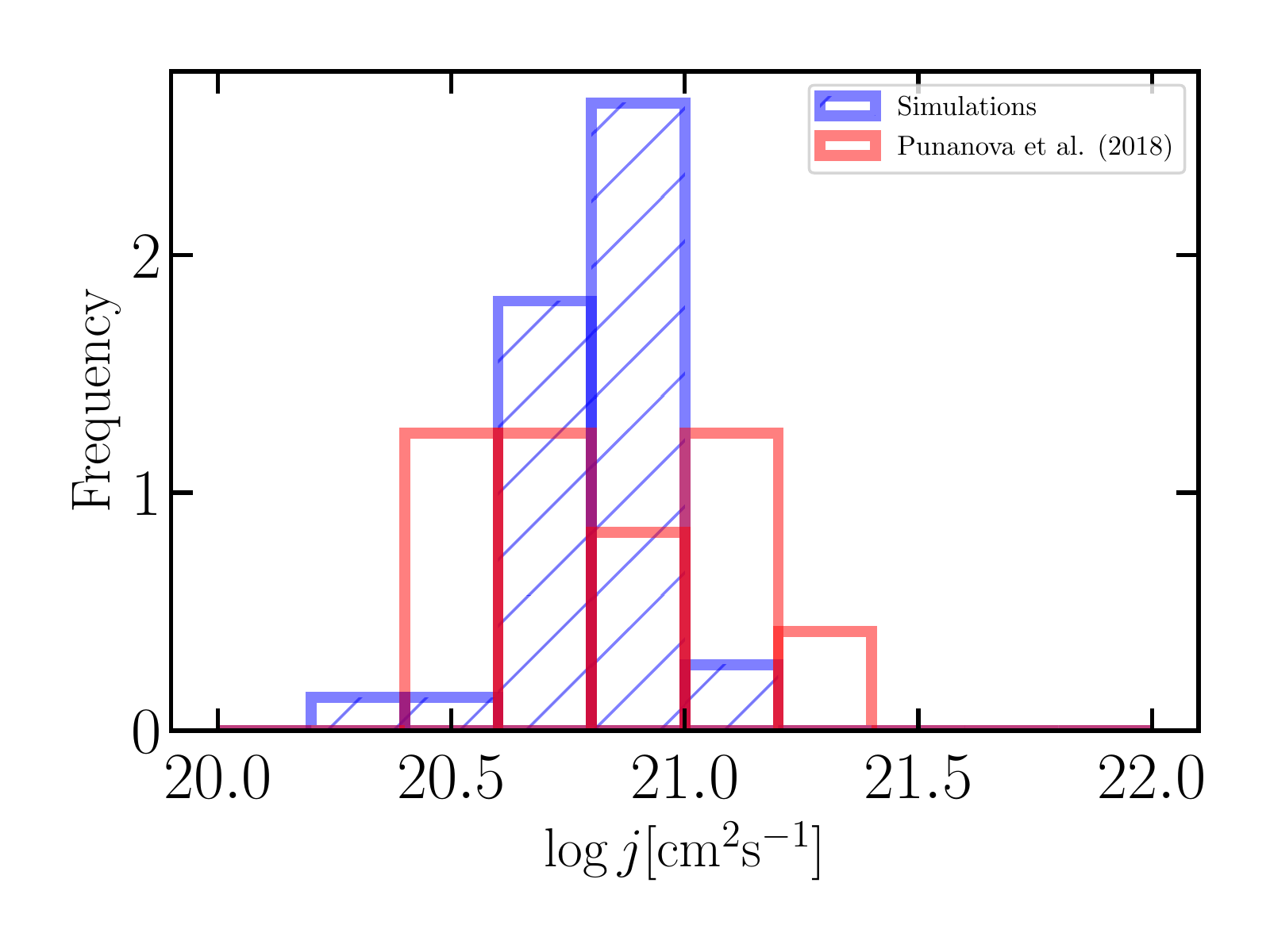}
\end{minipage}

\end{tabular}
\caption{Comparison of $j_{\rm 2D}$ and the observational data. The left and right panels are for not-inclined and for inclined models, respectively. The blue and red histograms are distributions of $j_{\rm 2d}$ and measured specific angular momentum in \cite{Punanova2018} towards a sample of cores in the Taurus molecular cloud using dense gas tracers.  }
\label{fig:j2dhist}
\end{figure}

Figure \ref{fig:obsmap} displays an example of the resultant column density and the mean velocity maps for a given core. In these plots, the longitudinal axis of the filament is parallel to the plane of the sky. Although we mimic the observation of the core specific angular momentum below, the longitudinal axis of the filament ($z$-axis) is not always parallel to the plane of the sky. For this reason, we also analyze the mean velocity maps after rotating the filament around the $x$-axis by $30^{\circ}$, i.e., the filament is inclined by $30^{\circ}$ with respect to the plane of the sky. Using the mean velocity map, we derive the two-dimensional specific angular momentum as follows. First, we fit the mean velocity map using the following equation: 
\begin{eqnarray}
v_{\rm los,fit}(x,z) = v_{\rm c,fit} + \Omega_{x}z - \Omega_{z}x,
\label{eq:fitv2d}
\end{eqnarray}
where $v_{\rm c,fit}$, $\Omega_{x}$, and $\Omega_{y}$ are free parameters of the fitting. This fitting method is used in a lot of previous observational works \citep{Goodman1993, Caselli2002, Tatematsu2016, Punanova2018}. We determine these parameters by minimizing
\begin{eqnarray}
\sigma_{\rm error} = \int \int (v_{\rm los}(x,z)-v_{\rm los,fit}(x,z))^2 dx dz.
\label{eq:mini}
\end{eqnarray}
Next, we calculate $\theta_{\rm 2d,fit} = \arctan{(\Omega_{z}/\Omega_{x})}$ and define $\theta_{\rm 2d,fit}$ as the core rotational axis. Then, we measure the distance $r$ from the rotational axis. Finally, we calculate the average angular momentum in each $dr$ bin. We also derive the total specific angular momentum in the line of sight velocity map as follows: 
\begin{eqnarray}
j_{\rm 2d} = \frac{2}{5}r_{\rm core}^2 \Omega,
\label{eq:j2d}
\end{eqnarray}
here we use $r_{\rm core}=0.05$ pc for simplicity, and $\Omega=\sqrt{\Omega_x^2+\Omega_z^2 }$ derived from the fitting using Equation \ref{eq:fitv2d} and Equation \ref{eq:mini}. $j_{\rm 2d}$ is compared with the observational results in Figure \ref{fig:j2dhist}. The observational results from \cite{Punanova2018} are well reproduced by our synthetic data when the filaments are inclined by $30^{\circ}$ with respect to the plane of the sky.

\begin{figure}[t]
\begin{tabular}{cc}
\begin{minipage}[t]{.35\textwidth}
\centering
\includegraphics[width=9cm]{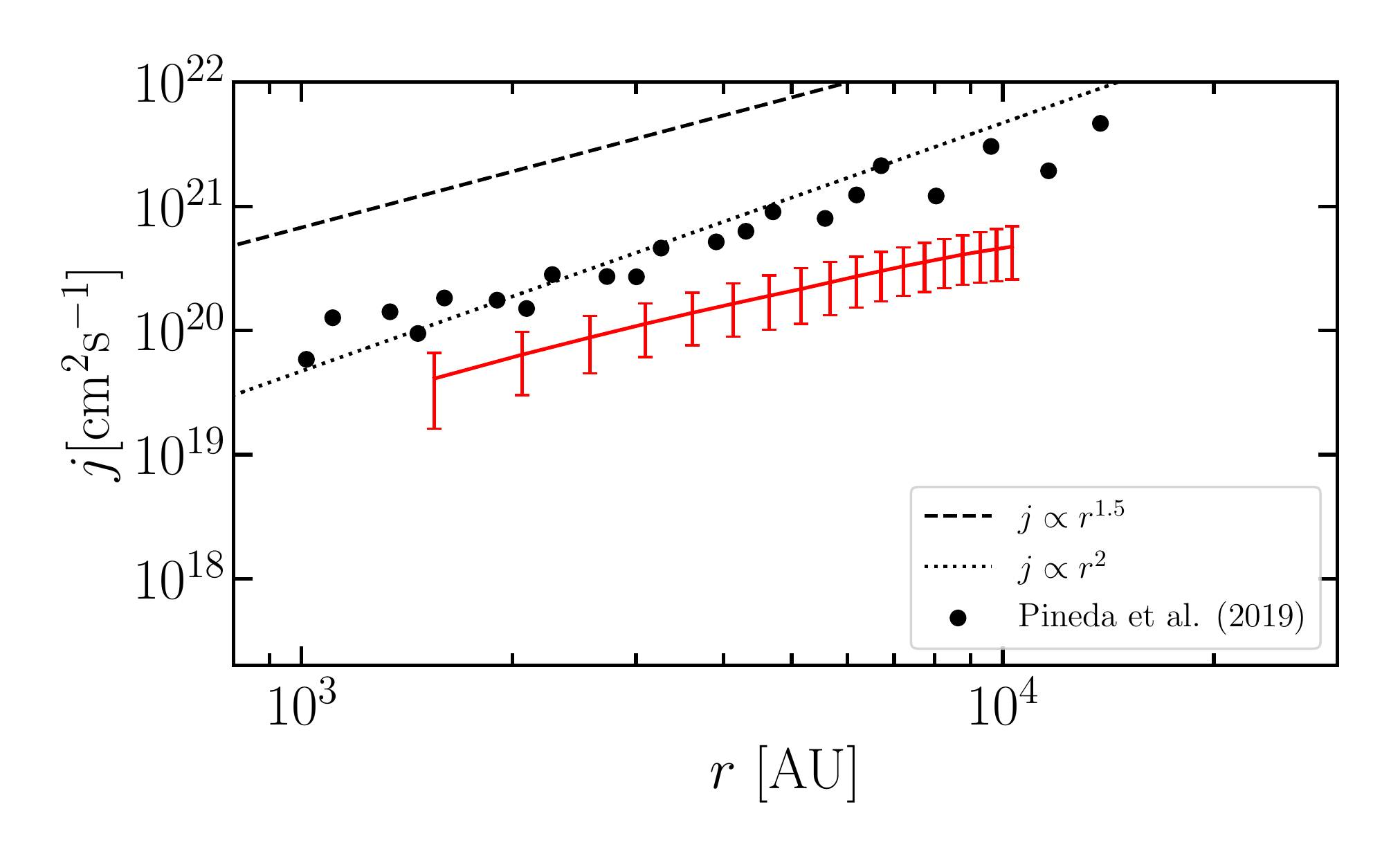}
\end{minipage}
\begin{minipage}{.20\textwidth}
\hspace{10mm}
\end{minipage}

\begin{minipage}[t]{.35\textwidth}
\centering
\includegraphics[width=9cm]{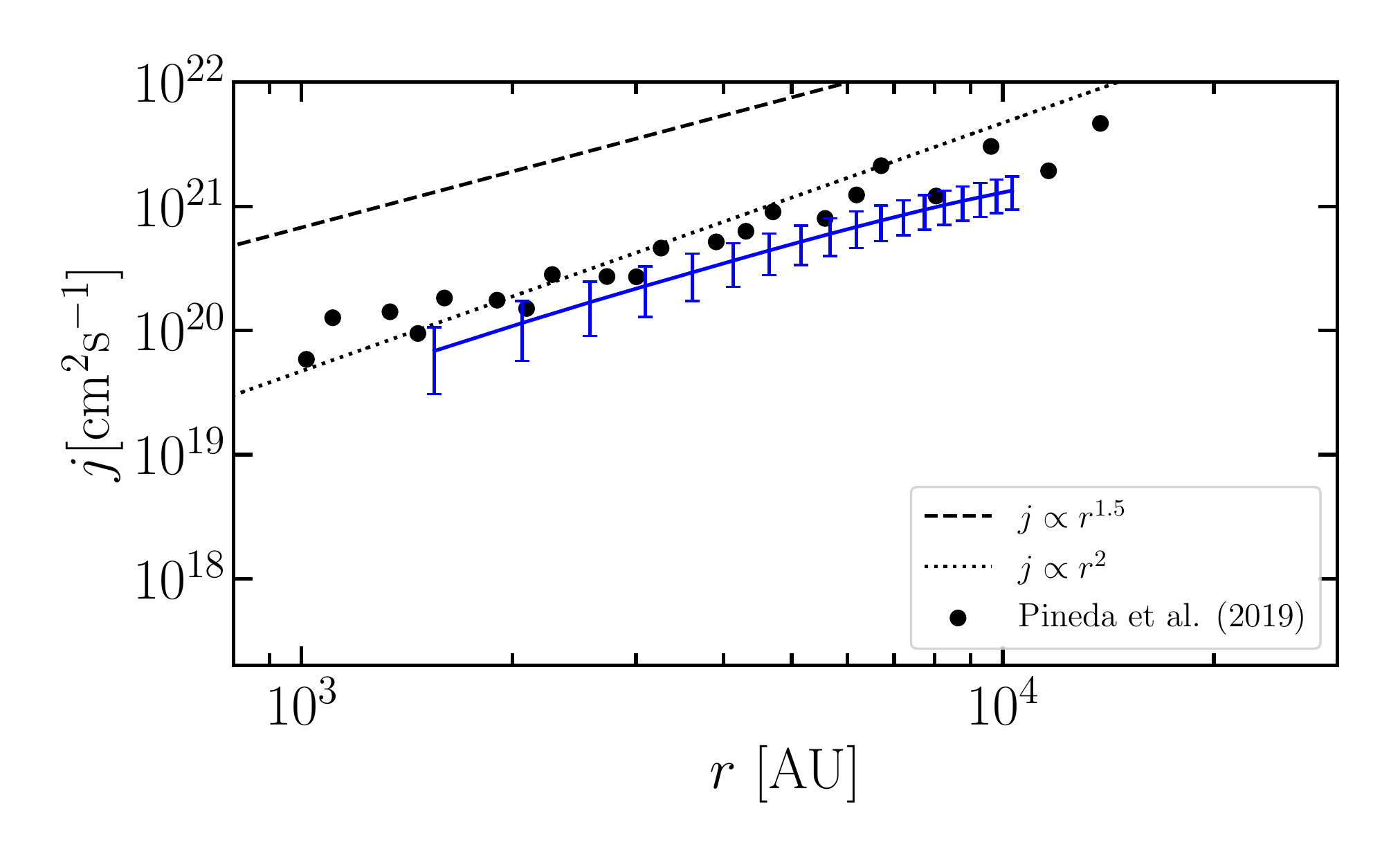}
\end{minipage}

\end{tabular}
\caption{Averaged $j$-$r$ relation derived from synthetic maps without (left) and with filament inclination (right). The black circles are observational data from \cite{Pineda2019}. The green and purple solid lines are $j \propto r^{1.5}$ and $j \propto r^{2}$ for comparison. The error bars refer to the standard deviation for all the cores. The slope indexes without inclination and with inclination are $1.28 \pm 0.02$ and $1.56 \pm 0.02$, respectively }
\label{fig:obsjrave}
\end{figure}

Figure \ref{fig:obsjrave} displays the averaged $j$-$r$ relation for all cores with and without inclination. The $j$-$r$ profiles of individual cores measured in the synthetic line of sight maps are shown in Appendix B. The error bars refer to the standard deviation. The slope indexes without and with inclination are $1.28 \pm 0.02$ and $1.56 \pm 0.02$, respectively. Here we compare our results with the observational results from \cite{Pineda2019}, who studied the radial angular momentum profile towards two Class 0 protostars and a first hydrostatic core candidate in the Perseus cloud. We found that our slopes are only $\sim20\%$ shallower than the $j \propto r^{1.8}$ reported by \cite{Pineda2019} and the amplitude of the $j$-$r$ relation with inclination is compatible with the observations. Considering other effects, e.g., magnetic fields, that may affect the theoretically derived slopes will be studied in future works. In addition, a larger statistics of sources is also needed to confirm the observationally derived slope.  \par

\begin{figure}[t]
\centering
\includegraphics[width=12cm]{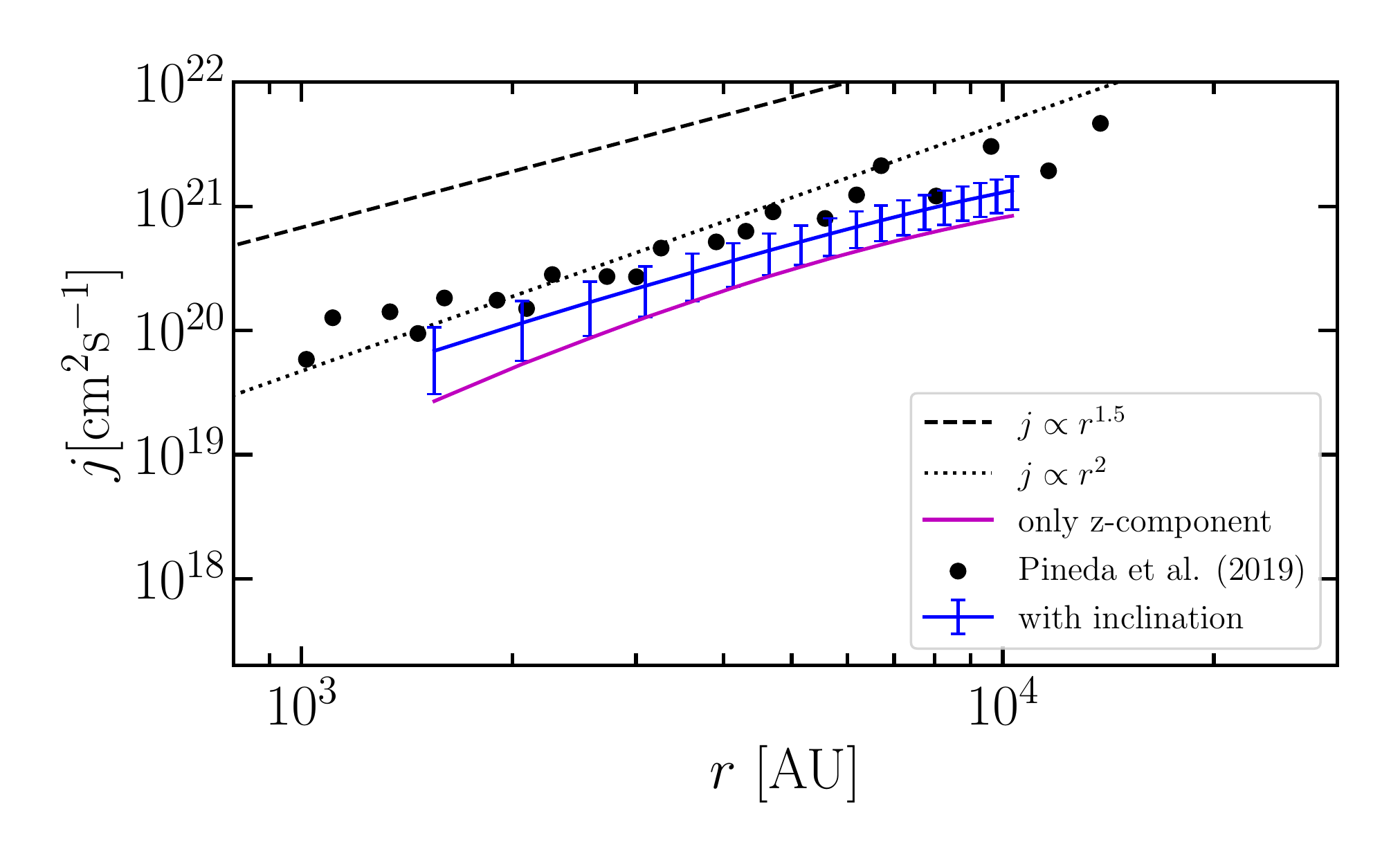}
\caption{Effect of accretion onto cores on the $j$-$r$ diagram. The blue solid line is the averaged $j$-$r$ relation when considering an inclination of $30^{\circ}$ of the filament with respect to the plane of the sky. The purple solid line is the $j$-$r$ relation with only the $z$-component of fluctuations and with inclination. The others are the same as Figure \ref{fig:obsjrave}.}
\label{filacc}
\end{figure}

\begin{figure}[t]
\centering
\includegraphics[width=12cm]{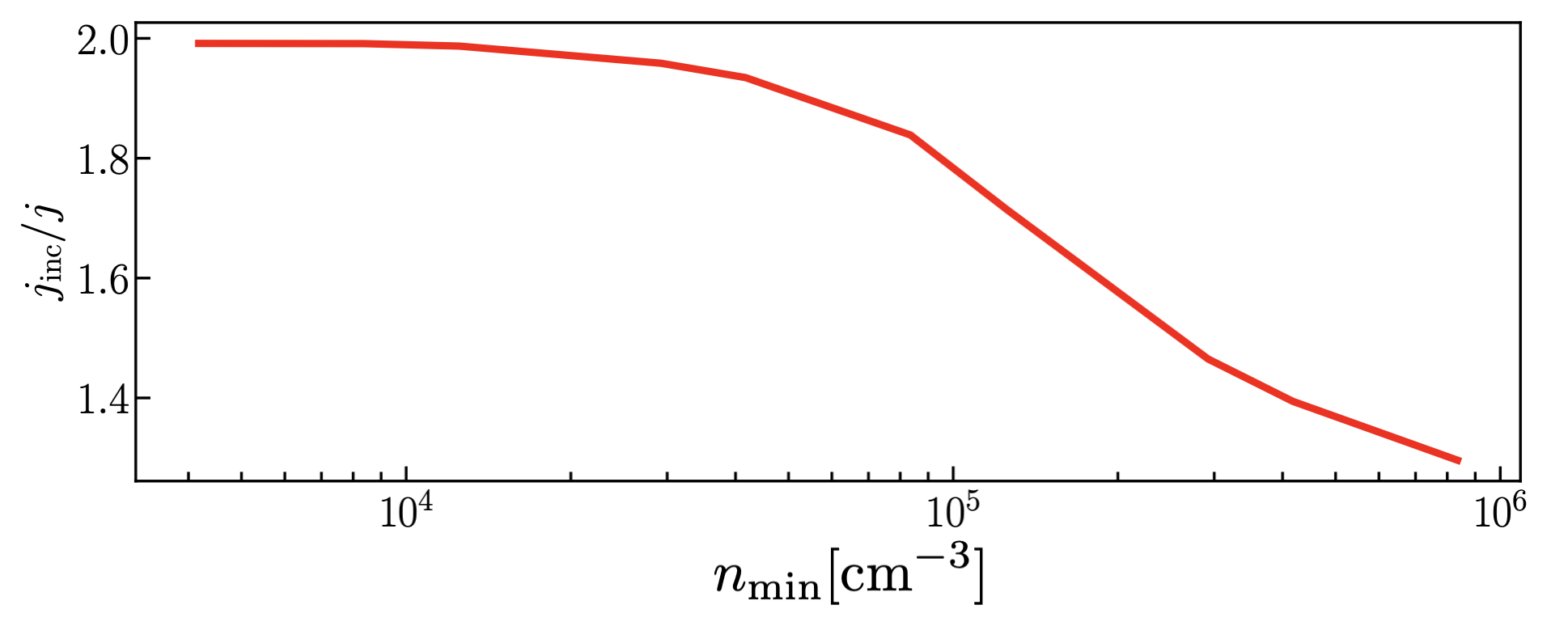}
\caption{Effect of accretion along the filament axis on the apparent measured angular momentum of the core. The vertical axis is the ratio of the core angular momentum measured in the line of sight velocity map with inclination with respect to that without inclination. The horizontal axis is the minimum density for the definition of core. Please note that the core definition adopted here differs from that used in Section 3.2.  }
\label{filacc_critdenn}
\end{figure}

Figure \ref{fig:j2dhist} and \ref{fig:obsjrave} show that the observed absolute value of the specific angular momentum and the slope index of the $j$-$r$ relation in the filaments with inclination are larger than without inclination. This may be due to the accretion onto the cores along the filaments and to the line of sight observable component of the velocity. In the case where the filament is on the plane of the sky the longitudinal accretion motions do not contribute to the apparent angular momentum, hence smaller values are derived. When the filament is inclined the longitudinal motions along the filament axis contribute to the observed line of sight velocity and hence to the larger values of $j$.  To confirm this, we run the simulation in which the filament has only the $z$-component of fluctuations, $k_z=2\pi/(0.4 {\rm pc})$, at the initial state. We stop the simulation just before the first core formation, and then we measure the angular momentum using the same analysis described above. The result with $30^{\circ}$ inclination is shown in Figure \ref{filacc}. Figure \ref{filacc} shows that, with the inclination, the accretion onto the core affects the observed $j$-$r$ relation. To investigate this effect in more detail, we also calculate the mean velocity map changing the minimum core density. The mean velocity map is derived from only SPH particles with density larger than the minimum core density. The dependence of the effect of accretion along the filament axis on the apparent measured rotation observed on the mean velocity map is shown in Figure \ref{filacc_critdenn}. The vertical axis is the ratio of angular momentum of the core measured in the mean velocity map with inclination to that without inclination. Note that the core definition adopted here differs from that used in Section 3.2. Figure \ref{filacc_critdenn} clearly shows that the contamination from the accretion along the filament decreases with the minimum density to calculate the mean velocity map. In high density regions, the flow tends to be spherical accretion rather than accretion along the filament longitudinal axis.\par

\begin{figure}[t]
\centering
\includegraphics[width=12cm]{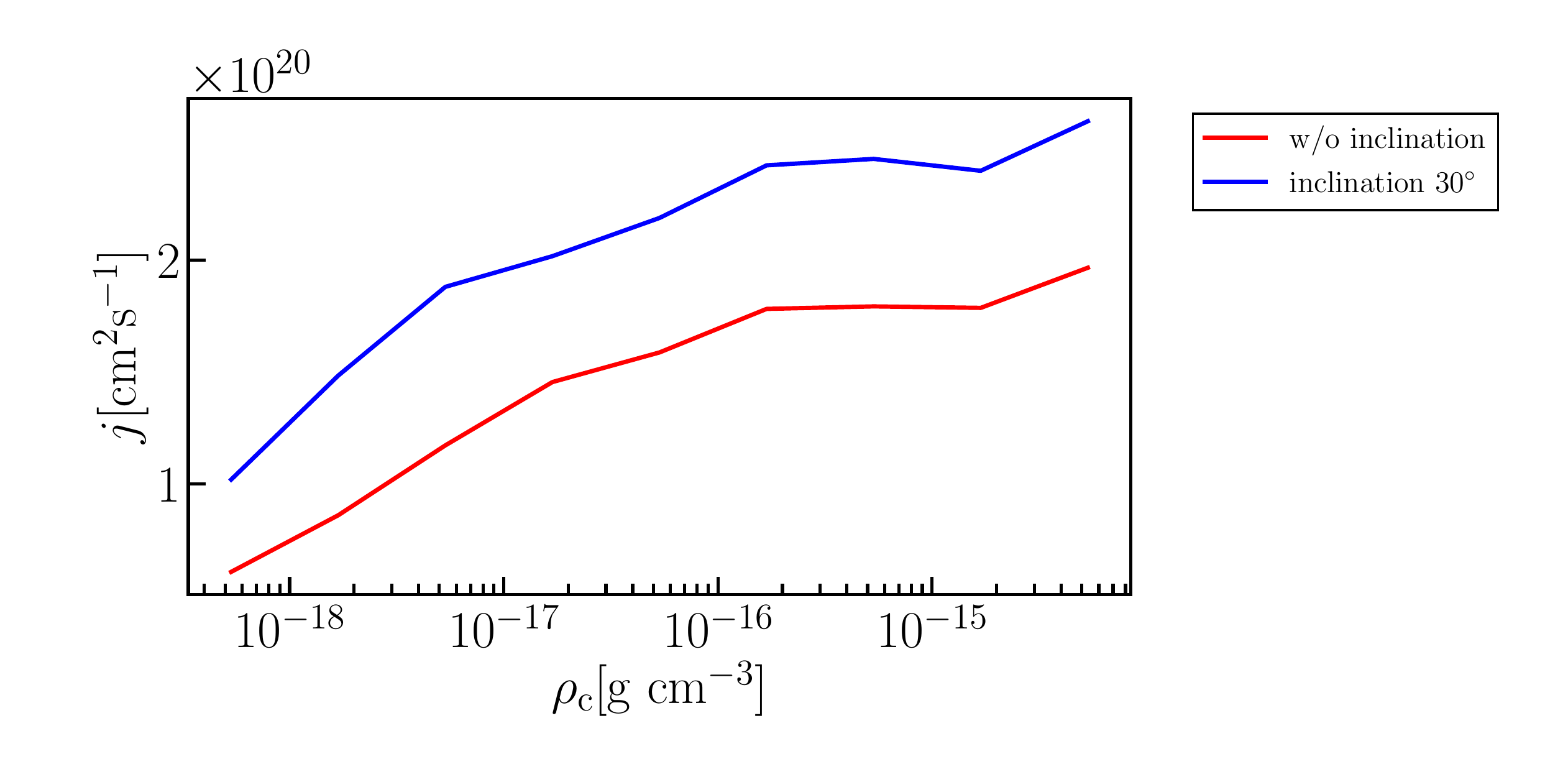}
\caption{Averaged time evolution of the specific angular momentum measured in the line of sight velocity maps such as Figure \ref{fig:obsmap}. In this plot, we measure the 2D angular momentum in the line of sight velocity map whose area is 0.05pc $\times$ 0.05pc at each time step. Note that, since the size of area of the line of sight velocity map is fixed in this measurement, the mass contained in the region increases with time. The horizontal axis is the maximum density of the core, and the vertical axis is the specific angular momentum measured in the line of sight velocity maps. The solid line is the specific angular momentum evolution that is averaged over all cores. The red and blue solid lines are the results with and without inclination, respectively.}
\label{j2dave}
\end{figure}

Figure \ref{j2dave} shows the time evolution of the angular momentum observed in the line of sight velocity map (Figure \ref{fig:obsmap}). In this plot, we measure the 2D angular momentum in the line of sight velocity map whose area is 0.05pc $\times$ 0.05pc at each time step. Note that, since the size of the area of the line of sight velocity map is fixed in this measurement, the mass contained in the region increases with time. Figure \ref{j2dave} shows that the observed angular momentum of the cores increases with time. This is because the gas with larger angular momentum accretes onto the core at later stages. \par

\subsection{Dependence on the Turbulence of the Velocity Field}
\begin{figure}[t]
\centering
\includegraphics[width=12cm]{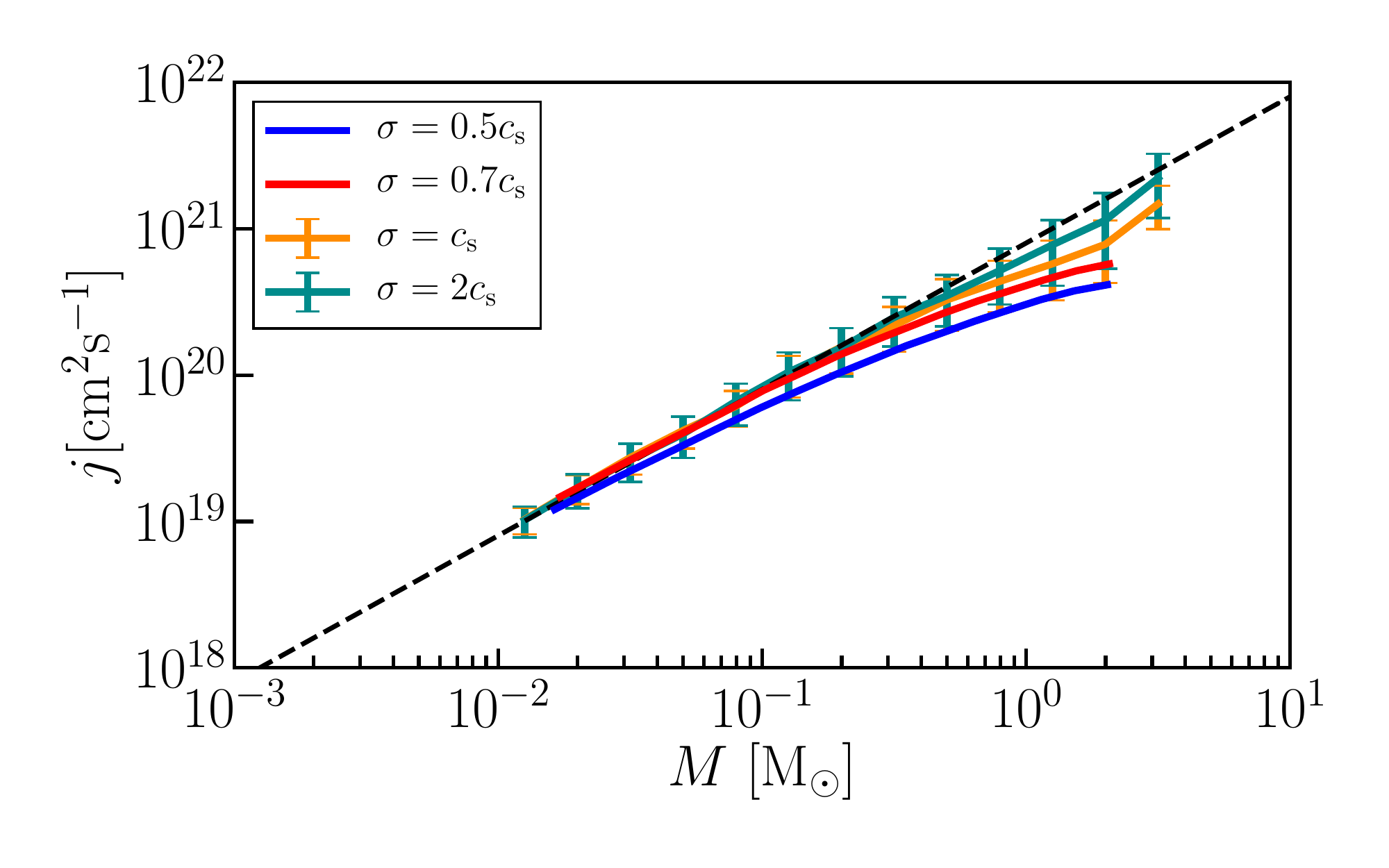}
\caption{Comparison of averaged $j$-$M$ profiles. The blue, red, orange, and green solid lines are $j$-$M$ relations with $\sigma=0.5c_{\rm s}$, $\sigma=0.7c_{\rm s}$, $\sigma=c_{\rm s}$, and $\sigma=2c_{\rm s}$, respectively. The black dashed line is $j \propto M$. }
\label{fig:jMprocomp}
\end{figure}

\begin{figure}[t]
\begin{tabular}{c}
\begin{minipage}[t]{.35\textwidth}
\centering
\includegraphics[width=9cm]{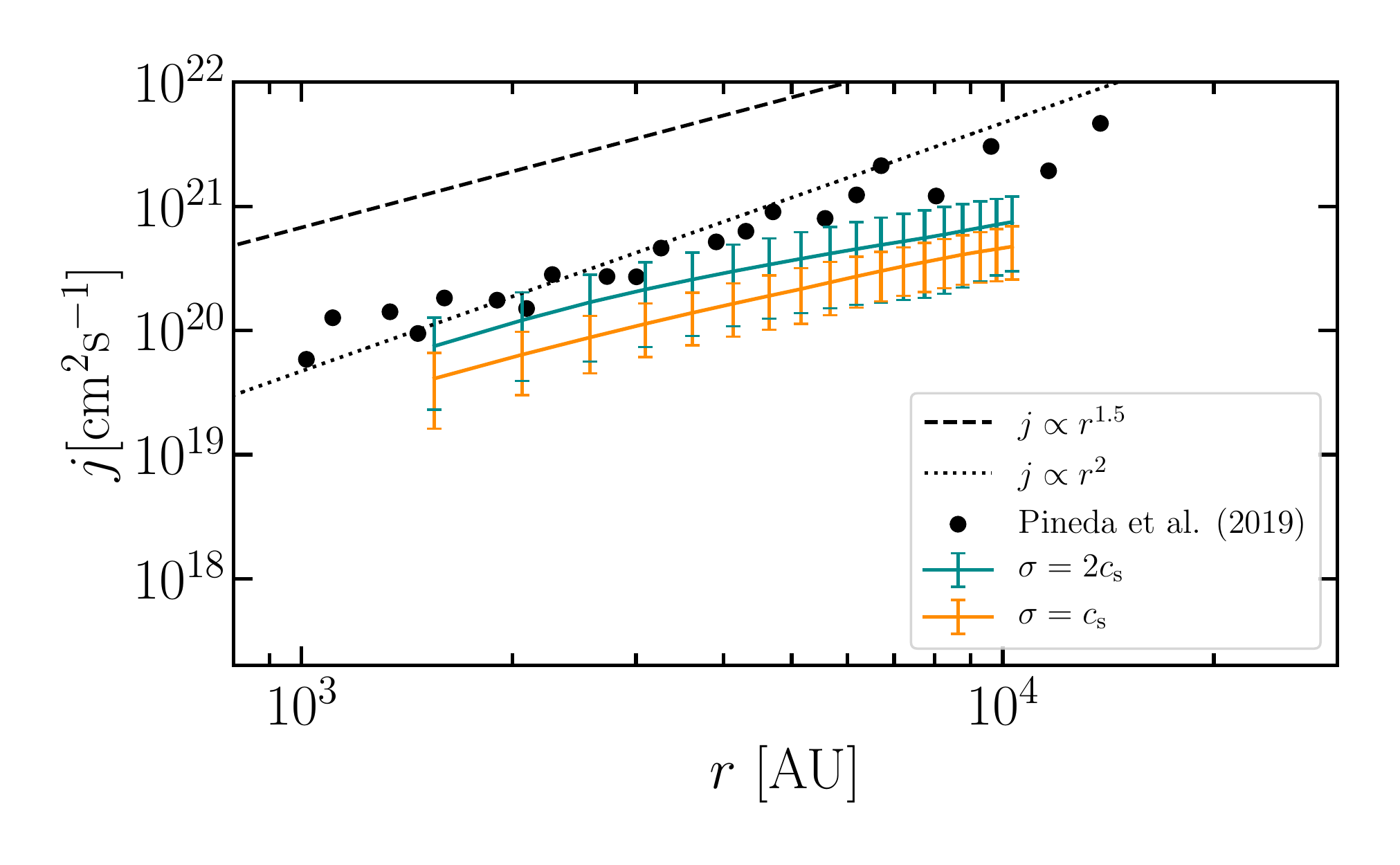}
\end{minipage}
\begin{minipage}{.2\textwidth}
\hspace{10mm}
\end{minipage}

\begin{minipage}[t]{.35\textwidth}
\centering
\includegraphics[width=9cm]{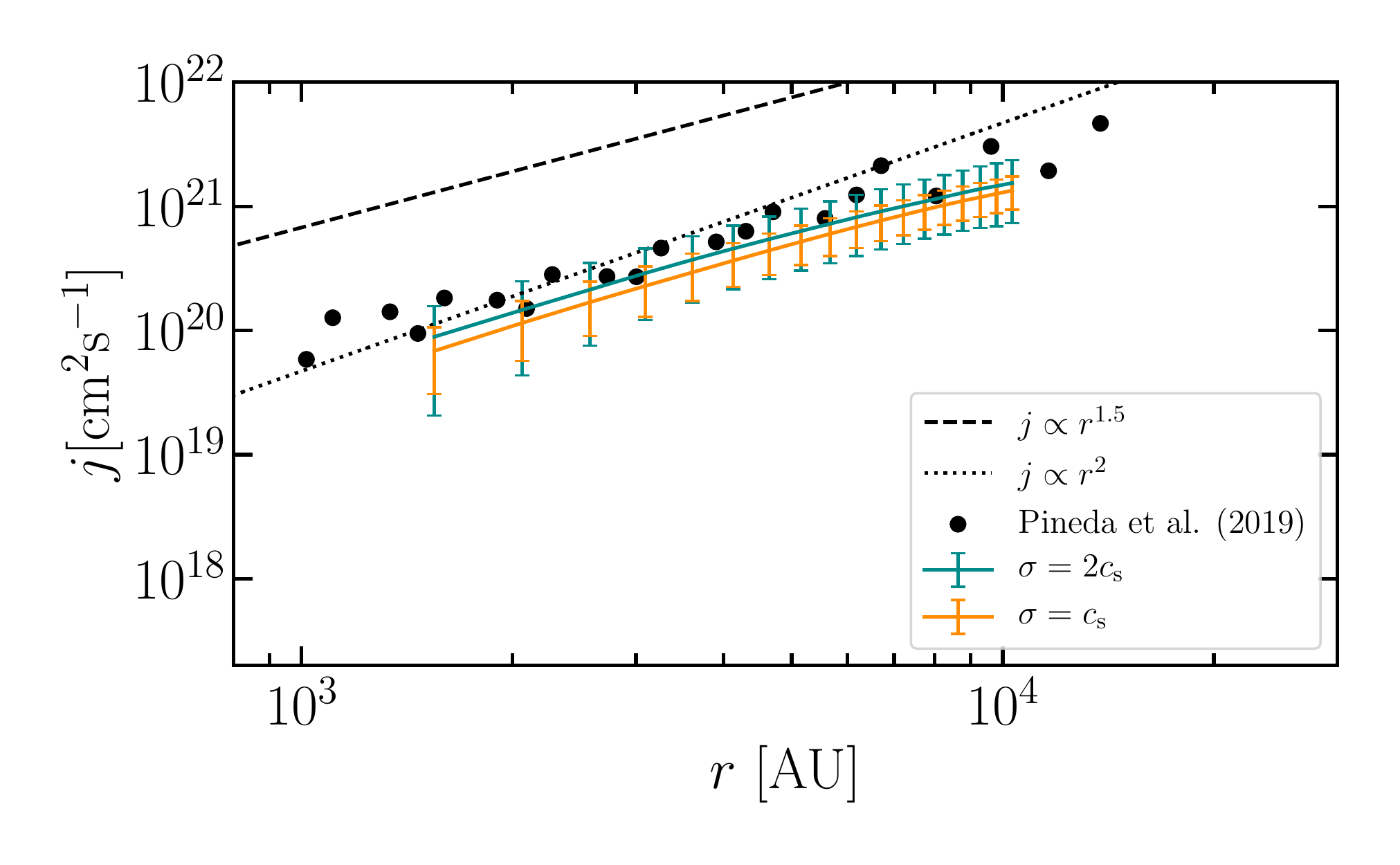}
\end{minipage}

\end{tabular}
\caption{Comparison of the observed $j$-$r$ relation without inclination of filament (left) and with inclination of filament (right). The orange and green solid lines are the $j$-$r$ relations derived from the line of sight velocity  with $\sigma=c_{\rm s}$ and $\sigma=2c_{\rm s}$, respectively. The black dashed line and dotted line are $j \propto r^{1.5}$ and $j \propto r^{2}$ for comparison.}
\label{fig:jM2dcomp}
\end{figure}

In this subsection, we discuss the dependence of the results shown in Section 3.4.1 and Section 4.3 on the initial velocity dispersion of turbulence. First, we discuss the dependence of the $j$-$M$ relation on the initial 3D turbulent velocity dispersion $\sigma$ in the filament. Figure \ref{fig:jMprocomp} displays the comparison of the averaged $j$-$M$ profiles for $\sigma=0.5c_{\rm s}$, $\sigma=0.7c_{\rm s}$, $\sigma=c_{\rm s}$, and $\sigma=2c_{\rm s}$. Figure \ref{fig:jMprocomp} shows that, with larger initial velocity dispersion, the $j$-$M$ profile converges to the self-similar solution over a larger range of enclosed mass. Figure \ref{fig:jM2dcomp} displays the comparison of the observed $j$-$r$ relation without and with filament inclination. Figure \ref{fig:jM2dcomp} clearly shows that larger specific angular momentum is observed in the filaments with larger initial velocity dispersion. These results suggest that the observed $j$-$r$ relations may be affected by both the inclination of the filaments with respect to the plane of the sky and the velocity dispersion of the parent filament.

\subsection{Properties of Cores Defined by Density Contours}

So far, we use only one core from each filament to compare the core properties at the same evolutionary stage of the cores. However, in reality, we observe all cores formed in each filament at possibly different evolutionary stages. In this subsection, we show the properties of all cores formed along the filaments at the same evolutionary stage of the filaments using the same density contour value in all simulations. The cores shown in this subsection have different masses. By adopting 1 $\rho_{\rm c0}$ as the minimum density for the core definition, we identify 115, 124, 101, and 52 cores at 2, 3, 4, and 5 $t_{\rm ff}$, respectively. Here, $\rho_{\rm c0}$ is the initial peak density of the filament. For the case of 2$\rho_{\rm c0}$, we identify 51, 79, 93, and 45 cores at 2, 3, 4, and 5 $t_{\rm ff}$, respectively. When we choose the higher density threshold of 3 $\rho_{\rm c0}$, we identify 20, 44, 69, and 39 cores at 2, 3, 4, and 5 $t_{\rm ff}$, respectively. The reason why the number of cores is small at 5 $t_{\rm ff}$ is that we stop the simulations when the maximum density of the core reaches the first core formation density. Some of the simulations finish before 5 $t_{\rm ff}$. 
\par

\begin{figure}[t]
\centering
\includegraphics[width=12cm]{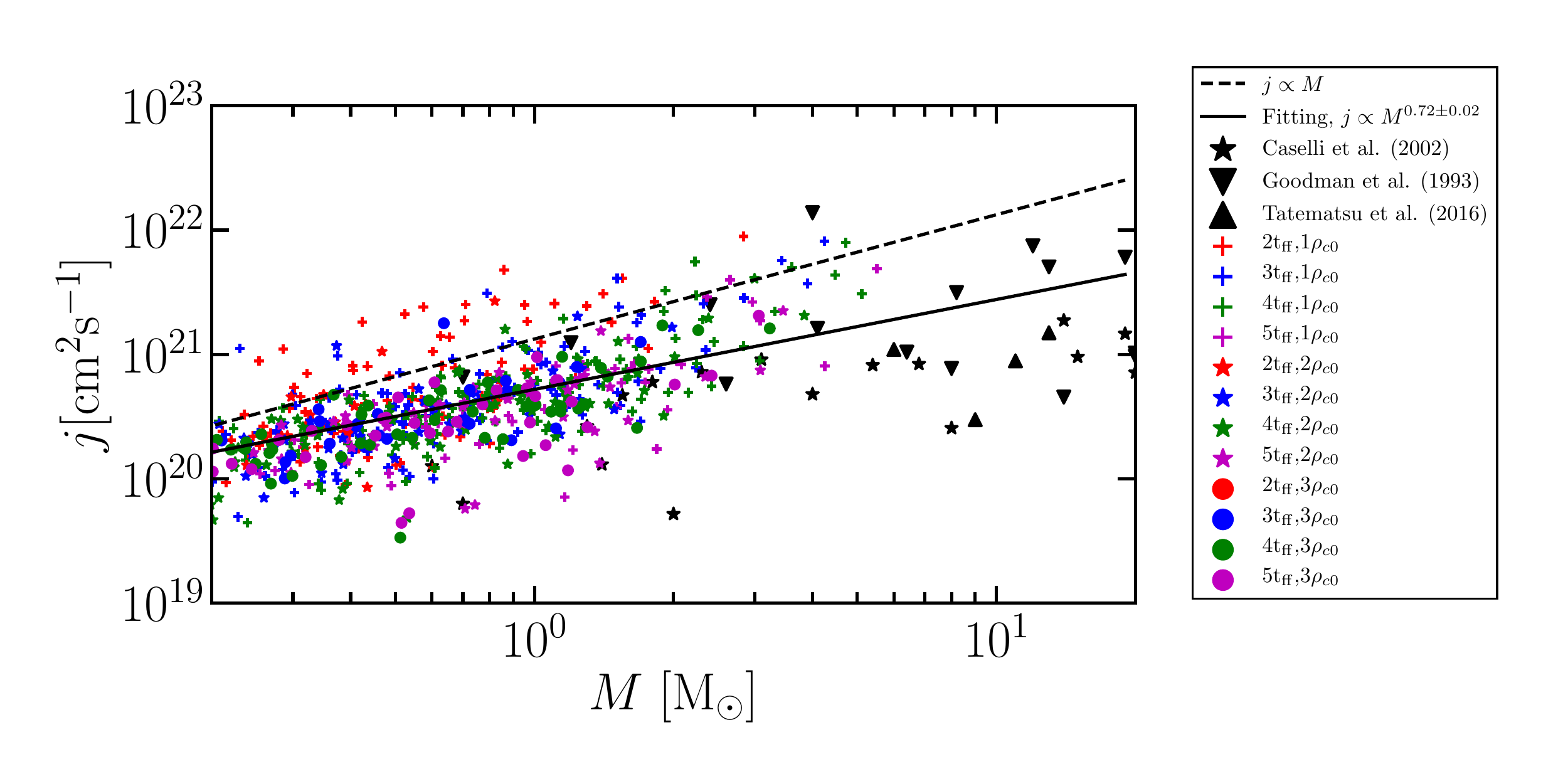}
\caption{$j$-$M$ diagram for all cores in our simulations. The vertical axis and horizontal axis are the core specific angular momentum and the core mass, respectively. The red, blue, green, and magenta symbols represent the cores defined at $t=2, 3, 4,$ and $5_t{ \rm ff}$, respectively. The plus, star, and circle symbols represent the critical density used to define the core ($1,2,$ and $3\rho_{\rm c0}$). The black dots are observational data. The black dashed line is $j \propto M$. The black solid line is the fitting result for all cores in our simulations. The fitting result is $j \propto M^{0.72 \pm 0.02}$. }
\label{fig:jmallrho}
\end{figure}

\begin{figure}[t]
\centering
\includegraphics[width=12cm]{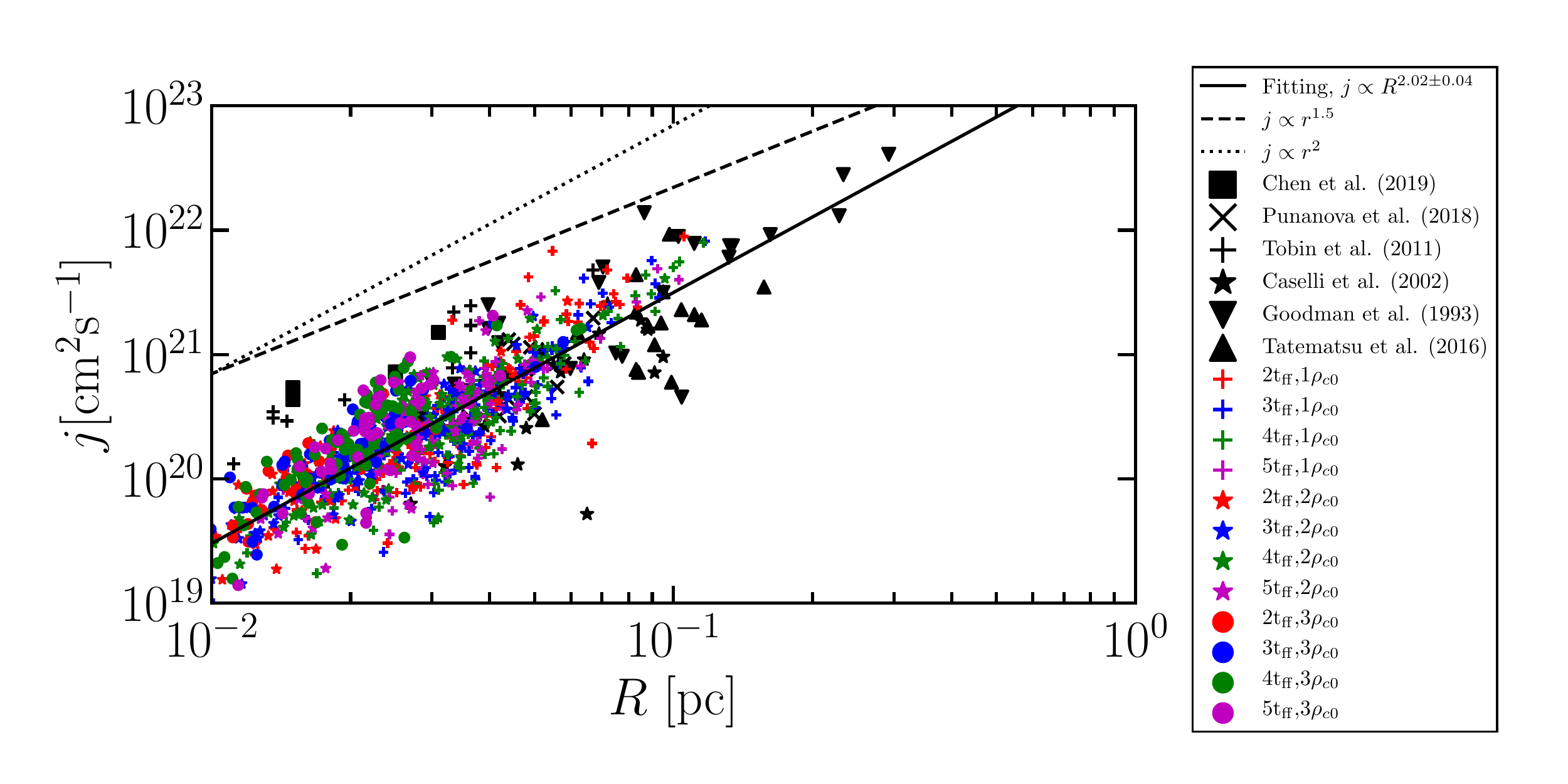}
\caption{$j$-$R$ diagram for all cores in our simulations. The vertical axis and horizontal axis are the specific angular momentum and the core radius, respectively. The black dashed line and dotted line are $j \propto R^{1.5}$ and $j \propto R^{2}$, respectively. The fitting result is $j \propto R^{2.02 \pm 0.04}$. }
\label{fig:jrallrho}
\end{figure}

\begin{figure}[t]
\centering
\includegraphics[width=12cm]{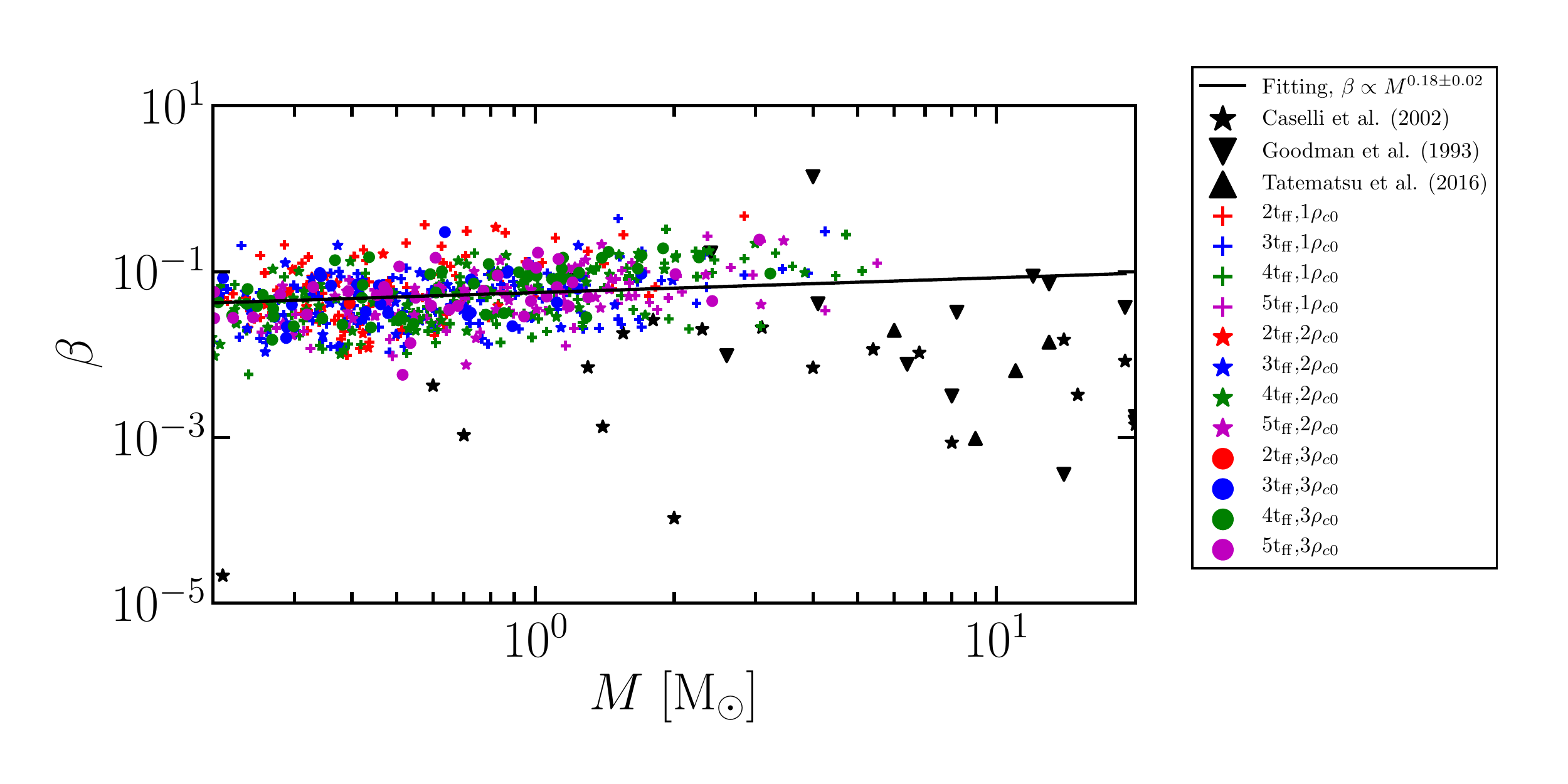}
\caption{$\beta$-$M$ diagram for all cores in our simulations. The fitting result is $\beta \propto M^{0.18 \pm 0.02}$. }
\label{fig:betaallrho}
\end{figure}

Figure \ref{fig:jmallrho} displays the $j$-$M$ diagram for all cores in our simulations at $t=2, 3, 4,$ and $5 \ t_{ \rm ff}$. Here we compare our results with observations presented in multiple papers from \cite{Caselli2002}, \cite{Goodman1993}, and \cite{Tatematsu2016}. Since the specific angular momentum of the cores is not derived in \cite{Caselli2002}, we calculated them using $j=pRv_{\rm rot}$, with the values of $v_{\rm rot}$ from Table 5 in \cite{Caselli2002}, $R$ from their Table 3, and $p = 2/5$, which is the same value used in \cite{Goodman1993}. For the core masses, we used the $M_{\rm ex}$ values given in Table 4 of \cite{Caselli2002}. We also rescaled the values given in \cite{Goodman1993} by adopting FWHM/2 as the radius of the cores.  Although the observational methods are not fully compatible, the compilation of observations is useful to compare with theoretical results and often shown in $j$-$R$ diagrams \cite[e.g.,][]{Chen2019}. The fitting result for all the simulated cores is $j \propto M^{0.72 \pm 0.02}$. Note that the fitting result shows a shallower slope than that of the self similar profile. This is because the cores have a shallower $j$-$M$ profile inherited from initial turbulent velocity field at early evolutionary stage. Even at the later evolutionary stage of the filaments (e.g., $5 \ t_{\rm ff}$), we find a shallower slope since the sample of cores is dominated by young cores. Figure \ref{fig:jrallrho} is the $j$-$R$ diagram for all cores in our simulations. The black dashed and dotted lines are $j \propto R^{1.5}$ and $j \propto R^{2}$, respectively. The fitting result is $j \propto R^{2.02 \pm 0.04}$. Figure \ref{fig:betaallrho} shows $\beta$-$M$ diagram for all the cores in our simulations. $\beta=R^3\Omega^2/3GM$ is the ratio of the rotational energy to the gravitational energy \cite[e.g.,][]{Belloche2013}. $\Omega$ is calculated using Equation \ref{eq:fitv2d}. The fitting result is $\beta \propto M^{0.18 \pm 0.02}$. These slope values are different from the results obtained in Section 3.4.1, especially for the $j$-$R$ diagram. In addition, Figures \ref{fig:jmallrho} and \ref{fig:jrallrho} indicate that $M \propto R^3$. In Figures \ref{fig:jmallrho}, \ref{fig:jrallrho}, and \ref{fig:betaallrho}, since we plot all the cores found in our 40 simulated filaments at the same elapsed time of the simulation ($t=2, 3, 4,$ and $5t_{ \rm ff}$), we observe the cores at different evolutionary stages (different central maximum density).  Since, at the early evolutionary stage, the cores have a flat inner density profile, the relation between mass and radius is expected to follow $M  \propto R^3$. Observations also found that prestellar cores have flat inner density regions \citep{Ward-Thompson1994,Caselli2019}. The least square fitting for the observational data gives $M \propto R^{2.9 \pm 0.4}$, $M \propto R^{1.8 \pm 0.3}$, and $M \propto R^{2.5 \pm 0.3}$ on the $M$-$R$ diagram for \cite{Caselli2002}, \cite{Goodman1993}, and \cite{Tatematsu2016}, respectively. However, as time progresses, the inner flat region becomes smaller and smaller since the cores collapse following the self-similar solution. This evolution leads to $M \propto R$.  The evolutionary timescale of young cores is longer than that of the evolved cores, thus the observational samples might also be dominated by young cores. That is why we find a steeper slope in Figure \ref{fig:jrallrho} compared to that in Figure \ref{fig:interAMradiusprof} which shows the internal $j$-$r$ profile. If we observe the inner profile of the cores, we may find the $M \propto R$ relations, unlike the averaged results of \cite{Caselli2002}, \cite{Goodman1993}, and \cite{Tatematsu2016}  showing $M \propto R^3$. The fitting result is a bit steeper than $j \propto R^{1.6}$ reported in \cite{Goodman1993}. However, as we discussed above, the evolutionary stage of the cores is important to discuss the slope index of the $j$-$R$ diagram. We have to carefully choose a sample of cores (e.g., same evolutionary stage) to discuss the properties of the angular momentum of cores. Therefore, the discussion on the internal angular momentum profiles of individual cores is more useful than the $j$-$R$ diagram for a sample of cores to understand the core angular momentum properties.\par

\begin{figure}[t]
\centering
\includegraphics[width=12cm]{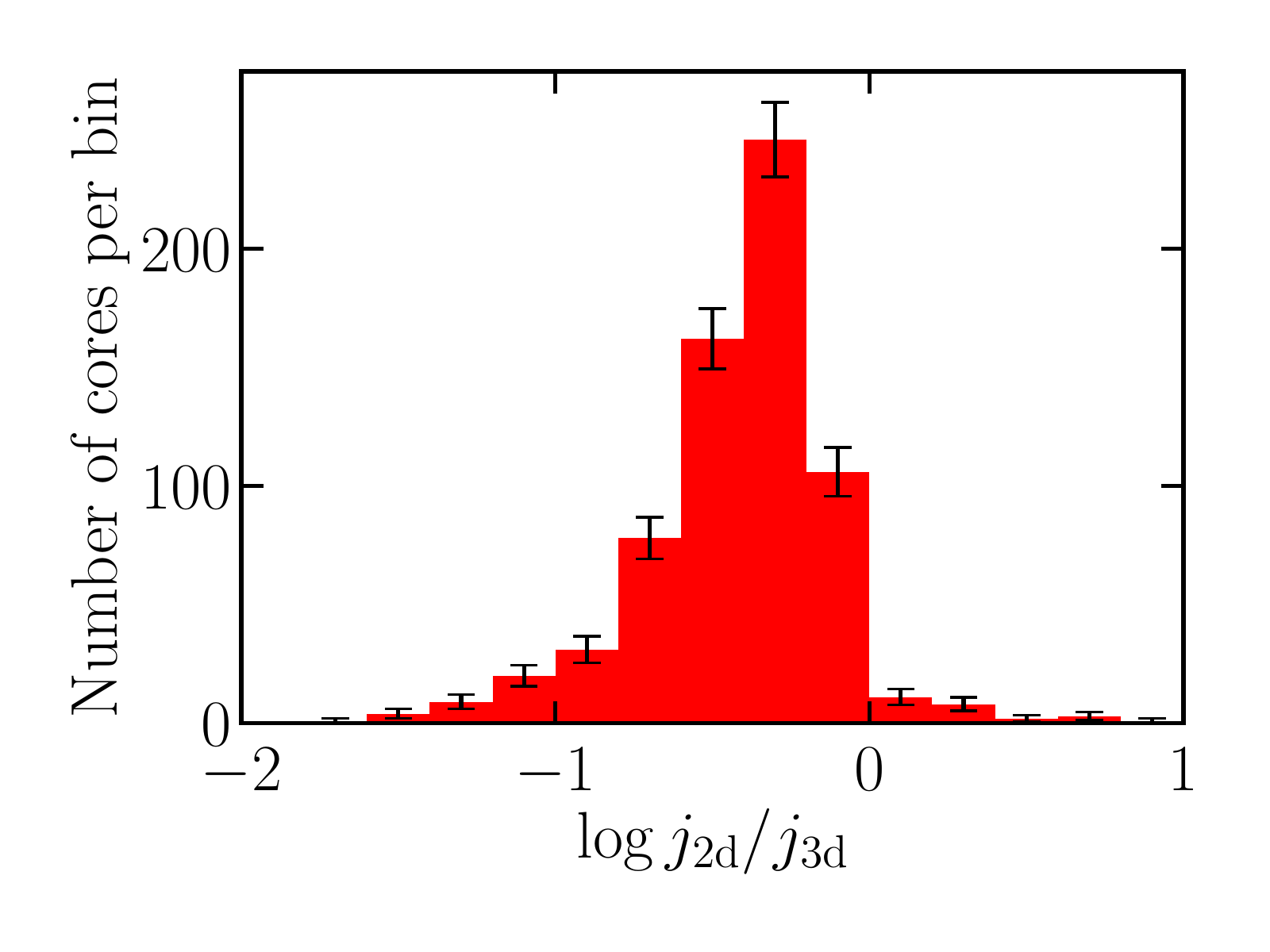}
\caption{Histogram of ratio of specific angular momentum derived from line of sight velocity map to that measured in 3D. The horizontal axis is logarithmic. }
\label{fig:j3d2dhist}
\end{figure}

Figure \ref{fig:j3d2dhist} is a histogram of the ratio of the specific angular momentum derived from the line of sight velocity map to that measured in 3D. \cite{Dib2010} claims that the specific angular momentum is overestimated by an order of magnitude when measured from the line of sight velocity map, although \cite{Zhang2018} pointed out that the specific angular momentum is underestimated by a factor of 2 or 3 in synthetic observations. Our result shown in Figure \ref{fig:j3d2dhist} supports the results in \cite{Zhang2018}. The average and standard deviation of the distribution shown in Figure \ref{fig:j3d2dhist} are 0.49 and 0.55, respectively. This is because we cannot observe the component of the angular momentum parallel to the line of sight direction. This effect leads to an underestimation of the angular momentum.

\section{Summary}
In this paper we study the time evolution of the core angular momentum using three-dimensional hydrodynamics SPH simulations. Our results are summarized as follows.\par
\begin{enumerate}
\item We find that a core tends to lose about 30\% of the angular momentum initially gained from the turbulent fluctuations along the filament by the time the maximum density of core reaches the density of a first hydrostatic core, $\rho_{\rm crit}=2.8\times 10^{-14}\  {\rm g \ cm^{-3}}$. This transfer of angular momentum takes place at the early stage of the collapse when the core reaches a density of  $\rho_{\rm c} \lesssim 10^{-17}\  {\rm g \ cm^{-3}}$. For densities larger than $\rho_{\rm c} \gtrsim 10^{-17} \ {\rm g \ cm^{-3}}$ the free fall time is too short for any transfer of angular momentum.  

\item We also analyze the mechanism of angular momentum transfer by calculating the gravitational and pressure torques in our simulations. We find that the core angular momentum is transferred mainly by the pressure torque.
\item We also find that the rotation axis of most cores are nearly perpendicular to the filament axis. This is because the initial core shape is ellipsoidal and its longer axis is along the filament axis.
\item The analysis of the internal structure of the angular momentum of the cores shows that the profile of the angular momentum in the cores converges to the self-similar solution ($j \propto M$) with time. When the velocity dispersion in the parent filament increases, the core collapse converges to the self-similar solution at larger radii. In this way the core gains larger total angular momentum if the initial velocity dispersion is larger despite the fact that the angular momentum profile converges to the self-similar solution.
\item The degree of complexity in the core slightly decreases with time. However, the complex velocity field survives even just before the first core formation. 
\item We also derive synthetic observations from our simulations. In the case where the parent filaments are in the plane of the sky, we find that the angular momentum measured from the line of sight velocity map is underestimated by a factor of 2 compared to the angular momentum measured in 3D.
\item When the parent filaments are inclined by $30^{\circ}$ with respect to the plane of the sky, the core angular momentum measured from the line of sight velocity map exceeds by a factor of 2 that obtained in the case without inclination. The results derived from inclined filaments are compatible with the observed core angular momentum.
\item The effect of the inclination of the filament on the measured angular momentum from line of sight velocity maps can be reduced by observing only high density regions in a core ($ \gtrsim 10^5 \  {\rm cm^{-3}}$).
\end{enumerate}
\par
In this paper, we do not include the effects of magnetic fields and the accretion onto the filament. Moreover, we adopt the equilibrium profile ($\rho \propto r^{-4}$) and do not test the shallower density profile  ($\rho \propto r^{-2}$) found in observations \cite[e.g.,][]{Arzoumanian_2011,Arzoumanian2019,Palmeirim2013}. These effects that may have an impact on the evolution of the cores and their angular momentum will be taken into account in subsequent papers. In addition, observations of larger core samples, derived in a consistent way, using the same tracers and angular resolution, are required to better infer the observed core properties and constrain the origin of the observed scatter of the $j$-$r$ and $j$-$M$ relations, i.e., discriminate between intrinsic variations of core properties and variations due to differing analysis methods. Comparing the properties found in our simulations with observations gives us a hint of the role of filament fragmentation in the origin of core angular momentum properties, which is important for our understanding of the initial conditions of planet formation. 

\ack
We thank the referee for carefully reading this paper and for their constructive comments. Y. Misugi thanks T. Inoue, H. Kobayashi, and K. Kurosaki for their advice and encouragement. Numerical computations were carried out on Cray XC50 at Center for Computational Astrophysics, National Astronomical Observatory of Japan. The computation was also carried out using the supercomputer ``Flow'' at Information Technology Center, Nagoya University. This work was financially supported by JST SPRING, Grant Number JPMJSP2125. Y. Misugi would like to take this opportunity to thank the ``Interdisciplinary Frontier Next-Generation Researcher Program of the Tokai Higher Education and Research System.''

\appendix

\section{Derivation of Equation \ref{eq:Janafinal}}
In this appendix, we detail the derivation of Equation \ref{eq:Janafinal}.  We consider the turbulent velocity field
\begin{eqnarray}
\bi{v}(\bi{x})=\sum_{\bi{k}} \bi{V}(\bi{k}) \sin(\bi{k}\cdot \bi{x} + \phi_{\bi{k}}),
\label{eqapp:velofieldfourier}
\end{eqnarray}
where $ \bi{V}(\bi{k})$ is the Fourier transform. Using Equation \ref{eqapp:velofieldfourier}, Equation \ref{eq:Jana} can be written as follows:
\begin{eqnarray}
\bi{J} &= \rho \sum_{\bi{k}} \int \bi{r} \times \bi{V}(\bi{k})\sin(\bi{k}\cdot \bi{x} + \phi_{\bi{k}}) d^3x\\
&=\rho \sum_{\bi{k}} \bi{V}(\bi{k}) \times \frac{\partial}{\partial \bi{k}}\int \cos(\bi{k}\cdot \bi{x} + \phi_{\bi{k}}) d^3x.
\label{eqapp:Jana1}
\end{eqnarray}
Using the addition theorem and change of variables $\bi{x}'=(a_1x,a_2y,a_3z)$, Equation \ref{eqapp:Jana1} can be written as
\begin{eqnarray}
\bi{J} =\rho a_1a_2a_3 \sum_{\bi{k}} \cos\phi_{\bi{k}}\bi{V}(\bi{k}) \times \frac{\partial}{\partial \bi{k}}\int \cos(\bi{k}'\cdot \bi{x}')  d^3x',
\label{eqapp:Jana2}
\end{eqnarray}
where $\bi{k}'=(k_xa_1,k_ya_2,k_za_3)$. We can easily calculate the integration,
\begin{eqnarray}
\bi{J} = 4 \pi \rho a_1a_2a_3 \sum_{\bi{k}} \cos\phi_{\bi{k}}\bi{V}(\bi{k}) \times \frac{\partial}{\partial \bi{k}} \frac{\sin y -y \cos y}{y^3},
\label{eqapp:Jana3}
\end{eqnarray}
where $y=|\bi{k}'|$. Then we can evaluate the derivative in Equation \ref{eqapp:Jana3},
\begin{eqnarray}
\bi{J} &= -12 \pi \rho a_1a_2a_3 \sum_{\bi{k}} \cos\phi_{\bi{k}}\bi{V}(\bi{k}) \times  \left(\frac{\sin y}{y^4} - \frac{\cos y}{y^3} - \frac{\sin y}{3y^2} \right) \frac{\partial k'}{\partial \bi{k}}\\
&= -\frac{M}{5}\sum_{\bi{k}}\bi{V}(\bi{k}) \times \bi{k}''f(y) \cos \phi_{\bi{k}}.
\label{eqapp:Jana3}
\end{eqnarray}
This is the same as Equation \ref{eq:Janafinal}. 

\section{$j$-$M$ Profile of Individual Core}

\begin{figure}[t]
\centering
\includegraphics[width=12cm]{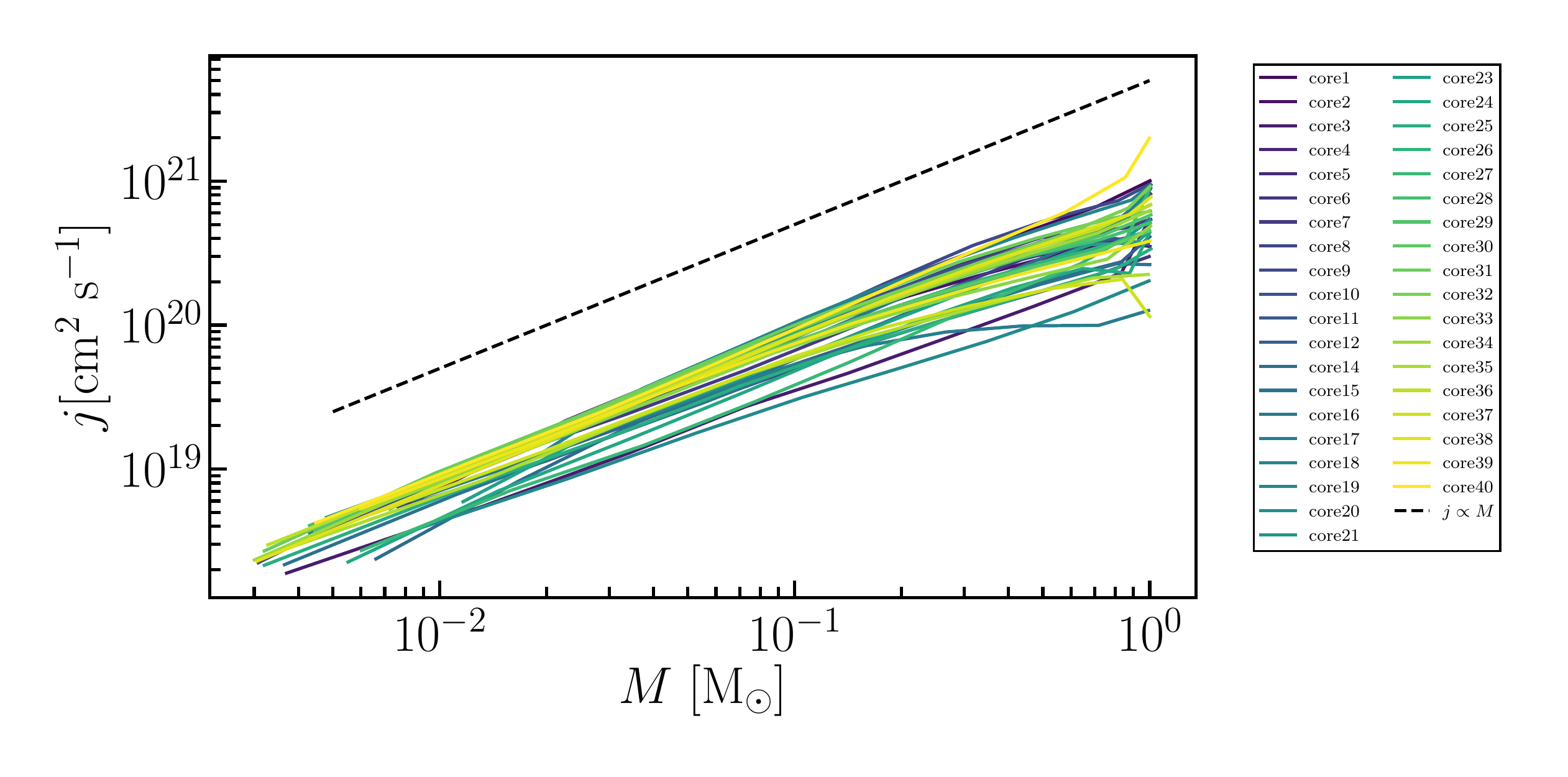}
\caption{Internal angular momentum structure at the final state. The vertical axis is the specific angular momentum of a shell. The horizontal axis is the total mass enclosed by the shell. The different colors of solid lines correspond to the different cores. The black dashed line represents $j \propto M$.}
\label{fig:interAMprof}
\end{figure}

Figure \ref{fig:interAMprof} displays the specific angular momentum profiles of the cores at the final state.  Figure \ref{fig:interAMprof} shows that the the specific angular momentum profile converges to the self-similar solution although the amplitude of $j$-$M$ profile varies among the 38 cores due to the initial seed of the turbulence.
 
\section{$j$-$r$ Profile of Individual Core}

\begin{figure}[t]
\begin{tabular}{cc}
\begin{minipage}[t]{.35\textwidth}
\centering
\includegraphics[width=9cm]{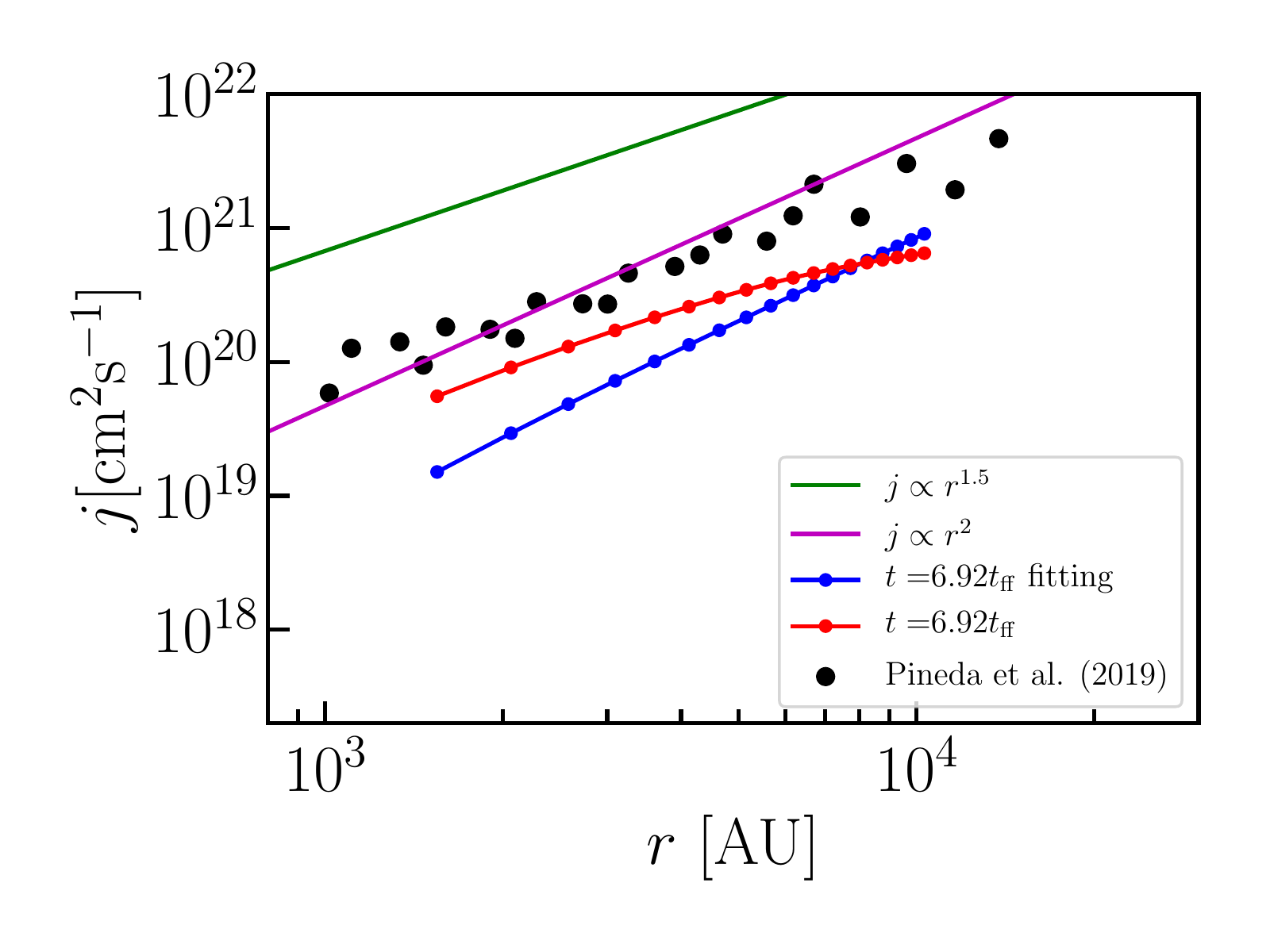}
\end{minipage}
\begin{minipage}{.20\textwidth}
\hspace{10mm}
\end{minipage}

\begin{minipage}[t]{.35\textwidth}
\centering
\includegraphics[width=9cm]{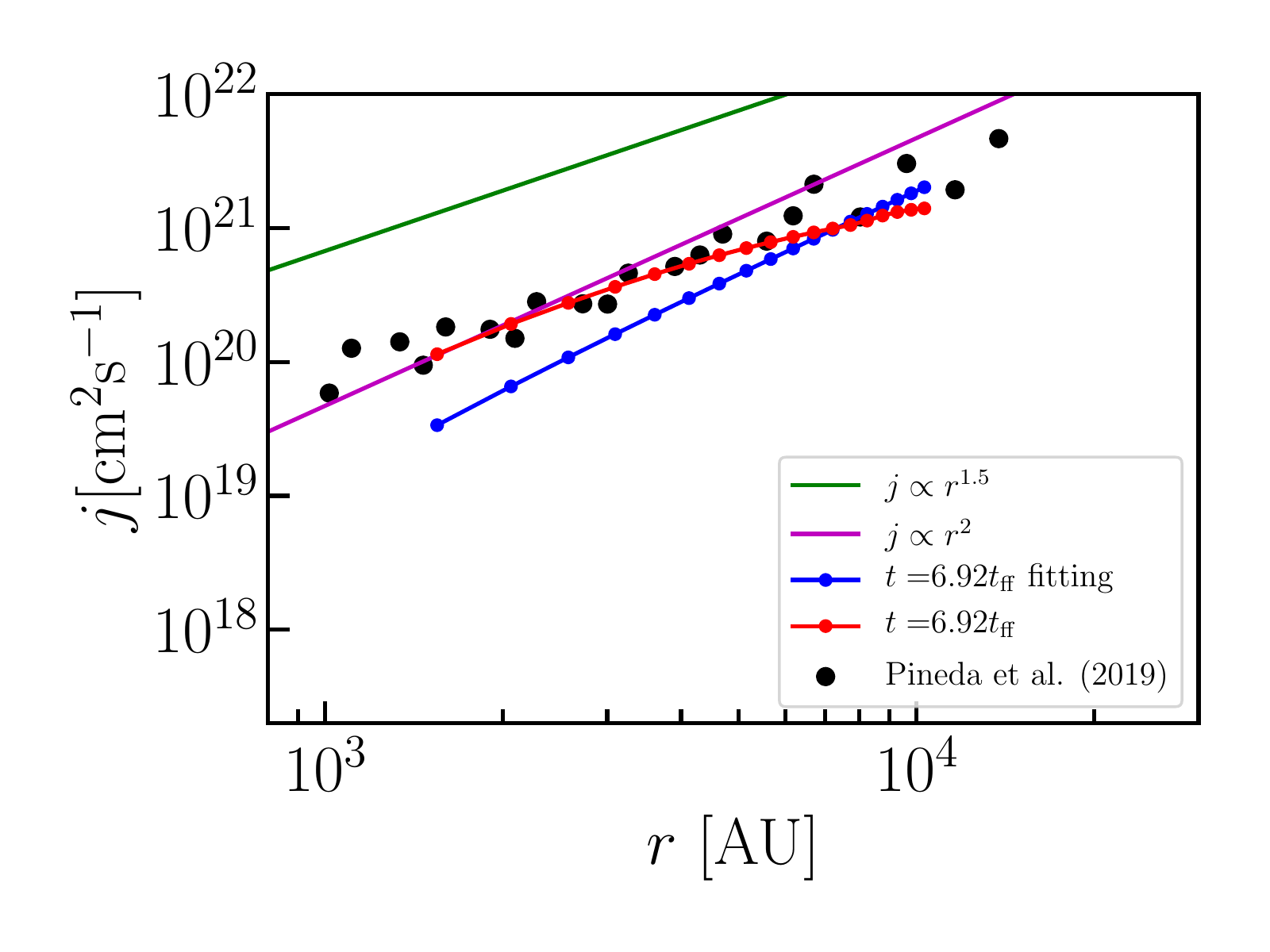}
\end{minipage}

\end{tabular}
\caption{Relation between the specific angular momentum and the radius from the core center derived from the line of sight map without and with inclination. The red solid line is the relation between specific angular momentum and radius from the core center derived from the line of sight velocity map. The blue solid line is result from the rigid body rotation fitting using Equation \ref{eq:fitv2d}. The left and right panels are for the not inclined and the inclined models, , respectively. The others are same as Figure \ref{fig:obsjrave}.}
\label{fig:obsjr}
\end{figure}

\begin{figure}[t]
\begin{tabular}{cc}
\begin{minipage}[t]{.35\textwidth}
\centering
\includegraphics[width=9cm]{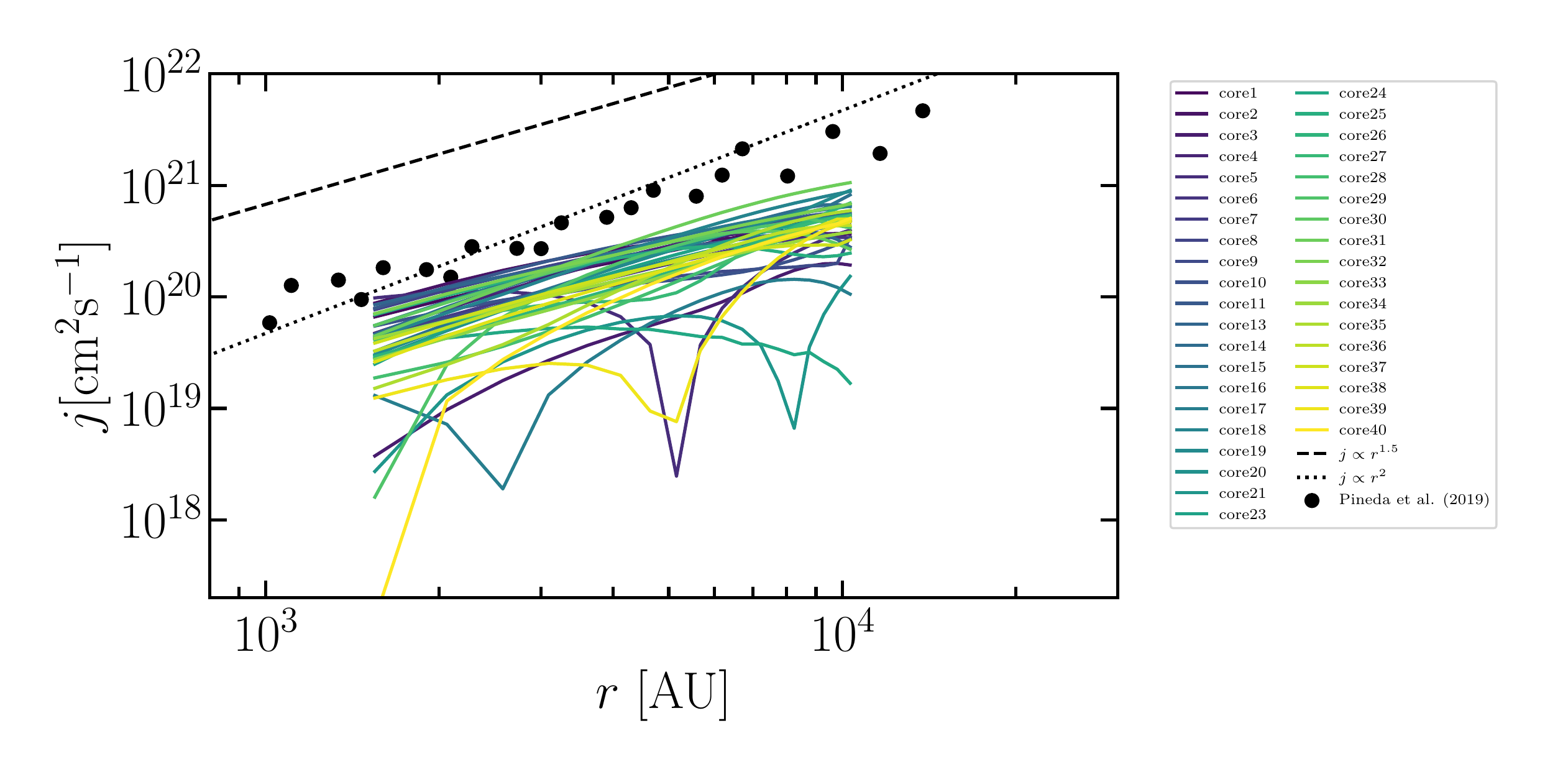}
\end{minipage}
\begin{minipage}{.20\textwidth}
\hspace{10mm}
\end{minipage}

\begin{minipage}[t]{.35\textwidth}
\centering
\includegraphics[width=9cm]{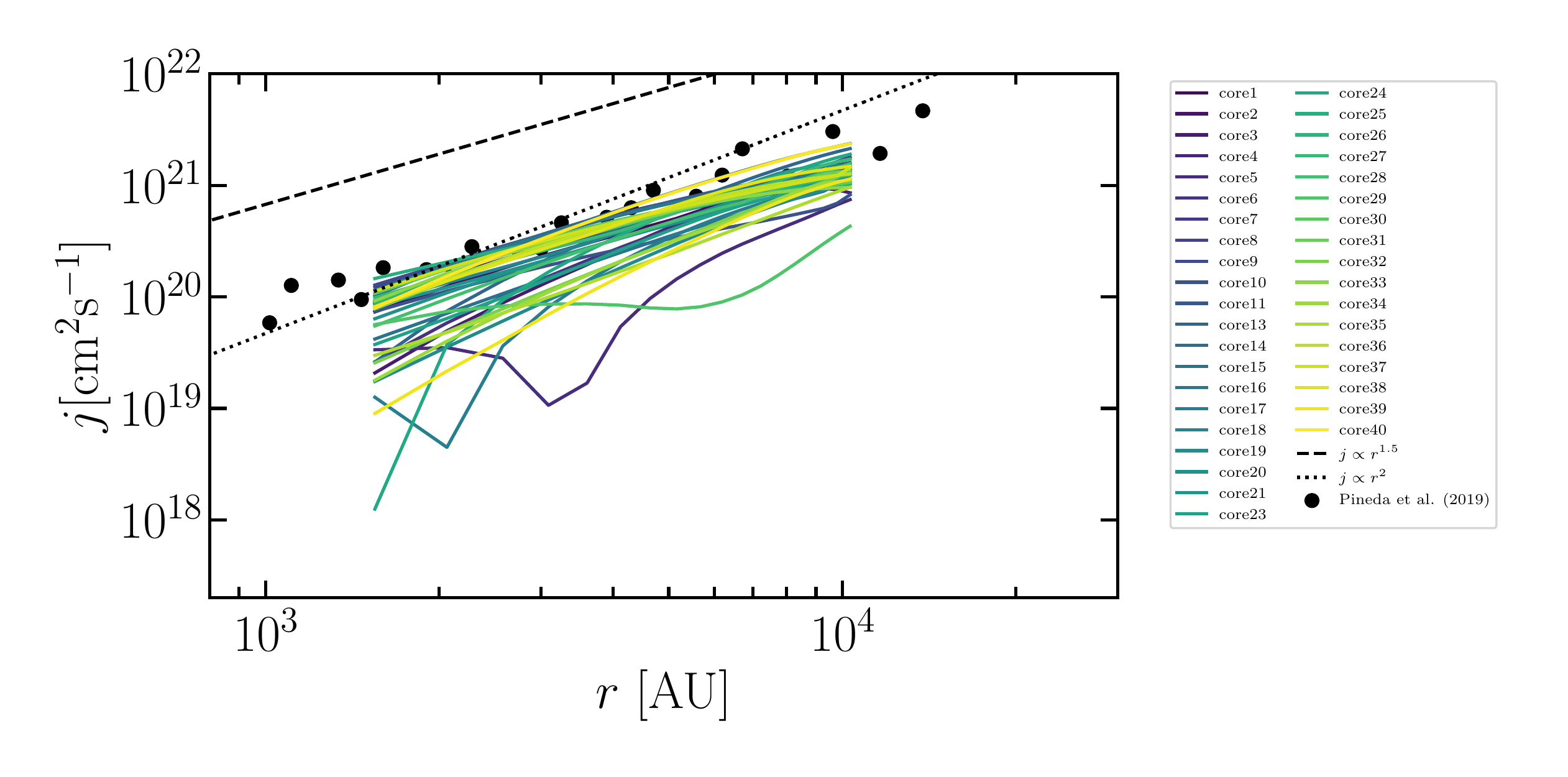}
\end{minipage}

\end{tabular}
\caption{$j$-$r$ relation for all cores. The different colors correspond to the different cores. The others are same as Figure \ref{fig:obsjr}.}
\label{fig:obsjrall}
\end{figure}

Figure \ref{fig:obsjr} shows the $j$-$r$ relation derived from the line of sight velocity map. The red solid line is the relation between the specific angular momentum and the distance from the rotation axis derived from the line of sight velocity map. The blue solid line is the result from the rigid body rotation fitting using Equation \ref{eq:fitv2d}. The left and right panels are for not-inclined and inclined models, respectively. Figure \ref{fig:obsjrall} displays the $j$-$r$ relation of all cores. The different colors correspond to the different cores. Figure \ref{fig:obsjrall} shows that $j$-$r$ profile of the cores has a variety.

\bibliography{Angularmomentum_paper2}

\begin{thebibliography}{}
\expandafter\ifx\csname natexlab\endcsname\relax\def\natexlab#1{#1}\fi

\bibitem[{{Abe} {et~al.}(2021){Abe}, {Inoue}, {Inutsuka}, \&
  {Matsumoto}}]{Abe2021}
{Abe}, D., {Inoue}, T., {Inutsuka}, S.-i., \& {Matsumoto}, T. 2021, \apj, 916,
  83

\bibitem[{{Andr{\'e}} {et~al.}(2014){Andr{\'e}}, {Di Francesco},
  {Ward-Thompson}, {Inutsuka}, {Pudritz}, \& {Pineda}}]{Andre2014}
{Andr{\'e}}, P., {Di Francesco}, J., {Ward-Thompson}, D., {et~al.} 2014, in
  Protostars and Planets VI, ed. H.~{Beuther}, R.~S. {Klessen}, C.~P.
  {Dullemond}, \& T.~{Henning}, 27

\bibitem[{{Andr{\'e}} {et~al.}(2010){Andr{\'e}}, {Men'shchikov}, {Bontemps},
  {K{\"o}nyves}, {Motte}, {Schneider}, {Didelon}, {Minier}, {Saraceno},
  {Ward-Thompson}, {di Francesco}, {White}, {Molinari}, {Testi}, {Abergel},
  {Griffin}, {Henning}, {Royer}, {Mer{\'{\i}}n}, {Vavrek}, {Attard},
  {Arzoumanian}, {Wilson}, {Ade}, {Aussel}, {Baluteau}, {Benedettini},
  {Bernard}, {Blommaert}, {Cambr{\'e}sy}, {Cox}, {di Giorgio}, {Hargrave},
  {Hennemann}, {Huang}, {Kirk}, {Krause}, {Launhardt}, {Leeks}, {Le Pennec},
  {Li}, {Martin}, {Maury}, {Olofsson}, {Omont}, {Peretto}, {Pezzuto}, {Prusti},
  {Roussel}, {Russeil}, {Sauvage}, {Sibthorpe}, {Sicilia-Aguilar}, {Spinoglio},
  {Waelkens}, {Woodcraft}, \& {Zavagno}}]{Andre2010}
{Andr{\'e}}, P., {Men'shchikov}, A., {Bontemps}, S., {et~al.} 2010, \aap, 518,
  L102

\bibitem[{{Andr{\'e}} {et~al.}(2022){Andr{\'e}}, {Palmeirim}, \&
  {Arzoumanian}}]{Andre2022}
{Andr{\'e}}, P.~J., {Palmeirim}, P., \& {Arzoumanian}, D. 2022, \aap, 667, L1

\bibitem[{{Arroyo-Ch{\'a}vez} \& {V{\'a}zquez-Semadeni}(2021)}]{Arroyo2021}
{Arroyo-Ch{\'a}vez}, G., \& {V{\'a}zquez-Semadeni}, E. 2021, arXiv e-prints,
  arXiv:2106.10381

\bibitem[{{Arzoumanian} {et~al.}(2013){Arzoumanian}, {Andr{\'e}}, {Peretto}, \&
  {K{\"o}nyves}}]{Arzoumanian2013}
{Arzoumanian}, D., {Andr{\'e}}, P., {Peretto}, N., \& {K{\"o}nyves}, V. 2013,
  \aap, 553, A119

\bibitem[{{Arzoumanian} {et~al.}(2018){Arzoumanian}, {Shimajiri}, {Inutsuka},
  {Inoue}, \& {Tachihara}}]{Arzoumanian2018}
{Arzoumanian}, D., {Shimajiri}, Y., {Inutsuka}, S.-i., {Inoue}, T., \&
  {Tachihara}, K. 2018, \pasj, 70, 96

\bibitem[{{Arzoumanian} {et~al.}(2011){Arzoumanian}, {Andr{\'e}}, {Didelon},
  {K{\"o}nyves}, {Schneider}, {Men'shchikov}, {Sousbie}, {Zavagno}, {Bontemps},
  {di Francesco}, {Griffin}, {Hennemann}, {Hill}, {Kirk}, {Martin}, {Minier},
  {Molinari}, {Motte}, {Peretto}, {Pezzuto}, {Spinoglio}, {Ward-Thompson},
  {White}, \& {Wilson}}]{Arzoumanian_2011}
{Arzoumanian}, D., {Andr{\'e}}, P., {Didelon}, P., {et~al.} 2011, \aap, 529, L6

\bibitem[{{Arzoumanian} {et~al.}(2019){Arzoumanian}, {Andr{\'e}},
  {K{\"o}nyves}, {Palmeirim}, {Roy}, {Schneider}, {Benedettini}, {Didelon}, {Di
  Francesco}, {Kirk}, \& {Ladjelate}}]{Arzoumanian2019}
{Arzoumanian}, D., {Andr{\'e}}, P., {K{\"o}nyves}, V., {et~al.} 2019, \aap,
  621, A42

\bibitem[{{Arzoumanian} {et~al.}(2021){Arzoumanian}, {Furuya}, {Hasegawa},
  {Tahani}, {Sadavoy}, {Hull}, {Johnstone}, {Koch}, {Inutsuka}, {Doi}, {Hoang},
  {Onaka}, {Iwasaki}, {Shimajiri}, {Inoue}, {Peretto}, {Andr{\'e}}, {Bastien},
  {Berry}, {Chen}, {Di Francesco}, {Eswaraiah}, {Fanciullo}, {Fissel}, {Hwang},
  {Kang}, {Kim}, {Kim}, {Kirchschlager}, {Kwon}, {Lee}, {Liu}, {Lyo}, {Pattle},
  {Soam}, {Tang}, {Whitworth}, {Ching}, {Coud{\'e}}, {Wang}, {Ward-Thompson},
  {Lai}, {Qiu}, {Bourke}, {Byun}, {Chen}, {Chen}, {Chen}, {Cho}, {Choi},
  {Choi}, {Chrysostomou}, {Chung}, {Dai}, {Diep}, {Duan}, {Duan}, {Eden},
  {Fiege}, {Franzmann}, {Friberg}, {Fuller}, {Gledhill}, {Graves}, {Greaves},
  {Griffin}, {Gu}, {Han}, {Hatchell}, {Hayashi}, {Houde}, {Jeong}, {Kang},
  {Kang}, {Kataoka}, {Kawabata}, {Kemper}, {Kim}, {Kim}, {Kim}, {Kim}, {Kirk},
  {Kobayashi}, {K{\"o}nyves}, {Kusune}, {Kwon}, {Lacaille}, {Law}, {Lee},
  {Lee}, {Lee}, {Lee}, {Lee}, {Li}, {Li}, {Li}, {Liu}, {Liu}, {Liu}, {Lu},
  {Mairs}, {Matsumura}, {Matthews}, {Moriarty-Schieven}, {Nagata}, {Nakamura},
  {Nakanishi}, {Ngoc}, {Ohashi}, {Park}, {Parsons}, {Pyo}, {Qian}, {Rao},
  {Rawlings}, {Rawlings}, {Retter}, {Richer}, {Rigby}, {Saito}, {Savini},
  {Scaife}, {Seta}, {Shinnaga}, {Tamura}, {Tang}, {Tomisaka}, {Tram},
  {Tsukamoto}, {Viti}, {Wang}, {Xie}, {Yen}, {Yoo}, {Yuan}, {Yun}, {Zenko},
  {Zhang}, {Zhang}, {Zhang}, {Zhou}, {Zhu}, {de Looze}, {Dowell}, {Eyres},
  {Falle}, {Friesen}, {Robitaille}, \& {van Loo}}]{Arzoumanian2021}
{Arzoumanian}, D., {Furuya}, R.~S., {Hasegawa}, T., {et~al.} 2021, \aap, 647,
  A78

\bibitem[{{Arzoumanian} {et~al.}(2022){Arzoumanian}, {Russeil}, {Zavagno},
  {Chun-Yuan Chen}, {Andr{\'e}}, {Inutsuka}, {Misugi}, {S{\'a}nchez-Monge},
  {Schilke}, {Men'shchikov}, \& {Kohno}}]{Arzoumanian2022}
{Arzoumanian}, D., {Russeil}, D., {Zavagno}, A., {et~al.} 2022, \aap, 660, A56

\bibitem[{{Barnes} \& {Hut}(1986)}]{Barnes1986}
{Barnes}, J., \& {Hut}, P. 1986, \nat, 324, 446

\bibitem[{{Basu}(1997)}]{Basu1997}
{Basu}, S. 1997, \apj, 485, 240

\bibitem[{{Baug} {et~al.}(2020){Baug}, {Wang}, {Liu}, {Tang}, {Zhang}, {Li},
  {Eswaraiah}, {Liu}, {Tej}, {Goldsmith}, {Bronfman}, {Qin}, {T{\'o}th}, {Li},
  \& {Kim}}]{Baug2020}
{Baug}, T., {Wang}, K., {Liu}, T., {et~al.} 2020, \apj, 890, 44

\bibitem[{{Belloche}(2013)}]{Belloche2013}
{Belloche}, A. 2013, in EAS Publications Series, Vol.~62, EAS Publications
  Series, ed. P.~{Hennebelle} \& C.~{Charbonnel}, 25--66

\bibitem[{{Caselli} {et~al.}(2002){Caselli}, {Benson}, {Myers}, \&
  {Tafalla}}]{Caselli2002}
{Caselli}, P., {Benson}, P.~J., {Myers}, P.~C., \& {Tafalla}, M. 2002, \apj,
  572, 238

\bibitem[{{Caselli} {et~al.}(2019){Caselli}, {Pineda}, {Zhao}, {Walmsley},
  {Keto}, {Tafalla}, {Chac{\'o}n-Tanarro}, {Bourke}, {Friesen}, {Galli}, \&
  {Padovani}}]{Caselli2019}
{Caselli}, P., {Pineda}, J.~E., {Zhao}, B., {et~al.} 2019, \apj, 874, 89

\bibitem[{{Chen} \& {Ostriker}(2018)}]{Chen2018}
{Chen}, C.-Y., \& {Ostriker}, E.~C. 2018, \apj, 865, 34

\bibitem[{{Chen} {et~al.}(2019){Chen}, {Storm}, {Li}, {Mundy}, {Frayer}, {Li},
  {Church}, {Friesen}, {Harris}, {Looney}, {Offner}, {Ostriker}, {Pineda},
  {Tobin}, \& {Chen}}]{Chen2019}
{Chen}, C.-Y., {Storm}, S., {Li}, Z.-Y., {et~al.} 2019, \mnras, 490, 527

\bibitem[{{Dib} {et~al.}(2010){Dib}, {Hennebelle}, {Pineda}, {Csengeri},
  {Bontemps}, {Audit}, \& {Goodman}}]{Dib2010}
{Dib}, S., {Hennebelle}, P., {Pineda}, J.~E., {et~al.} 2010, \apj, 723, 425

\bibitem[{{Dubinski} {et~al.}(1995){Dubinski}, {Narayan}, \&
  {Phillips}}]{Dubinski1995}
{Dubinski}, J., {Narayan}, R., \& {Phillips}, T.~G. 1995, \apj, 448, 226

\bibitem[{{Feddersen} {et~al.}(2020){Feddersen}, {Arce}, {Kong}, {Suri},
  {S{\'a}nchez-Monge}, {Ossenkopf-Okada}, {Dunham}, {Nakamura}, {Shimajiri}, \&
  {Bally}}]{Feddersen2020}
{Feddersen}, J.~R., {Arce}, H.~G., {Kong}, S., {et~al.} 2020, \apj, 896, 11

\bibitem[{{Gaudel} {et~al.}(2020){Gaudel}, {Maury}, {Belloche}, {Maret},
  {Andr{\'e}}, {Hennebelle}, {Galametz}, {Testi}, {Cabrit}, {Palmeirim},
  {Ladjelate}, {Codella}, \& {Podio}}]{Gaudel2020}
{Gaudel}, M., {Maury}, A.~J., {Belloche}, A., {et~al.} 2020, \aap, 637, A92

\bibitem[{{Goodman} {et~al.}(1993){Goodman}, {Benson}, {Fuller}, \&
  {Myers}}]{Goodman1993}
{Goodman}, A.~A., {Benson}, P.~J., {Fuller}, G.~A., \& {Myers}, P.~C. 1993,
  \apj, 406, 528

\bibitem[{{Hacar} {et~al.}(2016){Hacar}, {Kainulainen}, {Tafalla}, {Beuther},
  \& {Alves}}]{Hacar2016}
{Hacar}, A., {Kainulainen}, J., {Tafalla}, M., {Beuther}, H., \& {Alves}, J.
  2016, \aap, 587, A97

\bibitem[{{Hacar} \& {Tafalla}(2011)}]{Hacar2011}
{Hacar}, A., \& {Tafalla}, M. 2011, \aap, 533, A34

\bibitem[{{Hanawa} \& {Nakayama}(1997)}]{Hanawa1997}
{Hanawa}, T., \& {Nakayama}, K. 1997, \apj, 484, 238

\bibitem[{{Inoue} {et~al.}(2018){Inoue}, {Hennebelle}, {Fukui}, {Matsumoto},
  {Iwasaki}, \& {Inutsuka}}]{Inoue2018}
{Inoue}, T., {Hennebelle}, P., {Fukui}, Y., {et~al.} 2018, \pasj, 70, S53

\bibitem[{{Inutsuka}(2001)}]{Inutsuka2001}
{Inutsuka}, S.-i. 2001, \apjl, 559, L149

\bibitem[{{Inutsuka}(2002)}]{Inutsuka2002}
---. 2002, Journal of Computational Physics, 179, 238

\bibitem[{{Inutsuka} \& {Miyama}(1997)}]{Inutsuka_Miyama1997}
{Inutsuka}, S.-i., \& {Miyama}, S.~M. 1997, \apj, 480, 681

\bibitem[{{Iwasawa} {et~al.}(2016){Iwasawa}, {Tanikawa}, {Hosono}, {Nitadori},
  {Muranushi}, \& {Makino}}]{Iwasawa2016}
{Iwasawa}, M., {Tanikawa}, A., {Hosono}, N., {et~al.} 2016, \pasj, 68, 54

\bibitem[{{Koch} \& {Rosolowsky}(2015)}]{Koch2015}
{Koch}, E.~W., \& {Rosolowsky}, E.~W. 2015, \mnras, 452, 3435

\bibitem[{{Kong} {et~al.}(2019){Kong}, {Arce}, {Maureira}, {Caselli}, {Tan}, \&
  {Fontani}}]{Kong2019}
{Kong}, S., {Arce}, H.~G., {Maureira}, M.~J., {et~al.} 2019, \apj, 874, 104

\bibitem[{{K{\"o}nyves} {et~al.}(2015){K{\"o}nyves}, {Andr{\'e}},
  {Men'shchikov}, {Palmeirim}, {Arzoumanian}, {Schneider}, {Roy}, {Didelon},
  {Maury}, {Shimajiri}, {Di Francesco}, {Bontemps}, {Peretto}, {Benedettini},
  {Bernard}, {Elia}, {Griffin}, {Hill}, {Kirk}, {Ladjelate}, {Marsh}, {Martin},
  {Motte}, {Nguy{\^e}n Luong}, {Pezzuto}, {Roussel}, {Rygl}, {Sadavoy},
  {Schisano}, {Spinoglio}, {Ward-Thompson}, \& {White}}]{Konyves2015}
{K{\"o}nyves}, V., {Andr{\'e}}, P., {Men'shchikov}, A., {et~al.} 2015, \aap,
  584, A91

\bibitem[{{Kuznetsova} {et~al.}(2019){Kuznetsova}, {Hartmann}, \&
  {Heitsch}}]{Kuznetsova2019}
{Kuznetsova}, A., {Hartmann}, L., \& {Heitsch}, F. 2019, \apj, 876, 33

\bibitem[{{Kuznetsova} {et~al.}(2020){Kuznetsova}, {Hartmann}, \&
  {Heitsch}}]{Kuznetsova2020}
---. 2020, \apj, 893, 73

\bibitem[{{Larson}(1969)}]{Larson1969}
{Larson}, R.~B. 1969, \mnras, 145, 271

\bibitem[{{Machida} {et~al.}(2020){Machida}, {Hirano}, \&
  {Kitta}}]{Machida2020}
{Machida}, M.~N., {Hirano}, S., \& {Kitta}, H. 2020, \mnras, 491, 2180

\bibitem[{{Machida} {et~al.}(2005){Machida}, {Matsumoto}, {Hanawa}, \&
  {Tomisaka}}]{Machida2005}
{Machida}, M.~N., {Matsumoto}, T., {Hanawa}, T., \& {Tomisaka}, K. 2005,
  \mnras, 362, 382

\bibitem[{{Machida} {et~al.}(2008){Machida}, {Tomisaka}, {Matsumoto}, \&
  {Inutsuka}}]{Machida2008}
{Machida}, M.~N., {Tomisaka}, K., {Matsumoto}, T., \& {Inutsuka}, S.-i. 2008,
  \apj, 677, 327

\bibitem[{{Masunaga} \& {Inutsuka}(2000)}]{Masunaga2000}
{Masunaga}, H., \& {Inutsuka}, S.-i. 2000, \apj, 531, 350

\bibitem[{{Matsumoto} \& {Hanawa}(2003)}]{Matsumoto2003}
{Matsumoto}, T., \& {Hanawa}, T. 2003, \apj, 595, 913

\bibitem[{{Matsumoto} \& {Hanawa}(2011)}]{Matsumoto2011}
---. 2011, \apj, 728, 47

\bibitem[{{Matsumoto} {et~al.}(1997){Matsumoto}, {Hanawa}, \&
  {Nakamura}}]{Matsumoto1997}
{Matsumoto}, T., {Hanawa}, T., \& {Nakamura}, F. 1997, \apj, 478, 569

\bibitem[{{Misugi} {et~al.}(2019){Misugi}, {Inutsuka}, \&
  {Arzoumanian}}]{Misugi2019}
{Misugi}, Y., {Inutsuka}, S.-i., \& {Arzoumanian}, D. 2019, \apj, 881, 11

\bibitem[{{Miville-Desch{\^e}nes} {et~al.}(2010){Miville-Desch{\^e}nes},
  {Martin}, {Abergel}, {Bernard}, {Boulanger}, {Lagache}, {Anderson},
  {Andr{\'e}}, {Arab}, {Baluteau}, {Blagrave}, {Bontemps}, {Cohen},
  {Compiegne}, {Cox}, {Dartois}, {Davis}, {Emery}, {Fulton}, {Gry}, {Habart},
  {Huang}, {Joblin}, {Jones}, {Kirk}, {Lim}, {Madden}, {Makiwa}, {Menshchikov},
  {Molinari}, {Moseley}, {Motte}, {Naylor}, {Okumura}, {Pinheiro Gon{\c
  c}alves}, {Polehampton}, {Rod{\'o}n}, {Russeil}, {Saraceno}, {Schneider},
  {Sidher}, {Spencer}, {Swinyard}, {Ward-Thompson}, {White}, \&
  {Zavagno}}]{Miville2010}
{Miville-Desch{\^e}nes}, M.-A., {Martin}, P.~G., {Abergel}, A., {et~al.} 2010,
  \aap, 518, L104

\bibitem[{{Ntormousi} \& {Hennebelle}(2019)}]{Ntormousi2019}
{Ntormousi}, E., \& {Hennebelle}, P. 2019, \aap, 625, A82

\bibitem[{{Offner} {et~al.}(2008){Offner}, {Klein}, \& {McKee}}]{Offner2008}
{Offner}, S. S.~R., {Klein}, R.~I., \& {McKee}, C.~F. 2008, \apj, 686, 1174

\bibitem[{{Okoda} {et~al.}(2021){Okoda}, {Oya}, {Francis}, {Johnstone},
  {Inutsuka}, {Ceccarelli}, {Codella}, {Chandler}, {Sakai}, {Aikawa}, {Alves},
  {Balucani}, {Bianchi}, {Bouvier}, {Caselli}, {Caux}, {Charnley}, {Choudhury},
  {De Simone}, {Dulieu}, {Dur{\'a}n}, {Evans}, {Favre}, {Fedele}, {Feng},
  {Fontani}, {Hama}, {Hanawa}, {Herbst}, {Hirota}, {Imai}, {Isella},
  {J{\'\i}menez-Serra}, {Kahane}, {Lefloch}, {Loinard}, {L{\'o}pez-Sepulcre},
  {Maud}, {Maureira}, {Menard}, {Mercimek}, {Miotello}, {Moellenbrock}, {Mori},
  {Murillo}, {Nakatani}, {Nomura}, {Oba}, {O'Donoghue}, {Ohashi},
  {Ospina-Zamudio}, {Pineda}, {Podio}, {Rimola}, {Sakai}, {Segura-Cox},
  {Shirley}, {Svoboda}, {Taquet}, {Testi}, {Vastel}, {Viti}, {Watanabe},
  {Watanabe}, {Witzel}, {Xue}, {Zhang}, {Zhao}, \& {Yamamoto}}]{Okoda2021}
{Okoda}, Y., {Oya}, Y., {Francis}, L., {et~al.} 2021, \apj, 910, 11

\bibitem[{{Ostriker}(1964)}]{Ostriker1964}
{Ostriker}, J. 1964, \apj, 140, 1056

\bibitem[{{Palmeirim} {et~al.}(2013){Palmeirim}, {Andr{\'e}}, {Kirk},
  {Ward-Thompson}, {Arzoumanian}, {K{\"o}nyves}, {Didelon}, {Schneider},
  {Benedettini}, {Bontemps}, {Di Francesco}, {Elia}, {Griffin}, {Hennemann},
  {Hill}, {Martin}, {Men'shchikov}, {Molinari}, {Motte}, {Nguyen Luong},
  {Nutter}, {Peretto}, {Pezzuto}, {Roy}, {Rygl}, {Spinoglio}, \&
  {White}}]{Palmeirim2013}
{Palmeirim}, P., {Andr{\'e}}, P., {Kirk}, J., {et~al.} 2013, \aap, 550, A38

\bibitem[{{Peebles}(1969)}]{Peebles1969}
{Peebles}, P.~J.~E. 1969, \apj, 155, 393

\bibitem[{{Penston}(1969)}]{Penston1969}
{Penston}, M.~V. 1969, \mnras, 144, 425

\bibitem[{{Pineda} {et~al.}(2019){Pineda}, {Zhao}, {Schmiedeke}, {Segura-Cox},
  {Caselli}, {Myers}, {Tobin}, \& {Dunham}}]{Pineda2019}
{Pineda}, J.~E., {Zhao}, B., {Schmiedeke}, A., {et~al.} 2019, \apj, 882, 103

\bibitem[{{Pineda} {et~al.}(2022){Pineda}, {Arzoumanian}, {Andr{\'e}},
  {Friesen}, {Zavagno}, {Clarke}, {Inoue}, {Chen}, {Lee}, {Soler}, \&
  {Kuffmeier}}]{Pineda2022}
{Pineda}, J.~E., {Arzoumanian}, D., {Andr{\'e}}, P., {et~al.} 2022, arXiv
  e-prints, arXiv:2205.03935

\bibitem[{{Punanova} {et~al.}(2018){Punanova}, {Caselli}, {Pineda}, {Pon},
  {Tafalla}, {Hacar}, \& {Bizzocchi}}]{Punanova2018}
{Punanova}, A., {Caselli}, P., {Pineda}, J.~E., {et~al.} 2018, \aap, 617, A27

\bibitem[{{Roy} {et~al.}(2015){Roy}, {Andr{\'e}}, {Arzoumanian}, {Peretto},
  {Palmeirim}, {K{\"o}nyves}, {Schneider}, {Benedettini}, {Di Francesco},
  {Elia}, {Hill}, {Ladjelate}, {Louvet}, {Motte}, {Pezzuto}, {Schisano},
  {Shimajiri}, {Spinoglio}, {Ward-Thompson}, \& {White}}]{Roy2015}
{Roy}, A., {Andr{\'e}}, P., {Arzoumanian}, D., {et~al.} 2015, \aap, 584, A111

\bibitem[{{Roy} {et~al.}(2019){Roy}, {Andr{\'e}}, {Arzoumanian},
  {Miville-Desch{\^e}nes}, {K{\"o}nyves}, {Schneider}, {Pezzuto}, {Palmeirim},
  \& {Kirk}}]{Roy2019}
---. 2019, \aap, 626, A76

\bibitem[{{Saigo} \& {Hanawa}(1998)}]{Saigo1998}
{Saigo}, K., \& {Hanawa}, T. 1998, \apj, 493, 342

\bibitem[{{Sakai} {et~al.}(2019){Sakai}, {Hanawa}, {Zhang}, {Higuchi},
  {Ohashi}, {Oya}, \& {Yamamoto}}]{Sakai2019}
{Sakai}, N., {Hanawa}, T., {Zhang}, Y., {et~al.} 2019, \nat, 565, 206

\bibitem[{{Stephens} {et~al.}(2017){Stephens}, {Dunham}, {Myers}, {Pokhrel},
  {Sadavoy}, {Vorobyov}, {Tobin}, {Pineda}, {Offner}, {Lee}, {Kristensen},
  {J{\o}rgensen}, {Goodman}, {Bourke}, {Arce}, \& {Plunkett}}]{Stephens2017}
{Stephens}, I.~W., {Dunham}, M.~M., {Myers}, P.~C., {et~al.} 2017, \apj, 846,
  16

\bibitem[{{Stod{\'o}lkiewicz}(1963)}]{Stodolkiewicz1963}
{Stod{\'o}lkiewicz}, J.~S. 1963, \actaa, 13, 30

\bibitem[{{Tafalla} \& {Hacar}(2015)}]{Tafalla_2015}
{Tafalla}, M., \& {Hacar}, A. 2015, \aap, 574, A104

\bibitem[{{Tatematsu} {et~al.}(2016){Tatematsu}, {Ohashi}, {Sanhueza}, {Nguyen
  Luong}, {Umemoto}, \& {Mizuno}}]{Tatematsu2016}
{Tatematsu}, K., {Ohashi}, S., {Sanhueza}, P., {et~al.} 2016, \pasj, 68, 24

\bibitem[{{Tomisaka}(2000)}]{Tomisaka2000}
{Tomisaka}, K. 2000, \apjl, 528, L41

\bibitem[{{Tomisaka}(2002)}]{Tomisaka2002}
---. 2002, \apj, 575, 306

\bibitem[{{Ward-Thompson} {et~al.}(1994){Ward-Thompson}, {Scott}, {Hills}, \&
  {Andre}}]{Ward-Thompson1994}
{Ward-Thompson}, D., {Scott}, P.~F., {Hills}, R.~E., \& {Andre}, P. 1994,
  \mnras, 268, 276

\bibitem[{{Xu} {et~al.}(2022){Xu}, {Offner}, {Gutermuth}, \& {Tan}}]{Xu2022}
{Xu}, D., {Offner}, S. S.~R., {Gutermuth}, R., \& {Tan}, J.~C. 2022, arXiv
  e-prints, arXiv:2211.03781

\bibitem[{{Zhang} {et~al.}(2018){Zhang}, {Hartmann}, {Zamora-Avil{\'e}s}, \&
  {Kuznetsova}}]{Zhang2018}
{Zhang}, S., {Hartmann}, L., {Zamora-Avil{\'e}s}, M., \& {Kuznetsova}, A. 2018,
  \mnras, 480, 5495

\end{thebibliography}
\bibliographystyle{apj}

\end{document}